\documentclass[nofootinbib,twocolumn,preprintnumbers]{revtex4-1}
\pdfoutput=1
\usepackage{amsmath,amsthm,amssymb,multirow,psfrag}
\usepackage{epsfig}
\usepackage{color}

\graphicspath{{./Figures/}}

\begin{document}

\def\lsim{\mathrel{\rlap{\lower4pt\hbox{\hskip1pt$\sim$}}
    \raise1pt\hbox{$<$}}}
\def\gsim{\mathrel{\rlap{\lower4pt\hbox{\hskip1pt$\sim$}}
    \raise1pt\hbox{$>$}}}
\newcommand{\vev}[1]{ \left\langle {#1} \right\rangle }
\newcommand{\bra}[1]{ \langle {#1} | }
\newcommand{\ket}[1]{ | {#1} \rangle }
\newcommand{\ev}{ {\rm eV} }
\newcommand{\kev}{{\rm keV}}
\newcommand{\mev}{{\rm MeV}}
\newcommand{\gev}{{\rm GeV}}
\newcommand{\tev}{{\rm TeV}}
\newcommand{\mpl}{$M_{Pl}$}
\newcommand{\mw}{$M_{W}$}
\newcommand{\Ft}{F_{T}}
\newcommand{\Zparity}{\mathbb{Z}_2}
\newcommand{\BLambda}{\boldsymbol{\lambda}}
\newcommand{\be}{\begin{eqnarray}}
\newcommand{\ee}{\end{eqnarray}}
\newcommand{\met}{\;\not\!\!\!{E}_T}

\title{{\bf Extracting Effective Higgs Couplings in the Golden Channel}}

\author{Yi Chen}
\email{yi.chen@cern.ch}
\affiliation{Lauritsen Laboratory of Physics, California Institute of Technology, Pasadena, CA 92115, USA}

\author{Roberto Vega-Morales}
\email{roberto.vega@th.u-psud.fr\\robertovegamorales2010@u.northwestern.edu}
\affiliation{Laboratoire de Physique Th\'{e}orique d'Orsay, UMR8627-CNRS, Universit\'{e} Paris-Sud, Orsay, France}
\affiliation{Theoretical Physics Department, Fermilab, P.O. Box 500, Batavia, IL 60510, USA}
\affiliation{Department of Physics and Astronomy, Northwestern University, Evanston, IL 60208, USA}

\date{\today}

\begin{abstract}

Kinematic distributions in Higgs decays to four charged leptons, the so called `golden channel, are a powerful probe of the tensor structure of its couplings to neutral electroweak gauge bosons.~In this study we construct the first part of a comprehensive analysis framework designed to maximize the information contained in this channel in order to perform direct extraction of the various possible Higgs couplings.~To that end we first complete an earlier analytic calculation of the leading order fully differential cross sections for the golden channel signal and background to include the $4e$ and $4\mu$ final states with interference between identical final states.~We also examine the relative fractions of the different possible combinations of scalar-tensor couplings by integrating the fully differential cross section over all kinematic variables as well as show various doubly differential spectra for both the signal and background.~From these analytic expressions we then construct a `generator level' analysis framework based on the maximum likelihood method.~We demonstrate the ability of our framework to perform multi-parameter extractions of all the possible effective couplings of a spin-0 scalar to pairs of neutral electroweak gauge bosons including any correlations.~This framework provides a powerful method for study of these couplings and can be readily adapted to include the relevant detector and systematic effects which we demonstrate in an accompanying study to follow.

\end{abstract}
\preprint{FERMILAB-PUB-13-351-T, LPT-ORSAY 13-73, nuhep-th/13-03, CALT-68-2873}

\maketitle


\section{Introduction}
\label{sec:Intro}
With the recent discovery of the Higgs boson at the LHC~\cite{:2012gk,:2012gu} the focus now shifts to the determination of its detailed properties and in particular whether or not it possesses any anomalous couplings not predicted by the Standard Model.~Ideally, the constraining or measuring of these couplings should be done through direct parameter extraction with minimal theoretical assumptions.~The vast literature~\cite{Nelson:1986ki,Soni:1993jc,Chang:1993jy,Barger:1993wt,Arens:1994wd,Choi:2002jk,Buszello:2002uu,Godbole:2007cn,Kovalchuk:2008zz,Cao:2009ah,Gao:2010qx,DeRujula:2010ys,Gainer:2011xz,Coleppa:2012eh,Bolognesi:2012mm,Stolarski:2012ps,Boughezal:2012tz,Belyaev:2012qa,Avery:2012um,Campbell:2012ct,Campbell:2012cz,Modak:2013sb,Sun:2013yra,Gainer:2013rxa,Anderson:2013fba,Chen:2013waa,Buchalla:2013mpa} on Higgs decays to four charged leptons (electrons and muons) through neutral electroweak gauge bosons, the so called `golden channel', suggests that it can be a powerful channel in accomplishing this goal.

In addition, the high precision with which this channel is measured allows for one of the best opportunities to use analytic methods to analyze data.~As has already been suggested for the golden channel~\cite{Gao:2010qx,Bolognesi:2012mm,Anderson:2013fba} and to be further emphasized here, analytic methods are optimal for performing direct multi-parameter extraction within a minimal amount of computing time.~Furthermore, as we show in an accompanying study~\cite{Chen:2014pia,TechNote}, within an analytic framework one can also include the relevant detector effects in order to obtain a `detector level' likelihood in terms of the full set of observables available in the four lepton final state.~Of course other frameworks have also been recently constructed to study the golden channel (see for example recent Madgraph~\cite{Avery:2012um,Artoisenet:2013puc} or JHU generator~\cite{Anderson:2013fba} based implementations which also include the possibility to study other Higgs decay and production channels).~In this study we construct the first part of a comprehensive analysis framework, based on a largely analytic implementation, designed to maximize the information contained in the golden channel in order to perform direct extraction of the various effective Higgs couplings.

We begin by extending our previous leading order analytic calculations~\cite{Chen:2012jy}, for both the signal and background in the $2e2\mu$ final state, to now also include the $4e$ final state.~We include the interference between identical final states as well as interference between all intermediate states.~Explicitly we calculate for the signal process $\varphi \rightarrow ZZ + Z\gamma + \gamma\gamma \rightarrow 4e/4\mu$ where $\varphi$ is a spin-0 scalar and we have allowed for all possible tensor structures.~This covers all possible couplings of a spin-0 scalar to $ZZ$, $Z\gamma$, or $\gamma\gamma$ pairs.~For the dominant irreducible background we compute $q\bar{q} \rightarrow 4e/4\mu$ including both the t and s-channel process mediated by $Z$ and $\gamma$ vector bosons.~All vector bosons are allowed to be on or off-shell and we do not distinguish between them in what follows.

After presenting the calculation of the analytic fully differential cross sections, we then examine various aspects of the golden channel in more detail.~First, we isolate the individual contributions to the golden channel signal by obtaining the `partial fractions' for each possible combination of tensor structures which can contribution to the $\varphi \rightarrow ZZ + Z\gamma + \gamma\gamma \rightarrow 4\ell$ (where $4\ell = 2e2\mu, 4e, 4\mu$) process.~This is done by integrating the differential cross section over the set of kinematic variables for a given phase space.~These partial fractions give an indication of the relative contributions of each component to the golden channel and a rough picture of the potential sensitivity to the various tensor structures.~As part of this integration we also show a number of doubly differential spectra for signal and background in the Appendix. 

We then construct a maximum likelihood analysis using the analytic expressions of the fully differential cross sections to build the probability density functions (\emph{pdfs}).~This framework builds upon and extends recent studies which first introduce using analytic expressions to perform parameter extraction in the golden channel~\cite{Gao:2010qx,Bolognesi:2012mm,Anderson:2013fba}.~Using these analytic \emph{pdfs}, we study the ability of the golden channel to directly extract the couplings between a spin-0 scalar and $ZZ$, $Z\gamma$, and $\gamma\gamma$ pairs.~We validate our analysis framework by performing a number of simplified `generator level' studies.~To do this we choose an example parameter point in which all possible operators are simultaneously `turned on' in order to demonstrate the validity of our maximization procedure as well as our ability to simultaneously extract the various couplings as well as their correlations. 

Of course a proper treatment of the golden channel requires careful study of detector resolution and acceptance effects.~This also includes an adequate treatment of the production variables for both signal and background as well as taking into account higher order contributions.~We leave these issues to an accompanying paper~\cite{Chen:2014pia} where we construct a `detector level' analysis which includes a treatment of all these issues as well as systematic uncertainties while retaining the flexibility and speed in parameter extraction which we present at `generator level' in this study. 

The organization of this paper is as follows: in Sec.~\ref{sec:events} we briefly review the kinematics of the four lepton final state.~In Sec.~\ref{sec:Signal} we describe the calculation of the signal fully differential cross section while in Sec.~\ref{sec:Background} we describe the calculation of the background fully differential cross section. In Sec.~\ref{sec:scrutinize} we examine the relative fractions of all the possible operators which might contribute to $\varphi \rightarrow ZZ + Z\gamma + \gamma\gamma \rightarrow 4\ell$.~We then present our analysis framework and perform an example parameter extraction to motivate the possibility of extracting the various couplings directly.~We also comment on ongoing and future studies before concluding in Sec.~\ref{sec:conc}.~In the Appendix in Sec.~\ref{sec:Appendix} we also show various 2D projections for both the signal and background in the $4e$ channel as well as the relative fractions for a second set of phase space cuts.

\section{Kinematic Variables}
\label{sec:events}

\begin{figure}
\includegraphics[width=0.48\textwidth]{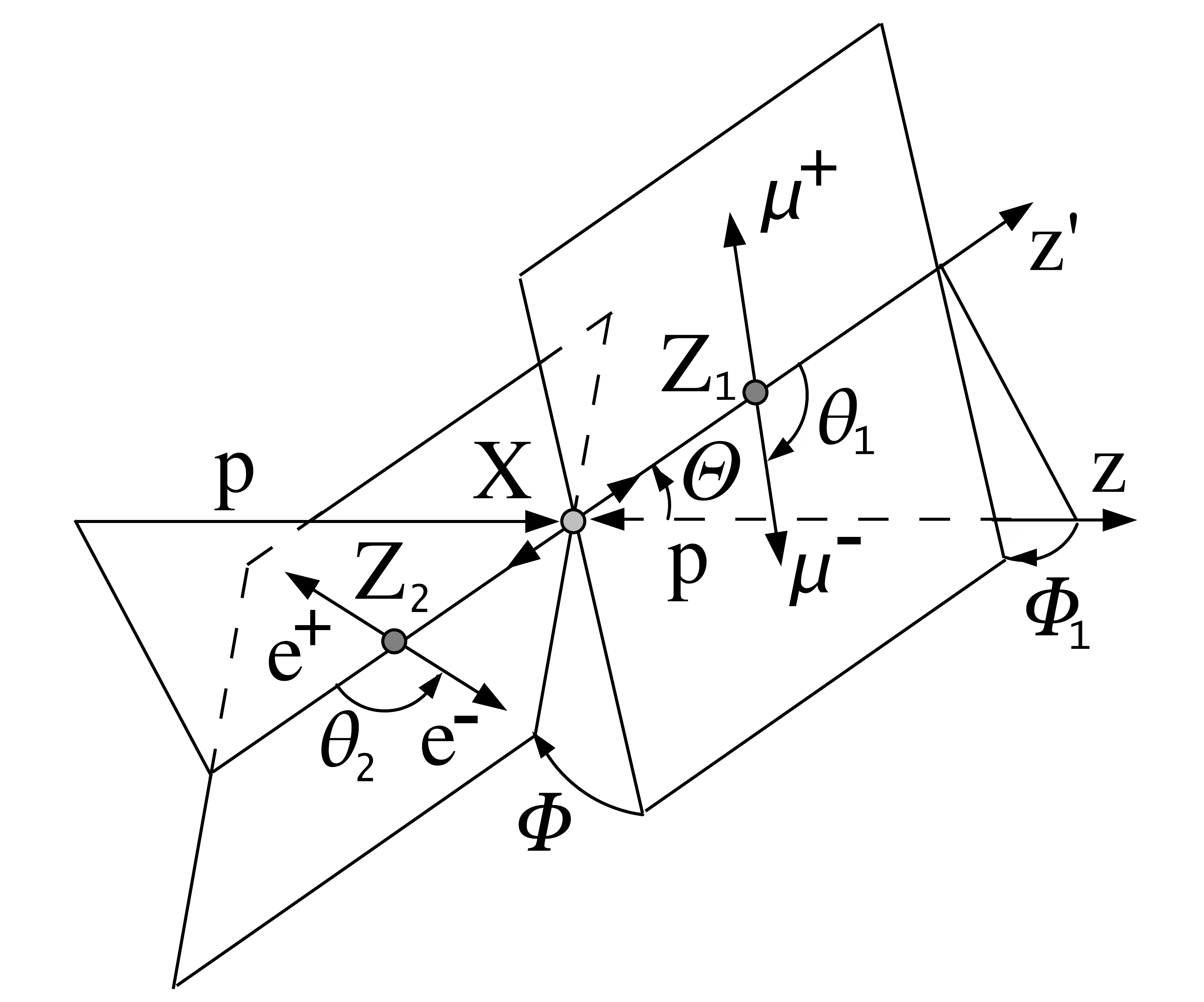}
\caption{Definition of angles in the four lepton CM frame $X$. Here $Z_1$ and $Z_2$ can be either $Z$ or $\gamma$.}
\label{fig:DecayPlanes}
\end{figure}

In this section we briefly discuss the set of observables used to parameterize the $\varphi \rightarrow ZZ + Z\gamma + \gamma\gamma \rightarrow 4\ell$ (where $4\ell = 2e2\mu, 4e, 4\mu$) and $q\bar{q} \rightarrow 4\ell$ fully differential cross sections.~The kinematics of four lepton events are described in detail in~\cite{Gao:2010qx} and are illustrated in Fig.~\ref{fig:DecayPlanes}.~The invariant masses in the system are defined as,
\begin{itemize}
\item $\sqrt{s} \equiv m_\varphi$ -- The invariant mass of the four lepton system or equivalently the Higgs mass for the signal case. 
\item $M_{1}$ -- The invariant mass of the lepton pair system which reconstructs closest to the $Z$ mass. 
\item $M_{2}$ -- The invariant mass of the other lepton pair system. 
\end{itemize}
These variables are all independent subject to the constraint $(M_1 + M_2) \leq \sqrt{s}$.~Note also that the $4e/4\mu$ final state can be reconstructed in two different ways due to the identical final state interference.~This is a quantum mechanical effect that occurs at the amplitude level and thus both reconstructions are valid.~The definitions $M_1$ and $M_2$ remained unchanged however.

The angular variables are defined as,
\begin{itemize}
\item $\Theta$ -- The `production angle' between the momentum vectors of the lepton pair which reconstructs to $M_1$ and the total $4\ell$ system momentum.
\item $\theta_{1,2}$ -- Polar angle of the momentum vectors of $e^-,\mu^-$ in the lepton pair rest frame.
\item $\Phi_1$ -- The angle between the plane formed by the $M_1$ lepton pair and the `production plane' formed out of the momenta of the incoming partons and the momenta of the two lepton pair systems.
\item $\Phi$ -- The angle between the decay planes of the final state lepton pairs in the rest frame of the $4\ell$ system.
\end{itemize}
We have ignored the off-set angle $\phi$, defining a global rotation of the event which is `flat' and thus not shown. We can group the angular variables as follows $\vec{\Omega} = (\Theta, \cos\theta_1, \cos\theta_2, \Phi_1, \Phi)$. 

There are also in principal the `production' variables associated with the initial partonic state four momentum.~This four momentum defines the invariant mass of the CM frame ($\sqrt{s}$), as well as the rapidity ($Y$) defined as the motion along the longitudinal direction, and the momentum in the transverse direction ($\vec{p}_T$).~In principal the inclusion of $Y$ and $\vec{p}_T$ as observables would increase the discriminating power of the golden channel, but as we are interested primarily in parameter extraction and these variables introduce additional systematic uncertainties we do not consider them in our set of observables.~When including detector effects, however, these production variables must be properly accounted for as we will do in~\cite{Chen:2014pia}.

\section{Signal}
\label{sec:Signal}
In this section we present the calculation of the signal fully differential cross section for the process $\varphi \rightarrow ZZ + Z\gamma + \gamma\gamma \rightarrow 4e/4\mu$.~We take $\varphi$ to be a general spin-0 scalar and consider all possible couplings to any combination of $Z$ and $\gamma$ pairs allowing for mixtures of both CP even and odd interactions.~We follow closely, with a slight variation in strategy and notation, the method used in~\cite{Chen:2012jy} for the calculation of the $2e2\mu$ final state and refer the reader there for many of the details.~Here the only additional calculation needed is that for the identical final state interference in the $4e/4\mu$ channels.~Various validations of the calculation can be found in the Appendix as well as~\cite{Chen:2012jy},\cite{Chen:2014pia}, and~\cite{TechNote}.

\subsection{Parametrization of Scalar-Tensor Couplings}
\label{subsec:Tensor_Couplings}

The general couplings of a scalar $\varphi$ to $ZZ, Z\gamma$ or $\gamma\gamma$ pairs can be parametrized by the following vertex\footnote{Note that the $A_2$ Lorentz structure differs by an overall sign from the definition found in~\cite{Chen:2012jy}.}, 
\begin{eqnarray}
\label{eqn:vertex}
\Gamma_{ij}^{\mu\nu}(k, k^\prime) &=& \frac{i}{v} 
\Big(A_{1ij} m_Z^2 g^{\mu\nu} + 
A_{2ij} \left(k^\nu k^{\prime \mu}  -  k\cdot k^\prime g^{\mu\nu}\right) \nonumber \\
&+& A_{3ij}  \epsilon^{\mu\nu\alpha\beta} k_{\alpha} k^\prime_{\beta} \Big),
\end{eqnarray}
where $ij=ZZ,Z\gamma$, or $\gamma\gamma$ and $k$ and $k^\prime$ represent the four momentum of the intermediate vector bosons with $v$ the Higgs vacuum expectation value (vev) which we have chosen as our overall normalization.~The $A_{1ij,2ij,3ij}$ are dimensionless arbitrary complex (momentum dependent) form factors.~For the purposes of this study however, we will approximate the couplings as constant as is done in other similar analysis~\cite{Gao:2010qx,Bolognesi:2012mm,Anderson:2013fba,Avery:2012um,Artoisenet:2013puc} though our framework can easily be made to include the full momentum dependence of the form factors.~For the case of a scalar coupling to $Z\gamma$ or $\gamma\gamma$, electromagnetic gauge invariance requires $A_1 = 0$, while for $ZZ$ it can be generated at tree level as in the SM or by higher dimensional operators.

We can also write Eq.(\ref{eqn:vertex}) as,
\begin{eqnarray}
\label{eqn:sum_vertex}
\Gamma_{ij}^{\mu\nu}(k, k^\prime) &=& \left(A_{1ij} V_{1}^{\mu\nu} + A_{2ij} V_{2}^{\mu\nu} +A_{3ij} V_{3}^{\mu\nu} \right) \nonumber \\
&=& \sum\limits_{n}^3 A_{nij} V_n^{\mu\nu} = \sum\limits_{n}^3 \Gamma_{ijn}^{\mu\nu},
\end{eqnarray}
where the coefficients $A_{nij}$ and Lorentz structure $V_n^{\mu\nu} $ are those found in Eq.(\ref{eqn:vertex}).~Although it is more general, the parametrization in Eq.(\ref{eqn:vertex}) can for example be mapped onto the Lagrangian\footnote{This vertex has been implemented into the FeynRules/Madgraph~\cite{Christensen:2008py,Alwall:2007st} framework for validation purposes.} given by,
\begin{eqnarray}
\label{eqn:siglag}
\mathcal{L} &\supset& \frac{1}{4v} 
\varphi \Big(2 g_h m_Z^2Z^\mu Z_\mu + g_Z Z^{\mu\nu}Z_{\mu\nu} + \tilde{g}_Z Z^{\mu\nu} \widetilde{Z}_{\mu\nu} \nonumber \\
&& +~2g_{Z\gamma} F^{\mu\nu}Z_{\mu\nu} +2\tilde{g}_{Z\gamma} F^{\mu\nu} \widetilde{Z}_{\mu\nu} \nonumber \\
&& +~g_{\gamma} F^{\mu\nu}F_{\mu\nu} + \tilde{g}_{\gamma} F^{\mu\nu} \widetilde{F}_{\mu\nu} \Big),
\end{eqnarray}
where we have allowed only up to dimension five operators and $Z_\mu$ is the $Z$ field while $V_{\mu\nu} = \partial_\mu V_\nu - \partial_\nu V_\mu$ is the usual bosonic field strengths.~The dual field strengths are defined as $\widetilde{V}_{\mu\nu} = \frac{1}{2} \epsilon_{\mu\nu\rho\sigma} V^{\rho \sigma}$.~Thus for this Lagrangian we would have, $A_{1ZZ}\equiv g_h, A_{2ZZ} \equiv g_Z, A_{3ZZ} \equiv \tilde{g}_Z$ and similarly for $Z\gamma$ and $\gamma\gamma$.~This makes Eq.(\ref{eqn:vertex}) a convenient parametrization for fitting to Lagrangian parameters that might be generated in various models at dimension five or less.~If $\varphi$ is purely the Standard Model Higgs, then $A_{1ZZ}=2$, while all other coefficients are taken as approximately zero\footnote{There is potentially non-negligible contributions from $A_{2Z\gamma}$ and $A_{2\gamma\gamma}$ which we discuss briefly in the Appendix.}.~Note also that in this parameterization we have not made any theoretical assumptions about the nature of $\varphi$ such as imposing that the couplings are related by $SU(2)_L \otimes U(1)_Y$ gauge invariance for example.

We note that it is important to include all possible Higgs couplings including the $Z\gamma$ and $\gamma\gamma$ contributions in the signal differential cross section.~This is because since the Higgs appears to be mostly `standard model like'~\cite{Falkowski:2013dza} this means we are primarily searching for small anomalous deviations from the standard model prediction.~Thus when attempting to extract specific couplings we must be sure that one small effect is not being mistaken for another (this is also why it is important to include the interference effects between the identical final state leptons).~This is especially true because as we will see many of the couplings are correlated.~Including all possible couplings and doing a simultaneous fit ensures we minimize the possibility of introducing a bias when attempting to extract these couplings.

We allow for all vertex structures in Eq.~(\ref{eqn:vertex}) to contribute simultaneously including all possible interference effects.~Of course Eq.(\ref{eqn:vertex}) can be mapped onto Lagrangians with dimension greater than five with appropriate translation of the parameters, but we work explicitly with the vertex in Eq.(\ref{eqn:vertex}) and Eq.(\ref{eqn:sum_vertex}) when calculating the fully differential cross section for $\varphi \rightarrow ZZ + Z\gamma + \gamma\gamma \rightarrow 4e/4\mu$.~Below we summarize the details of the calculation which is performed using the Tracer~\cite{Jamin:1993} package in Mathematica~\cite{mathematica} to perform the necessary Dirac algebra.

\subsection{Calculation}
\label{subsec:SigCalc}
\begin{figure*}
~~OF~~~~~~~~~~~~~~~~~~~~~~~~~~~~~~~~~~~~~~~~~~~~~~~~~~~~SF\\
~~~\includegraphics[width=0.3\textwidth]{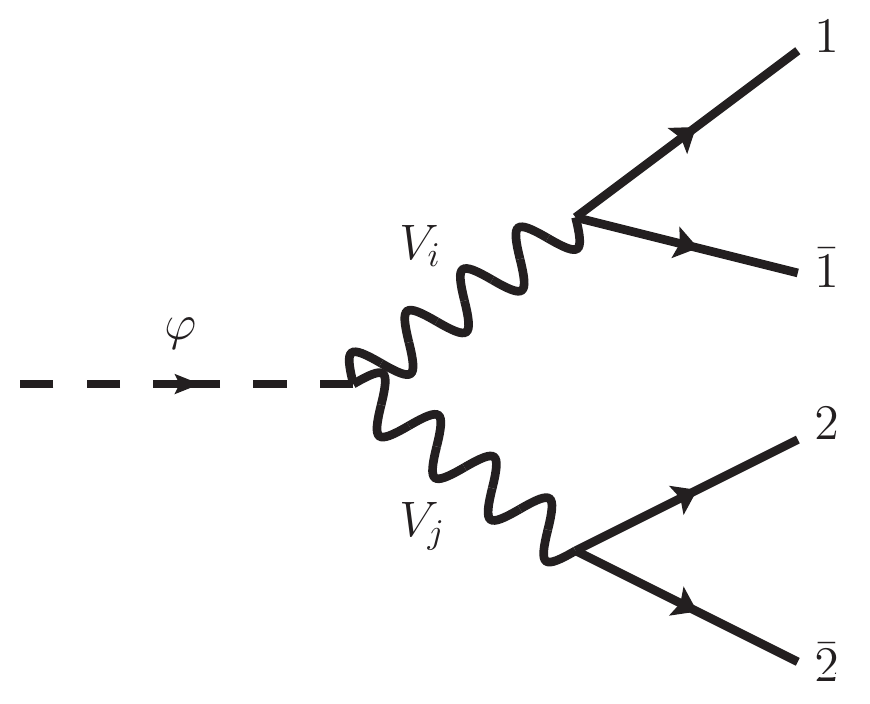}~~~~~
\includegraphics[width=0.3\textwidth]{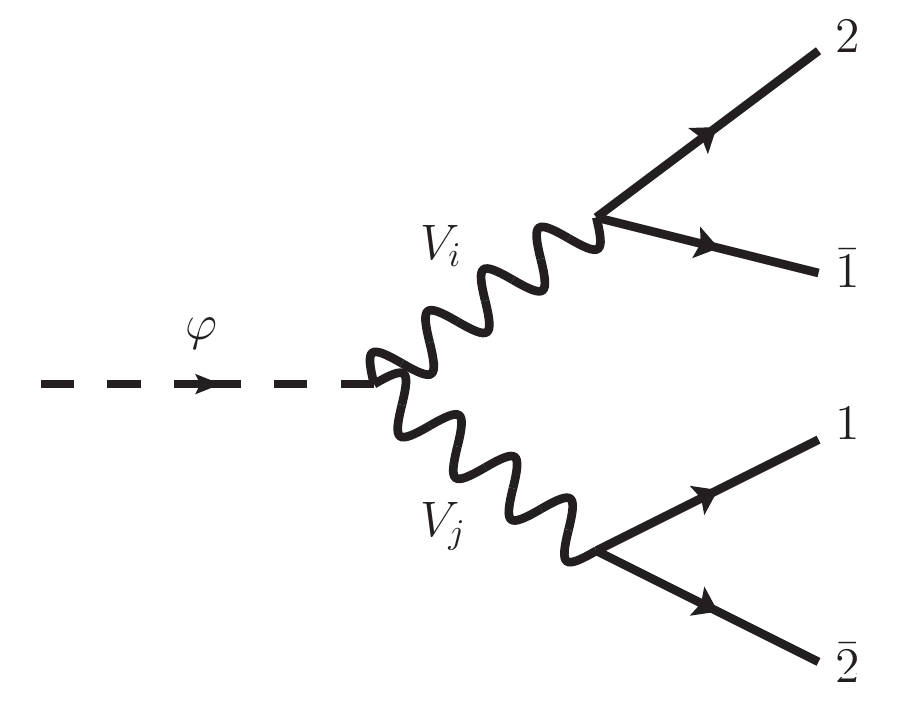}
\caption{Feynman diagrams~\cite{Binosi:2008ig} contributing to $\varphi \rightarrow V_i V_j \rightarrow 4\ell$ where $4\ell = 2e, 4e/4\mu$ and $V_{i,j} = Z, \gamma$.~The arrows are to indicate the direction of momentum flow and $1,2$ label the lepton momenta.~On the left we have the opposite (OF) flavor diagram present in both the $2e2\mu$ and $4e/4\mu$ channels.~On the right we have the same flavor (SF) flavor diagram present only in the $4e/4\mu$ channel.~Note also that the diagram on the right hand side implicitly comes with an overall minus sign to account for the switching of identical fermions ($1 \leftrightarrow 2$).}
\label{fig:sigdiags}
\end{figure*}

To compute the process $\varphi \rightarrow ZZ + Z\gamma + \gamma\gamma \rightarrow 4e/4\mu$ we include the diagrams shown in Fig.~\ref{fig:sigdiags} where $i,j = Z,\gamma$ and parameterize the scalar coupling to gauge bosons as in Eq.(\ref{eqn:sum_vertex}).~For any $ij$ intermediate state, the amplitude ${\mathcal{M}_{OF}}_{ij}$ exists for both the opposite flavor (OF) $2e2\mu$ final state as well as for the same flavor (SF) $4e/4\mu$ final state. The amplitude ${\mathcal{M}_{SF}}_{ij}$ which is obtained by exchanging the four momentum of the particles (or anti-particles), is only present for the $4e/4\mu$ final state.~The total amplitude for any particular intermediate state is the sum of the two diagrams and can be written as,
\begin{eqnarray}
\label{eqn:sigamp}
\mathcal{M}_{ij} &=& {\mathcal{M}_{OF}}_{ij} +{\mathcal{M}_{SF}}_{ij},
\end{eqnarray}
Assuming the final state leptons to be massless, we can write the OF amplitude as,
\begin{widetext}
\begin{eqnarray}
\label{eqn:SigOFAmplitude}
&&{\mathcal{M}_{OF}}_{ij} = 
\bar{u}_2 ( i\gamma^\gamma (g^j_{\ell R} P_R + g^j_{\ell L} P_L) ) v_2
\left( \frac{-ig_{\nu\gamma}}{k_{2\bar{2}}^2 - m_j^2 + im_j\Gamma_j} \right) 
\Gamma_{ij}^{\mu\nu}(k_{1\bar{1}}, k_{2\bar{2}}) \nonumber \\
&&\left( \frac{-ig_{\mu\sigma}}{k_{1\bar{1}}^2 - m_i^2 + im_i\Gamma_i} \right)
\bar{u}_1 ( i\gamma^\sigma (g^i_{\ell R} P_R + g^i_{\ell L} P_L) ) v_1, 
\end{eqnarray}
\end{widetext}
where $i,j$ label $Z$ or $\gamma$ while $1$ and $2$ label the final state leptons and can in principal be $e$ or $\mu$. The vector boson four momenta are given by $k_{xy} = (p_x + p_y)$ where $p_x$ are the four momentum of the final state leptons.~Note that we have also set $k = k_{1\bar{1}}$ and $k^\prime = k_{2\bar{2}}$ in the vertex function $\Gamma^{\mu\nu}_{ij}$.~The SF amplitude can be obtained from the OF amplitude by swapping $u_1 \leftrightarrow u_2$ as well as $p_1 \leftrightarrow p_2$ and can be written as,
\begin{widetext}
\begin{eqnarray}
\label{eqn:SigSFAmplitude}
&&{\mathcal{M}_{SF}}_{ij} = 
(-) \bar{u}_1 ( i\gamma^\gamma (g^j_{\ell R} P_R + g^j_{\ell L} P_L) ) v_2
\left( \frac{-ig_{\nu\gamma}}{k_{1\bar{2}}^2 - m_j^2 + im_j\Gamma_j} \right) 
\Gamma_{ij}^{\mu\nu}(k_{1\bar{2}}, k_{2\bar{1}}) \nonumber \\
&&\left( \frac{-ig_{\mu\sigma}}{k_{2\bar{1}}^2 - m_i^2 + im_i\Gamma_i} \right)
\bar{u}_2  ( i\gamma^\sigma (g^i_{\ell R} P_R + g^i_{\ell L} P_L) ) v_1 ,  
\end{eqnarray}
\end{widetext}
where note an overall minus sign is included to account for the swapping of identical fermions and now $k = k_{1\bar{2}}$ and $k^\prime = k_{2\bar{1}}$ in the vertex $\Gamma^{\mu\nu}_{ij}$.~Upon squaring Eq.(\ref{eqn:sigamp}) this gives for the amplitude squared,
\begin{eqnarray}
\label{eqn:sigamp2}
{\mathcal{M}_{ij}} {\mathcal{M}^\ast_{\bar{i}\bar{j}}} &=& {\mathcal{M}_{OF}}_{ij}{\mathcal{M}^\ast_{OF}}_{\bar{i}\bar{j}} + {\mathcal{M}_{SF}}_{ij}{\mathcal{M}^\ast_{SF}}_{\bar{i}\bar{j}} \nonumber \\
&+& 2Re\left({\mathcal{M}_{OF}}_{ij} {\mathcal{M}^\ast_{SF}}_{\bar{i}\bar{j}} \right).
\end{eqnarray}
The ${\mathcal{M}_{OF}}_{ij}{\mathcal{M}^\ast_{OF}}_{\bar{i}\bar{j}}$ term is equivalent to the $2e2\mu$ matrix element squared which was calculated in~\cite{Chen:2012jy}.~We repeat this part of the calculation here for clarity and consistency of notation.~After summing over final state lepton polarizations we can obtain a general amplitude squared which encompasses any combination of intermediate states and is given by,
\begin{eqnarray}
\label{eqn:OFsigamp2}
&& {\mathcal{M}_{OF}}_{ij}{\mathcal{M}^\ast_{OF}}_{\bar{i}\bar{j}} = \nonumber \\
&&(D_{1\bar{1}i} D_{2\bar{2}j} D^\ast_{1\bar{1}\bar{i}} D^\ast_{2\bar{2}\bar{j}})^{-1} 
(g_{\mu\sigma} g_{\nu\gamma} g_{\bar{\mu}\bar{\sigma}} g_{\bar{\nu}\bar{\gamma}})\nonumber \\
&&\mathcal{T}^\ell_{i\bar{i}}(p_1,\sigma, p_{\bar{1}}, {\bar{\sigma}})
\mathcal{T}^\ell_{j\bar{j}}(p_2,\gamma, p_{\bar{2}}, {\bar{\gamma}})
\left( \sum\limits_{n,\bar{n}}^{3}\Gamma_{ijn}^{\mu\nu}\Gamma_{\bar{i}\bar{j}\bar{n}}^{\ast\bar{\mu}\bar{\nu}} \right) ,
\end{eqnarray}
where $\Gamma_{ijn}^{\mu\nu}$ are given in Eq.(\ref{eqn:vertex}) and Eq.(\ref{eqn:sum_vertex}) and we have defined the objects,
\begin{eqnarray}
\label{eqn:trace4}
&&\mathcal{T}^X_{i\bar{i}}(p_a,\sigma, p_{b}, {\bar{\sigma}}) = \nonumber \\
&\frac{1}{2}& (g^i_{X R} g^{\bar{i}}_{X R} + g^i_{X L} g^{\bar{i}}_{X L}) \times
Tr(\not p_a\gamma^\sigma \not p_{b} \gamma^{\bar{\sigma}})+ \nonumber \\
&\frac{1}{2}&(g^i_{X R} g^{\bar{i}}_{X R} - g^i_{X L} g^{\bar{i}}_{X L}) \times
Tr(\not p_a\gamma^\sigma \not p_{b} \gamma^{\bar{\sigma}}\gamma^5), \end{eqnarray}
for the Dirac strings while for the propagators we have,
\begin{eqnarray}
\label{eqn:props}
D_{xyi} =  k_{xy}^2 - m_i^2 + i\Gamma_i m_i.
\end{eqnarray}
The $g^i_{\ell R,L}$ represent the lepton couplings to $Z$ and $\gamma$, but are in fact at this point general left and right handed couplings of a `$Z$-like' spin-1 vector boson to a pair of fermions.~The bars on Lorentz, $i,j$, and $n$ indices are to indicate that the corresponding index belongs to the conjugated amplitude and are distinct indices from the un-bared ones.~We treat all couplings at every vertex encountered when tracing over the Dirac strings as distinct as well as all Breit-Wigner factors so for any amplitude squared term there can in principal be four different vector bosons as intermediate states.~In the case of the photon we have of course $g^\gamma_{\ell R}=g^\gamma_{\ell L}=-e_{em}$ and $m_\gamma=\Gamma_\gamma=0$.

After expanding Eq.(\ref{eqn:OFsigamp2}) we obtain,
\begin{eqnarray}
\label{eqn:OFamp_squared}
&&{\mathcal{M}_{OF}}_{ij}{\mathcal{M}^\ast_{OF}}_{\bar{i}\bar{j}}  \nonumber \\
&=& \sum\limits_{n,\bar{n}}^{3} \Big(
\mathcal{C}^{++}_{n\bar{n}ij\bar{i}\bar{j}} L^{++}_{n\bar{n}} 
+\mathcal{C}^{+-}_{n\bar{n}ij\bar{i}\bar{j}} L^{+-}_{n\bar{n}} \nonumber \\
&+& \mathcal{C}^{-+}_{n\bar{n}ij\bar{i}\bar{j}} L^{-+}_{n\bar{n}}  
+ \mathcal{C}^{--}_{n\bar{n}ij\bar{i}\bar{j}} L^{--}_{n\bar{n}} \Big)  \nonumber \\
&=& \sum\limits_{n,\bar{n}}^{3} \sum\limits_{a,b}^2  
\mathcal{C}^{ab}_{n\bar{n}ij\bar{i}\bar{j}} L^{ab}_{n\bar{n}} ,
\end{eqnarray}
where $a,b = (\pm,\pm)$ with $a$ and $b$ corresponding to the fermion pairs labeled $1$ and $2$ respectively in the OF diagram of Fig.~\ref{fig:sigdiags} and have defined,
\begin{eqnarray}
\label{eqn:OFgaugestruc}
\mathcal{C}^{\pm\pm}_{n\bar{n}ij\bar{i}\bar{j}} &=& 
\frac{1}{4} A_{nij} A_{\bar{n}\bar{i}\bar{j}}^\ast (D_{1\bar{1}i} D_{2\bar{2}j} D^\ast_{1\bar{1}\bar{i}} D^\ast_{2\bar{2}\bar{j}})^{-1} \\
&\times&  (g^i_{\ell R} g^{\bar{i}}_{\ell R} \pm g^i_{\ell L} g^{\bar{i}}_{\ell L})(g^j_{\ell R} g^{\bar{j}}_{\ell R} \pm g^j_{\ell L} g^{\bar{j}}_{\ell L}) \nonumber \\
\nonumber \\
L^{\pm\pm}_{n\bar{n}} &=& (g_{\mu\sigma} T_{1\pm}^{\sigma\bar{\sigma}} g_{\bar{\mu}\bar{\sigma}})(g_{\nu\gamma} T_{2\pm}^{\gamma\bar{\gamma}} g_{\bar{\nu}\bar{\gamma}})V_{n}^{\mu\nu} V_{\bar{n}}^{\ast\bar{\mu}\bar{\nu}}. \nonumber
\end{eqnarray} 
The $T^{\sigma\bar{\sigma}}_{1\pm}$ are the Dirac traces found in Eq.(\ref{eqn:trace4}) and $\pm$ indicates whether the trace ends with a $\gamma^5$~($-$) or not~($+$).~From the objects in Eq.(\ref{eqn:OFgaugestruc}) we can go on to obtain the full amplitude squared for the $2e2\mu$ channel as done in~\cite{Chen:2012jy}.

For the $4e/4\mu$ final state we also have the second squared term ${\mathcal{M}_{SF}}_{ij}{\mathcal{M}^\ast_{SF}}_{\bar{i}\bar{j}}$, but this is obtained easily from ${\mathcal{M}_{OF}}_{ij}{\mathcal{M}^\ast_{OF}}_{\bar{i}\bar{j}}$ by swapping $p_1 \leftrightarrow p_2$.~Thus the only new term left to calculate in the $4e/4\mu$ case is the interference term ${\mathcal{M}_{OF}}_{ij}{\mathcal{M}^\ast_{SF}}_{\bar{i}\bar{j}}$.~Note also that the amplitudes in the $4e/4\mu$ case come with a symmetry factor of 1/2 for the identical final states, which we explicitly add at a later step.~After squaring the amplitude we find for the interference term,
\begin{eqnarray}
\label{eqn:int_term}
&&{\mathcal{M}_{OF}}_{ij}{\mathcal{M}^\ast_{SF}}_{\bar{i}\bar{j}} = \nonumber \\
&&- (D_{1\bar{1}i} D_{2\bar{2}j} D^\ast_{2\bar{1}\bar{i}} D^\ast_{1\bar{2}\bar{j}})^{-1} 
(g_{\mu\sigma} g_{\nu\gamma}g_{\bar{\nu}\bar{\gamma}} g_{\bar{\mu}\bar{\sigma}}) \nonumber \\
&&\mathcal{T}^{\ell}_{ij\bar{i}\bar{j}}(p_2,\gamma,p_{\bar{2}},{\bar{\gamma}},p_1,\sigma,p_{\bar{1}},{\bar{\sigma}}) 
\left( \sum\limits_{n,\bar{n}}^{3} \Gamma_{ijn}^{\mu\nu}\Gamma_{\bar{i}\bar{j}\bar{n}}^{\ast\bar{\mu}\bar{\nu}} \right),
\end{eqnarray}
where $\Gamma_{ijn}^{\mu\nu}$ are given in Eq.(\ref{eqn:vertex}) and Eq.(\ref{eqn:sum_vertex}) and we have defined,
\begin{eqnarray}
\label{eqn:trace8}
&& \mathcal{T}_{ij\bar{i}\bar{j}}^{X}(p_a,\gamma,p_b,\bar{\gamma},p_c,\sigma,p_d,\bar{\sigma}) = \nonumber \\ 
&\frac{1}{2}& (g^i_{X R}g^j_{X R}g^{\bar{i}}_{X R}g^{\bar{j}}_{X R} + g^i_{X L}g^j_{X L} g^{\bar{i}}_{X L}g^{\bar{j}}_{X L}) \nonumber \\
&\times& Tr(\not{p}_a \gamma^\gamma \not{p}_b \gamma^{\bar{\gamma}} \not{p}_c \gamma^{\sigma} \not{p}_d \gamma^{\bar{\sigma}}) + \nonumber \\ 
&\frac{1}{2}& (g^i_{X R}g^j_{X R}g^{\bar{i}}_{X R}g^{\bar{j}}_{X R} - g^i_{X L}g^j_{X L} g^{\bar{i}}_{X L}g^{\bar{j}}_{X L})  \nonumber \\
&\times& Tr(\not{p}_a \gamma^\gamma \not{p}_b \gamma^{\bar{\gamma}} \not{p}_c \gamma^{\sigma} \not{p}_d \gamma^{\bar{\sigma}}\gamma^5). 
\end{eqnarray}
Expanding out the terms in Eq.(\ref{eqn:int_term}) we can write the interference term as,
\begin{eqnarray}
\label{eqn:gensigamp2exp}
&&{\mathcal{M}_{OF}}_{ij}{\mathcal{M}^\ast_{SF}}_{\bar{i}\bar{j}}  \nonumber \\
&=& \sum\limits_{n,\bar{n}}^{3} \Big(
\mathcal{C}^{+}_{n\bar{n}ij\bar{i}\bar{j}} L^{+}_{n\bar{n}} +  
\mathcal{C}^{-}_{n\bar{n}ij\bar{i}\bar{j}} L^{-}_{n\bar{n}} \Big) \nonumber \\
&=& \sum\limits_{n,\bar{n}}^{3} \sum\limits_{a}^{2} 
\mathcal{C}^{a}_{n\bar{n}ij\bar{i}\bar{j}} L^{a}_{n\bar{n}}, 
\end{eqnarray}
where the coefficients and Lorentz structure are now,
\begin{eqnarray}
\label{eqn:SFgaugestruc}
\mathcal{C}^{\pm}_{n\bar{n}ij\bar{i}\bar{j}} &=& 
\frac{1}{2} A_{nij} A_{\bar{n}\bar{i}\bar{j}}^\ast 
(D_{1\bar{1}i} D_{2\bar{2}j} D^\ast_{2\bar{1}\bar{i}} D^\ast_{1\bar{2}\bar{j}})^{-1} \nonumber \\
&\times& (g^i_{\ell R} g^{\bar{i}}_{\ell R}g^j_{\ell R} g^{\bar{j}}_{\ell R} \pm g^i_{\ell L} g^{\bar{i}}_{\ell L} g^j_{\ell L} g^{\bar{j}}_{\ell L}) \nonumber \\
\nonumber \\
L^{\pm}_{n\bar{n}} &=& - (g_{\mu\sigma} g_{\bar{\mu}\bar{\sigma}} T^{\gamma\bar{\gamma}\sigma\bar{\sigma}}_{12\pm} g_{\nu\gamma} g_{\bar{\nu}\bar{\gamma}})
V_{n}^{\mu\nu}V_{\bar{n}}^{\ast\bar{\mu}\bar{\nu}}.
\end{eqnarray}
The $T^{\gamma\bar{\gamma}\sigma\bar{\sigma}}_{12\pm}$ are the Dirac traces found in Eq.(\ref{eqn:trace8}) and again $\pm$ indicates whether the trace ends with a $\gamma^5$~($-$) or not~($+$).~Note that again the vector boson momentum in $V_{n}^{\mu\nu}$ of Eq.(\ref{eqn:SFgaugestruc}) is given by $k = k_{1\bar{1}}$ and $k^\prime = k_{2\bar{2}}$, but now in $V_{\bar{n}}^{\ast\bar{\mu}\bar{\nu}}$ we have $k = k_{1\bar{2}}$ and $k^\prime = k_{2\bar{1}}$. 

We can now take advantage of the fact that $L^{\pm\pm}_{n\bar{n}}$ and $L^{\pm}_{n\bar{n}}$ are independent of the intermediate state vector bosons to perform the sum over $i,j = Z, \gamma$ and obtain general coefficients for the Lorentz structure which include all contributions from $Z$ and $\gamma$ gauge bosons,
\begin{eqnarray}
\label{eqn:gen_coeffs}
\mathcal{C}^{\pm\pm}_{n\bar{n}} =
\sum\limits_{ij\bar{i}\bar{j}} \mathcal{C}^{\pm\pm}_{n\bar{n}ij\bar{i}\bar{j}},~
\mathcal{C}^{\pm}_{n\bar{n}} =
\sum\limits_{ij\bar{i}\bar{j}} \mathcal{C}^{\pm}_{n\bar{n}ij\bar{i}\bar{j}}.
\end{eqnarray}
The full amplitude squared for  $\varphi \rightarrow ZZ + Z\gamma + \gamma\gamma \rightarrow 4e/4\mu$ can then be built out of the objects in Eqs.(\ref{eqn:OFgaugestruc}), (\ref{eqn:SFgaugestruc}), and (\ref{eqn:gen_coeffs}) as follows \footnote{Analytic expressions may be obtained by emailing the authors or at a website which is currently under construction~\cite{WEBSITE}.},
\begin{eqnarray}
\label{eqn:amp4e_squared}
&&|\mathcal{M}^\varphi_{4e,4\mu}|^2 = 
(\frac{1}{4}) \sum\limits_{n,\bar{n}}^{3} \sum\limits_{a,b}^2 
\Big(
\mathcal{C}^{ab}_{n\bar{n}} L^{ab}_{n\bar{n}} + \nonumber \\
&& \mathcal{C}^{ab}_{n\bar{n}} L^{ab}_{n\bar{n}} |_{p_1\leftrightarrow p_2}
+ 2 Re (\mathcal{C}^{a}_{n\bar{n}} L^{a}_{n\bar{n}}) 
\Big) ,
\end{eqnarray}
where we have included the 1/4 symmetry factor for the identical final state fermions.~One can also easily obtain the amplitude squared for any combination of vertex structures in Eq.(\ref{eqn:vertex}) by not taking the sum over $n$ and $\bar{n}$ and choosing the desired $n,\bar{n}$ combination.~We will take advantage of this property when performing integration and when we examine the interference effects between different operators below. 

The final fully differential cross section (which is treated at fixed $\sqrt{s}$) can then be obtained via,
\begin{eqnarray}
\label{eqn:final_diffcxn}
\frac{d\sigma_{\varphi\rightarrow 4e/4\mu}}{dM_1^2dM_2^2d\vec{\Omega}} = 
\Pi_{4\ell} |\mathcal{M}^\varphi_{4e/4\mu}|^2~,
\end{eqnarray}
where $d\vec{\Omega}=dc_\Theta dc_{\theta_1} dc_{\theta_2} d\Phi d\Phi_1$ ($c_\theta = \cos{\theta}$) and $ \Pi_{4\ell}$ is the final state massless lepton four body phase space derived following~\cite{Nakamura:2010zzi} and given by,
\begin{eqnarray}
\label{eqn:phasespace}
\Pi_{4\ell} &=&
(\frac{1}{2 \pi})^2 (\frac{1}{32 \pi^2})^2 (\frac{1}{32 \pi s}) \nonumber \\
&\times&\Big(1 + \frac{(M_1^2 - M_2^2)^2}{s^2} - \frac{2(M_1^2 + M_2^2)}{s}\Big)^{1/2}.
\end{eqnarray}
Unlike the $2e2\mu$ final state, the coefficients $\mathcal{C}^a_{n\bar{n}}$ in the interference term of Eq.(\ref{eqn:amp4e_squared}) depend on the polar angles $\cos\theta_{1,2}$ and in particular through the denominators of the vector boson propagators (see Eq.(\ref{eqn:SigSFAmplitude})).~This makes analytic integration difficult.~Thus analytic expressions for the doubly differential mass spectra are not obtained in the $4e/4\mu$ channel as they were for $2e2\mu$~\cite{Chen:2012jy}.~In Fig.~\ref{fig:sig_doubdists} of the Appendix we show plots for the differential mass spectra after performing the angular integration numerically as well as various other doubly differential distributions for the SM signal hypotheses.~Again, details of the validation procedure can be found in~\cite{Chen:2012jy,Chen:2014pia,TechNote}.


\section{Background}
\label{sec:Background}

The dominant irreducible background to the golden channel comes from $q\bar{q}$ annihilation into gauge bosons.  At energies $\sim 125~GeV$ the dominant contribution comes from t-channel $Z\gamma$ production~\cite{Chen:2012jy}.~However, as was seen in~\cite{Chen:2012jy} contributions from s-channel process diagrams can effect the angular distributions, such as the distribution of the azimuthal angle between the lepton decay planes $\Phi$ defined in Sec.\ref{sec:events}.~Furthermore, we include the $ZZ$ and $\gamma\gamma$ contributions since in principal these are always present and may have observable interference effects due to the fact that they add at the amplitude level when decaying to charged leptons and can mimic some of the effects of the signal tensor structures.~Of course higher order effects, including the $gg$ initiated process~\cite{Zecher:1994kb, Binoth:2008pr, Kauer:2012hd} will contribute as well, but these are expected to be sub-dominant and mainly only effect the `input' invariant mass (and overall normalization) for the fully differential cross sections.~Since we are not including production variables in our set of observables and are not concerned with the overall normalization, neglecting these contributions has a minimal effect on our analysis framework, but as mentioned previously should properly be taken into account when including detector effects.

In this section we extend a previous calculation of the $2e2\mu$ channel to include the $4e/4\mu$ final state.~The calculation follows in the same manner as for $2e2\mu$ (with some slight changes in notation) except that now one must include the contribution from interference between the final state identical particles, which in some kinematic regimes can have non-negligible effects~\cite{Zecher:1994kb,Avery:2012um}.~In this section we describe the calculation of this interference, while the parts of the calculation which are identical to the $2e2\mu$ case can be found in~\cite{Chen:2012jy}.~Again we use the Tracer~\cite{Jamin:1993} package in Mathematica~\cite{mathematica} to perform the necessary algebra.


\subsection{Calculation}
\label{subsec:BackCalc}

The background calculation is significantly more involved than the signal calculation due to a much larger number of Feynman diagrams (48 in total as opposed to 8 for signal) in addition to a more complicated Lorentz structure.~As in the signal case the amplitude can be written as sum of opposite flavor (OF) amplitude and a same flavor (SF) amplitude.~Thus the amplitude squared can again be written as,
\begin{eqnarray}
\label{eqn:bgamp2}
{\mathcal{M}_{ij}} {\mathcal{M}^\ast_{\bar{i}\bar{j}}} &=& {\mathcal{M}_{OF}}_{ij}{\mathcal{M}^\ast_{OF}}_{\bar{i}\bar{j}} + {\mathcal{M}_{SF}}_{ij}{\mathcal{M}^\ast_{SF}}_{\bar{i}\bar{j}} \nonumber \\
&+& 2Re\left({\mathcal{M}_{OF}}_{ij} {\mathcal{M}^\ast_{SF}}_{\bar{i}\bar{j}} \right).
\end{eqnarray}
The first term ${\mathcal{M}_{OF}}_{ij}{\mathcal{M}^\ast_{OF}}_{\bar{i}\bar{j}}$ is exactly equivalent to the $2e2\mu$ amplitude squared calculated in~\cite{Chen:2012jy} to which we refer the reader for details. The second term $ {\mathcal{M}_{SF}}_{ij}{\mathcal{M}^\ast_{SF}}_{\bar{i}\bar{j}}$ can be easily computed from the first by the simple exchange $p_1 \leftrightarrow p_2$ as was done in the signal case. Thus the only new term left to calculate is the identical final state interference term ${\mathcal{M}_{OF}}_{ij} {\mathcal{M}^\ast_{SF}}_{\bar{i}\bar{j}}$.

\begin{figure*}
~~~~OF~~~~~~~~~~~~~~~~~~~~~~~~~~~~~~~~~~~~~~~~~~~~~~SF\\
\includegraphics[width=0.3\textwidth]{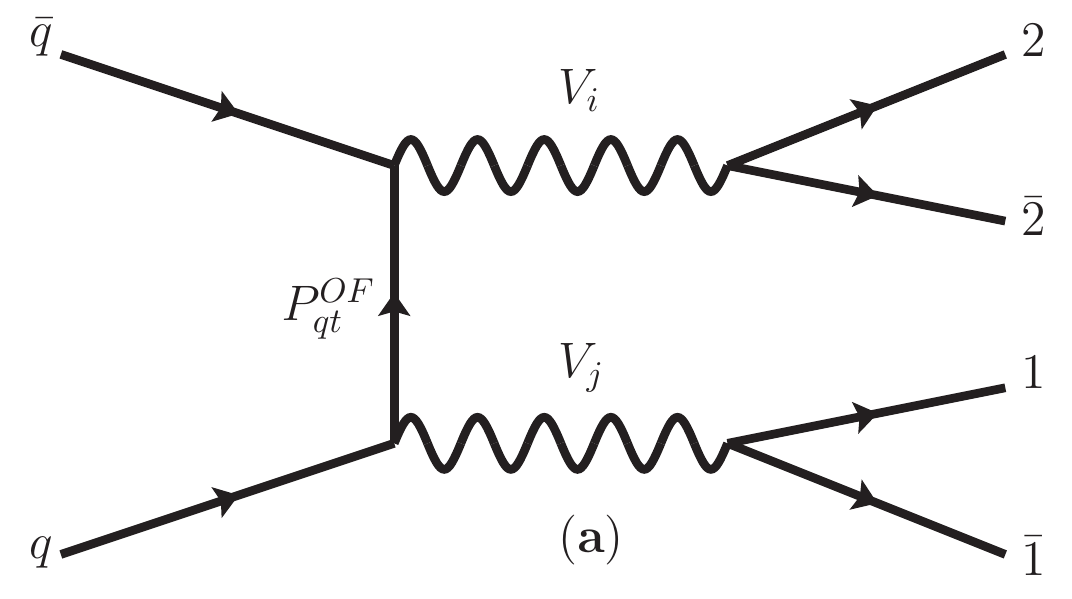}
\includegraphics[width=0.3\textwidth]{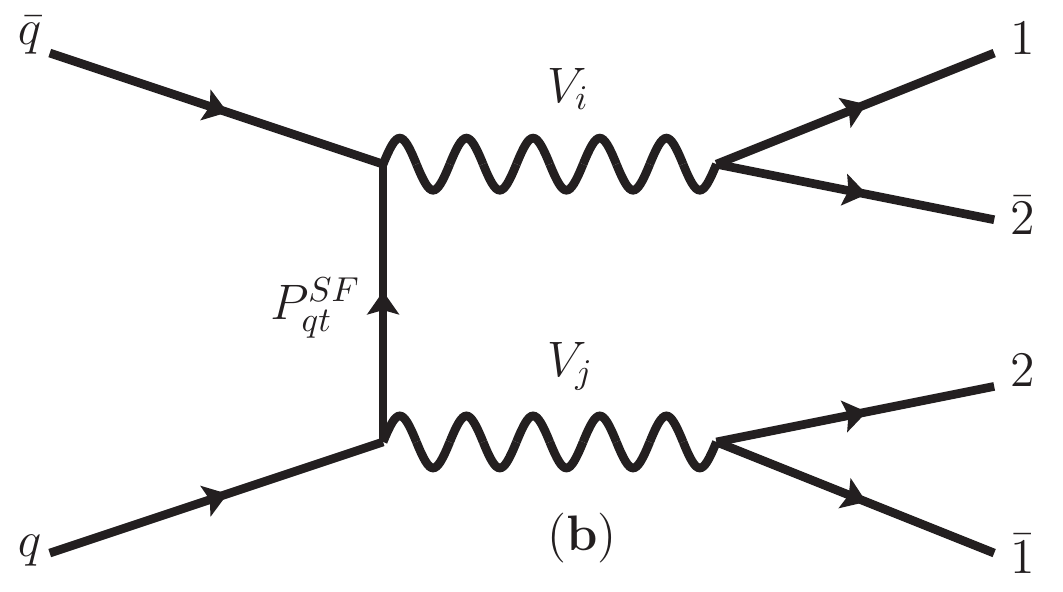}\\
\includegraphics[width=0.3\textwidth]{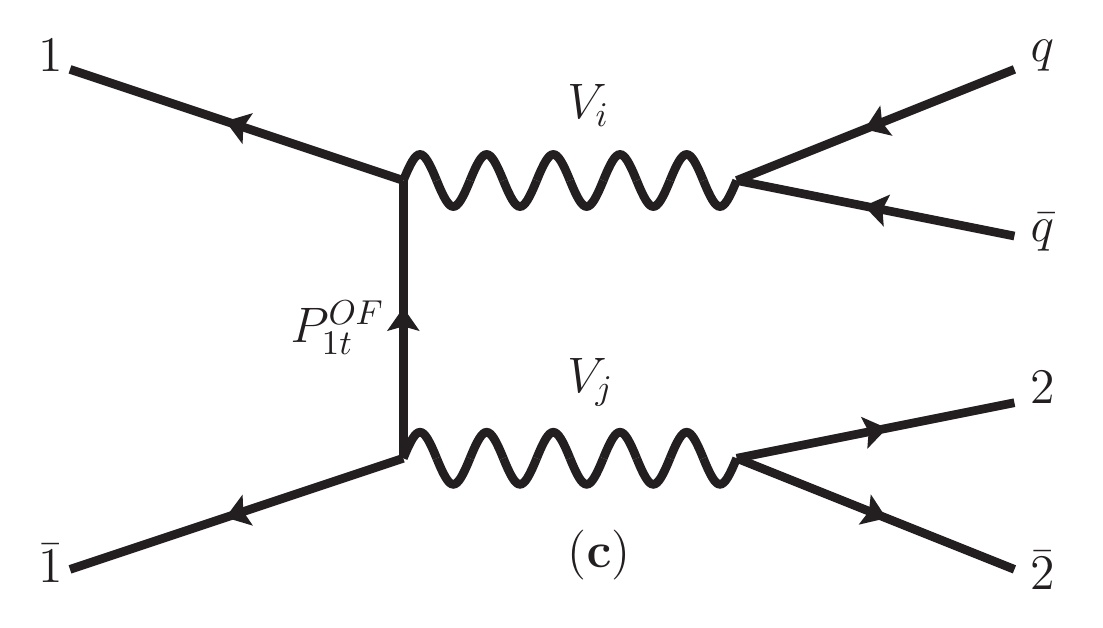}
\includegraphics[width=0.3\textwidth]{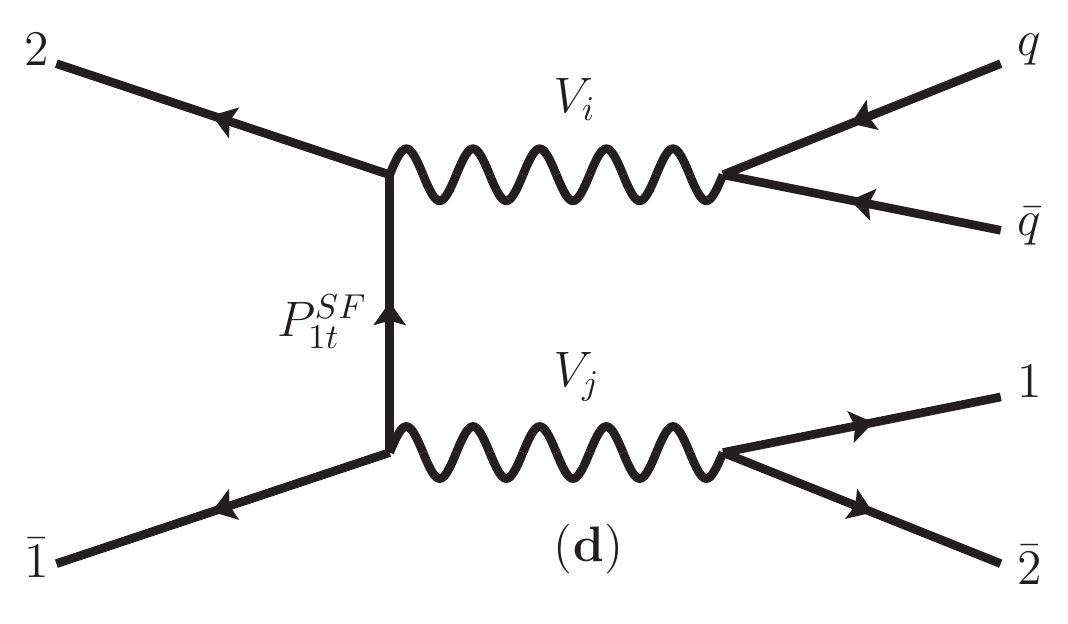}\\
\includegraphics[width=0.3\textwidth]{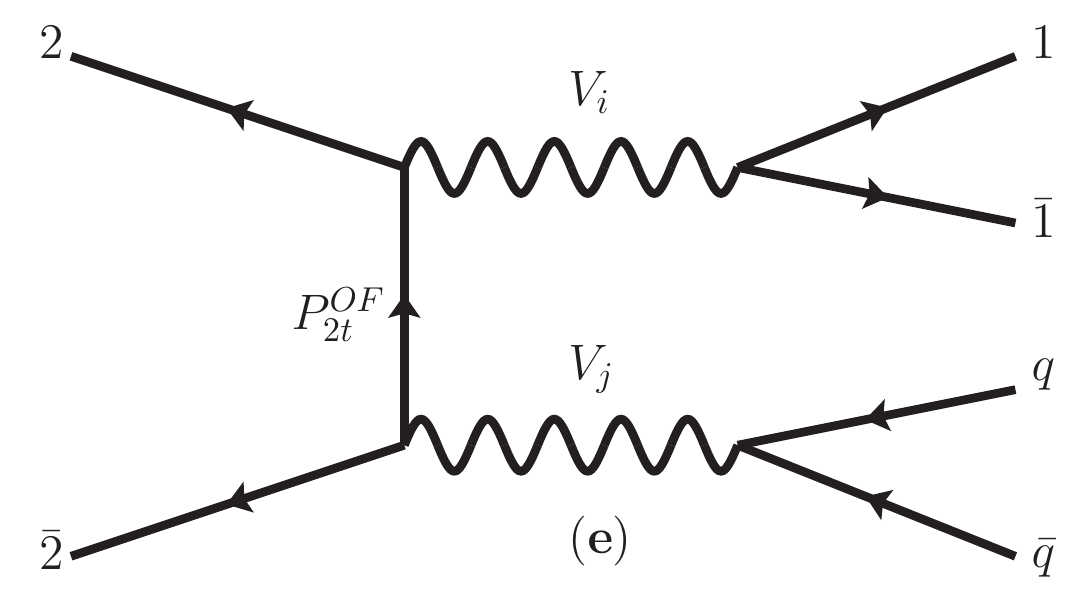}
\includegraphics[width=0.3\textwidth]{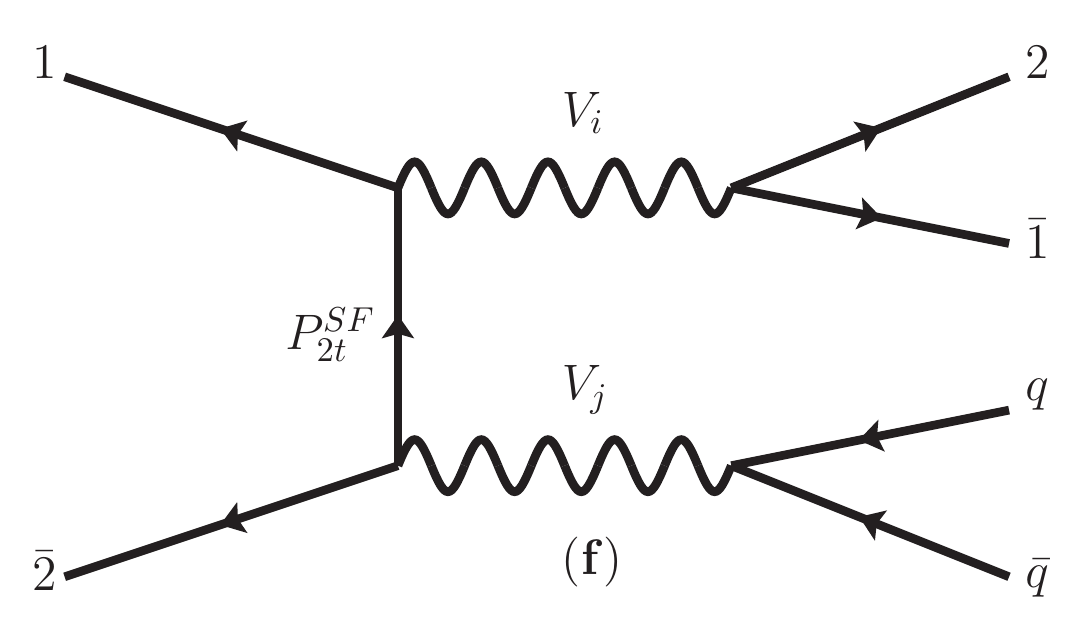}
\caption{Feynman diagrams~\cite{Binosi:2008ig} contributing to $q\bar{q} \rightarrow V_i V_j\rightarrow 4e/4\mu$ and $q\bar{q} \rightarrow V_{i,j} \rightarrow 4e/4\mu$ where $V_{i,j} = Z, \gamma$ and $1,2$ label the lepton momenta.~Note that although we define all diagrams as `t-channel' type, diagrams $(c)-(f)$ are in fact s-channel type in the usual convention so the fermions labeled by $1$ and $2$ are not to be confused as being in the initial state.~This is taken into account in how the various momenta are assigned as indicated by the arrows.~The diagrams on the left hand side are labeled by OF, while those on the right are labeled by SF.~Note also that the diagrams on the right hand side implicitly come with an overall minus sign to account for the switching of identical fermions ($1 \leftrightarrow 2$).}
\label{fig:XYZdiags}
\end{figure*}

Following the strategy in~\cite{Chen:2012jy} we organize the diagrams after `twisting' them into the form shown in Fig.~\ref{fig:XYZdiags} where we allow the intermediate vector bosons to take on any combination of $Z$ or $\gamma$, but once chosen are treated as fixed.~We use the conventions indicated in~\cite{Chen:2012jy} and in particular refer to all of the diagrams in Fig~\ref{fig:XYZdiags} as `t-channel' type diagrams while the `u-channel' diagrams are obtained by switching the vertex at which the vector bosons are attached.~This is not to be confused with the typical vocabulary for this process which refers to diagrams $(a)$ and $(b)$ as t-channel and diagrams $(c)-(f)$ as s-channel.~We find this re-naming convenient for organizing and reducing the many terms which need to be computed for the differential cross section (see~\cite{Chen:2012jy} for a more detailed explanation).~The diagrams on the left are labeled by OF, while those on the right are labeled by SF.~Note also that the diagrams on the right hand side implicitly come with an overall minus sign to account for the switching of identical fermions ($p_1 \leftrightarrow p_2$).

The Lorentz structure for all of these amplitudes is clearly the same.~One needs only to keep proper track of how the various momentum are routed through each diagram.~We can see this by considering the amplitude explicitly.~Using the massless initial quark and final state lepton approximation we can write any of the OF amplitudes on the left hand side in Fig.~\ref{fig:XYZdiags} as,
\begin{widetext}
\begin{eqnarray}
\label{eqn:BgOFAmplitude}
\mathcal{M}^{Xn}_{OFij} &=&
\bar{u}_Z ( i\gamma^\sigma (g_{ZR}^j P_R + g_{ZL}^j P_L) ) v_Z
\left( \frac{-ig_{\mu\sigma}}{k_{Z\bar{Z}}^2 - m_j^2 + im_j\Gamma_j} \right)\nonumber \\
&\times&\bar{v}_X 
( i\gamma^\mu (g_{XR}^i P_R + g_{XL}^i P_L) )
\left( \frac{i\not P^{OF}_{Xn}}{P_{Xn}^{OF2}} \right)
( i\gamma^\nu (g_{XR}^j P_R + g_{XL}^j P_L) )
u_X \nonumber \\
&\times&\left( \frac{-ig_{\nu\gamma}}{k_{Y\bar{Y}}^2 - m_i^2 + im_i\Gamma_i} \right)
\bar{u}_Y ( i\gamma^\gamma (g_{YR}^i P_R + g_{YL}^i P_L) ) v_Y ,
\end{eqnarray}
\end{widetext}
where we label the amplitude by the `long' Dirac string, in this case $X$.~The labels $X/Y/Z = 1,2,q$ where $1,2$ are for final state lepton pairs while $q$ is for the initial state quarks.~The $i,j = Z, \gamma$ label the vector bosons and $n = t, u$ labels the t and u-channel diagrams in our new vocabulary.~The internal vector boson momenta are again defined as $k_{xy} = (p_x + p_y)$, while the internal fermion momentum are given by,
\begin{eqnarray}
\label{eqn:OFfermionmom}
P^{OF}_{qt} &=& p_q - k_{1\bar{1}},~P^{OF}_{qu} = p_q - k_{2\bar{2}}\\
P^{OF}_{1t} &=& - (p_{\bar{1}} + k_{2\bar{2}}),~P^{OF}_{1u} = k_{q\bar{q}} - p_{\bar{1}}\nonumber\\
P^{OF}_{2t} &=& k_{q\bar{q}}- p_{\bar{2}},~P^{OF}_{2u} = - (p_{\bar{2}} + k_{1\bar{1}}).\nonumber
\end{eqnarray}  
For any of the SF amplitudes a similar formula as in Eq.(\ref{eqn:BgOFAmplitude}) applies except we take $p_1 \leftrightarrow p_2$ and multiply by an overall minus sign in the corresponding OF amplitude with the quark string in the same position (this simply corresponds to diagrams in the same row of Fig.~\ref{fig:XYZdiags}).~Thus we have for the SF amplitude,
\begin{eqnarray}
\label{eqn:BgSFAmplitude}
\mathcal{M}^{Xn}_{SFij} = -\mathcal{M}^{Xn}_{OFij} \Big|_{p_1 \leftrightarrow p_2} ,
\end{eqnarray}  
while the internal fermion momentum are now given by,
\begin{eqnarray}
\label{eq:SFfermionmom}
P^{SF}_{qt} &=& p_q - k_{2\bar{1}},~P^{SF}_{qu} = p_q - k_{1\bar{2}}\\
P^{SF}_{1t} &=& - (p_{\bar{1}} + k_{1\bar{2}}),~P^{SF}_{1u} = k_{q\bar{q}} - p_{\bar{1}}\nonumber\\
P^{SF}_{2t} &=& k_{q\bar{q}} - p_{\bar{2}},~P^{SF}_{2u} = - (p_{\bar{2}} + k_{2\bar{1}}).\nonumber
\end{eqnarray}  
To obtain any of the physical amplitudes one simply assigns the appropriate labels to Eq.(\ref{eqn:BgOFAmplitude}) or Eq.(\ref{eqn:BgSFAmplitude}) as well as the appropriate momenta.~Thus for example for diagram $(c)$ we have $X \rightarrow 1$, $Y \rightarrow q$, $Z \rightarrow 2$, and $n \rightarrow t$. To switch from t-channel type to u-channel diagrams one simply takes $t\rightarrow u$ and $\gamma^\sigma \leftrightarrow \gamma^\gamma$ while to obtain the corresponding $SF$ diagram simply take $OF\rightarrow SF$ and $1 \leftrightarrow 2$ and multiply by an overall sign.~Note that for the $Z$ propagators we drop the momentum dependent terms since they do not contribute in the massless lepton approximation.

As in the case of the signal, the next step is to find a generalized amplitude squared for any two of the six diagrams.~Since we are only concerned with obtaining the interference term ${\mathcal{M}_{OF}}_{ij} {\mathcal{M}^\ast_{SF}}_{\bar{i}\bar{j}}$ we need only consider the terms coming from multiplying the amplitudes on the left hand side (OF) with those on the right hand side (SF) in Fig.~\ref{fig:XYZdiags}.~These organize themselves into three distinct types of Lorentz structure.~The first type is found when multiplying the two diagrams in the top row of Fig.~\ref{fig:XYZdiags} (corresponding to t-channel di-boson production in the conventional language).~These give,
\begin{eqnarray}
\label{eqn:genbgampXX}
&&\mathcal{M}^{qn}_{OFij}\mathcal{M}^{q\bar{n}\ast}_{SF\bar{i}\bar{j}} = \nonumber \\
&\times& 
(g_{\mu\sigma} g_{\bar{\mu}\bar{\sigma}} g_{\nu\gamma}  g_{\bar{\nu}\bar{\gamma}})(D_{1\bar{1}i} D_{2\bar{2}j} D^\ast_{2\bar{1}\bar{i}} D^\ast_{1\bar{2}\bar{j}})^{-1} 
\nonumber \\
&\times&\mathcal{T}_{ij\bar{i}\bar{j}}^{q}(p_{\bar{q}},\nu,P^{OF}_{qn},\mu,p_q,\bar{\mu},P^{SF}_{q\bar{n}},\bar{\nu}) \nonumber \\
&\times&\mathcal{T}_{ij\bar{i}\bar{j}}^{\ell}(p_2,\gamma,p_{\bar{2}},\bar{\gamma},p_1,\sigma,p_{\bar{1}},\bar{\sigma}) ,
\end{eqnarray}
where the $D_{xyi}$ are defined in Eq.(\ref{eqn:props}) and the $\mathcal{T}_{ij\bar{i}\bar{j}}^{X}$ are defined in Eq.(\ref{eqn:trace8}).~Again the bars on Lorentz, $i,j$, and $n$ indices are to indicate that the corresponding index belongs to the conjugated amplitude and are distinct indices from the un-bared ones.~Expanding out the terms in Eq.(\ref{eqn:genbgampXX}) we can organize in a manner similar to Eqs.(\ref{eqn:OFamp_squared}) and (\ref{eqn:gensigamp2exp}) writing the amplitude squared as,
\begin{eqnarray}
\label{eq:genbgXXexp}
 \mathcal{M}^{qn}_{OFij} \mathcal{M}^{q\bar{n}\ast}_{SF\bar{i}\bar{j}} = 
 \sum\limits_{a,b} \mathcal{C}^{ab}_{qqij\bar{i}\bar{j}} L^{ab}_{qq n\bar{n}} ,
\end{eqnarray}
where again $a,b = (\pm,\pm)$ with $a$ and $b$ corresponding to the quark and lepton strings and we have defined the Lorentz structure coefficients,
\begin{eqnarray}
\label{eqn:qqLcoeffs}
&&\mathcal{C}^{\pm\pm}_{qq ij\bar{i}\bar{j}} = (4D_{1\bar{1}i} D_{2\bar{2}j} D^\ast_{2\bar{1}\bar{i}} D^\ast_{1\bar{2}\bar{j}})^{-1}  \nonumber \\
&\times&(g^i_{q R}g^j_{q R}g^{\bar{i}}_{q R}g^{\bar{j}}_{q R} \pm g^i_{q L}g^j_{q L} g^{\bar{i}}_{q L}g^{\bar{j}}_{q L}) \nonumber \\
&\times& (g^i_{\ell R}g^j_{\ell R}g^{\bar{i}}_{\ell R}g^{\bar{j}}_{\ell R} \pm g^i_{\ell L}g^j_{\ell L} g^{\bar{i}}_{\ell L}g^{\bar{j}}_{\ell L}), 
\end{eqnarray}
and Lorentz structure,
\begin{eqnarray}
\label{eqn:qqLstruc}
&&L^{\pm\pm}_{qq n\bar{n}} = 
(g_{\mu\sigma} T_{qqn\bar{n}\pm}^{\nu\mu\bar{\mu}\bar{\nu}} g_{\bar{\mu}\bar{\sigma}})
(g_{\nu\gamma} T_{\ell \pm}^{\gamma\bar{\gamma}\sigma\bar{\sigma}} g_{\bar{\nu}\bar{\gamma}}),
\end{eqnarray}
where the $T$ objects are the traces found in Eq.(\ref{eqn:genbgampXX}). 

The next type of Lorentz structure is found for any OF/SF pair of diagrams in $( c)-(f)$ (interference between s-channel diagrams in the usual language).~For those in the same row we can write,
\begin{eqnarray}
\label{eqn:genbgampYY}
&&\mathcal{M}^{Yn}_{OFij}\mathcal{M}^{Y\bar{n}\ast}_{SF\bar{i}\bar{j}} = 
(g_{\mu\sigma} g_{\bar{\mu}\bar{\sigma}} g_{\nu\gamma}  g_{\bar{\nu}\bar{\gamma}})
\nonumber \\
&\times& 
(D_{Z\bar{Z}i} D_{q\bar{q}j} D^\ast_{Y\bar{Z}\bar{i}} D^\ast_{q\bar{q}\bar{j}})^{-1} 
\mathcal{T}_{j\bar{j}}^{q}(p_{\bar{q}},\gamma,p_q,\bar{\gamma}) \\
&\times&\mathcal{T}_{iji\bar{i}\bar{j}\bar{i}}^{\ell}(p_Y,\nu,P^{OF}_{Yn},\mu,p_{\bar{Y}},\bar{\mu},P^{SF}_{Y\bar{n}},\bar{\nu},p_Z,\sigma,p_{\bar{Z}},\bar{\sigma}), \nonumber 
\end{eqnarray}
where here $Y/Z = 1,2$ while the $\mathcal{T}_{j\bar{j}}^{X}$ are defined in Eq.(\ref{eqn:trace4}) and we have also defined,
\begin{eqnarray}
\label{eqn:trace12}
&& \mathcal{T}_{iji\bar{i}\bar{j}\bar{i}}^{X}(p_a,\nu,p_b,\mu,p_c,\bar{\gamma},p_d,\sigma,p_e,\bar{\mu},p_f,\bar{\nu}) = \nonumber \\ 
&\frac{1}{2}& (g^{i2}_{X R}g^j_{X R}g^{\bar{i}2}_{X R}g^{\bar{j}}_{X R} + g^{i2}_{X L}g^j_{X L} g^{\bar{i}2}_{X L}g^{\bar{j}}_{X L}) \nonumber \\
&\times& Tr(\not{p}_a \gamma^\nu \not{p}_b \gamma^{\mu} \not{p}_c \gamma^{\bar{\gamma}} \not{p}_d \gamma^{\sigma},\not{p}_e,\bar{\mu},\not{p}_f,\bar{\nu}) + \nonumber \\ 
&\frac{1}{2}&(g^{i2}_{X R}g^j_{X R}g^{\bar{i}2}_{X R}g^{\bar{j}}_{X R} - g^{i2}_{X L}g^j_{X L} g^{\bar{i}2}_{X L}g^{\bar{j}}_{X L})  \nonumber \\
&\times& Tr(\not{p}_a \gamma^\nu \not{p}_b \gamma^{\mu} \not{p}_c \gamma^{\bar{\gamma}} \not{p}_d \gamma^{\sigma},\not{p}_e,\bar{\mu},\not{p}_f,\bar{\nu},\gamma^5). 
\end{eqnarray}
Expanding out Eq.(\ref{eqn:genbgampYY}) we can write the amplitude squared as,
\begin{eqnarray}
\label{eq:genbgYYexp}
\mathcal{M}^{Yn}_{OFij} \mathcal{M}^{Y\bar{n}\ast}_{SF\bar{i}\bar{j}} = 
\sum\limits_{a,b} \mathcal{C}^{ab}_{YYij\bar{i}\bar{j}} L^{ab}_{YY n\bar{n}} , 
\end{eqnarray}
where again $a,b = (\pm,\pm)$ and we have defined the Lorentz structure coefficients,
\begin{eqnarray}
\label{eqn:YYLcoeffs}
\mathcal{C}^{\pm\pm}_{YY ij\bar{i}\bar{j}} &=& 
(4D_{Z\bar{Z}i} D_{q\bar{q}j} D^\ast_{Y\bar{Z}\bar{i}} D^\ast_{q\bar{q}\bar{j}})^{-1}  \nonumber \\
&\times&(g^j_{q R}g^{\bar{j}}_{q R} \pm g^j_{q L}g^{\bar{j}}_{q L}) \nonumber \\
&\times& (g^{i2}_{\ell R}g^j_{\ell R}g^{\bar{i}2}_{\ell R}g^{\bar{j}}_{\ell R} \pm g^{i2}_{\ell L}g^j_{\ell L} g^{\bar{i}2}_{\ell L}g^{\bar{j}}_{\ell L}), 
\end{eqnarray}
and Lorentz structure,
\begin{eqnarray}
\label{eqn:YYLstruc}
&&L^{\pm\pm}_{YY n\bar{n}} = 
(g_{\mu\sigma} T_{q\pm}^{\gamma\bar{\gamma}} g_{\bar{\mu}\bar{\sigma}})
(g_{\nu\gamma} T_{YYn\bar{n}\pm}^{\nu\mu\bar{\mu}\bar{\nu}\sigma\bar{\sigma}} g_{\bar{\nu}\bar{\gamma}}),
\end{eqnarray}
where the $T$ objects are the traces found in Eq.(\ref{eqn:genbgampYY}).~For products of diagrams in different rows in $(c)-(f)$ we obtain the following,
\begin{eqnarray}
\label{eqn:genbgampYZ}
&&\mathcal{M}^{Yn}_{OFij}\mathcal{M}^{Z\bar{n}\ast}_{SF\bar{i}\bar{j}} = 
(g_{\mu\sigma} g_{\bar{\mu}\bar{\sigma}} g_{\nu\gamma}  g_{\bar{\nu}\bar{\gamma}})
\nonumber \\
&\times& 
(D_{Z\bar{Z}i} D_{q\bar{q}j} D^\ast_{q\bar{q}\bar{i}} D^\ast_{Z\bar{Y}\bar{j}})^{-1} 
\mathcal{T}_{j\bar{j}}^{q}(p_{\bar{q}},\gamma,p_q,\bar{\sigma}) \\
&\times&\mathcal{T}_{iji\bar{i}\bar{j}\bar{i}}^{\ell}(p_Y,\nu,P^{OF}_{Yn},\mu,p_{\bar{Y}},\bar{\gamma},p_Z,\sigma,p_{\bar{Z}},\bar{\mu},P^{SF}_{Z\bar{n}},\bar{\nu}). \nonumber 
\end{eqnarray}
Again expanding out Eq.(\ref{eqn:genbgampYZ}) we can write the amplitude squared as,
\begin{eqnarray}
\label{eq:genbgYZexp}
\mathcal{M}^{Yn}_{OFij} \mathcal{M}^{Z\bar{n}\ast}_{SF\bar{i}\bar{j}} = 
\sum\limits_{a,b} \mathcal{C}^{ab}_{YZij\bar{i}\bar{j}} L^{ab}_{YZ n\bar{n}} ,
\end{eqnarray}
where again $a,b = (\pm,\pm)$ and we have defined the Lorentz structure coefficients,
\begin{eqnarray}
\label{eqn:YZLcoeffs}
\mathcal{C}^{\pm\pm}_{YZ ij\bar{i}\bar{j}} &=& 
(4D_{Z\bar{Z}i} D_{q\bar{q}j} D^\ast_{q\bar{q}\bar{i}} D^\ast_{Z\bar{Y}\bar{j}})^{-1} \nonumber \\
&\times&(g^j_{q R}g^{\bar{j}}_{q R} \pm g^j_{q L}g^{\bar{j}}_{q L}) \nonumber \\
&\times& (g^{i2}_{\ell R}g^j_{\ell R}g^{\bar{i}2}_{\ell R}g^{\bar{j}}_{\ell R} \pm g^{i2}_{\ell L}g^j_{\ell L} g^{\bar{i}2}_{\ell L}g^{\bar{j}}_{\ell L}), 
\end{eqnarray}
and Lorentz structure,
\begin{eqnarray}
\label{eqn:YZLstruc}
&&L^{\pm\pm}_{YZ n\bar{n}} = 
(g_{\mu\sigma} T_{q\pm}^{\gamma\bar{\gamma}} g_{\bar{\mu}\bar{\sigma}})
(g_{\nu\gamma} T_{YZn\bar{n} \pm}^{\nu\mu\bar{\gamma}\sigma\bar{\mu}\bar{\nu}} g_{\bar{\nu}\bar{\gamma}}),
\end{eqnarray}
where the $T$ objects are the traces found in Eq.(\ref{eqn:genbgampYZ}).

The final type of Lorentz structure occurs when a diagram from the first row (t-channel quark exchange diagram) interferes with one of the diagrams in $( c)-(f)$ (s-channel process in the usual language).~For these we can write,
\begin{eqnarray}
\label{eqn:genbgampXY}
&&\mathcal{M}^{qn}_{OFij}\mathcal{M}^{Y\bar{n}\ast}_{SF\bar{i}\bar{j}} = 
(D_{Y\bar{Y}i} D_{Z\bar{Z}j} D^\ast_{Y\bar{Z}\bar{i}} D^\ast_{q\bar{q}\bar{j}})^{-1}
\nonumber \\
&\times& 
(g_{\mu\sigma} g_{\bar{\mu}\bar{\sigma}} g_{\nu\gamma}  g_{\bar{\nu}\bar{\gamma}}) 
\mathcal{T}_{ij\bar{j}}^{q}(p_{\bar{q}},\nu,P^{OF}_{qn},\mu,p_q,\bar{\gamma})\nonumber  \\
&\times&\mathcal{T}_{ij\bar{i}\bar{j}\bar{i}}^{\ell}(p_Z,\gamma,p_{\bar{Z}},\bar{\sigma},p_{Y},\sigma,p_{\bar{Y}},\bar{\mu},P^{SF}_{Y\bar{n}},\bar{\nu}),
\end{eqnarray}
where $Y=1,2$ and we have defined,
\begin{eqnarray}
\label{eqn:trace6}
&& \mathcal{T}_{ij\bar{j}}^{X}(p_a,\nu,p_b,\mu,p_c,\bar{\gamma}) = \nonumber \\ 
&\frac{1}{2}& (g^{i}_{X R}g^j_{X R}g^{\bar{j}}_{X R} + g^{i}_{X L}g^j_{X L}g^{\bar{j}}_{X L}) \nonumber \\
&\times& Tr(\not{p}_a \gamma^\nu \not{p}_b \gamma^{\mu} \not{p}_c \gamma^{\bar{\gamma}}) + \nonumber \\ 
&\frac{1}{2}&(g^{i}_{X R}g^j_{X R}g^{\bar{j}}_{X R} - g^{i}_{X L}g^j_{X L} g^{\bar{j}}_{X L})  \nonumber \\
&\times& Tr(\not{p}_a \gamma^\nu \not{p}_b \gamma^{\mu} \not{p}_c \gamma^{\bar{\gamma}},\gamma^5),
\end{eqnarray}
as well as,
\begin{eqnarray}
\label{eqn:trace10}
&& \mathcal{T}_{ij\bar{i}\bar{j}\bar{i}}^{X}(p_a,\nu,p_b,\mu,p_c,\bar{\gamma},p_d,\sigma,p_e,\bar{\mu}) = \nonumber \\ 
&\frac{1}{2}& (g^{i}_{X R}g^j_{X R}g^{\bar{i}2}_{X R}g^{\bar{j}}_{X R} + g^{i}_{X L}g^j_{X L} g^{\bar{i}2}_{X L}g^{\bar{j}}_{X L}) \nonumber \\
&\times& Tr(\not{p}_a \gamma^\nu \not{p}_b \gamma^{\mu} \not{p}_c \gamma^{\bar{\gamma}} \not{p}_d \gamma^{\sigma},\not{p}_e,\bar{\mu}) + \nonumber \\ 
&\frac{1}{2}&(g^{i}_{X R}g^j_{X R}g^{\bar{i}2}_{X R}g^{\bar{j}}_{X R} - g^{i}_{X L}g^j_{X L} g^{\bar{i}2}_{X L}g^{\bar{j}}_{X L})  \nonumber \\
&\times& Tr(\not{p}_a \gamma^\nu \not{p}_b \gamma^{\mu} \not{p}_c \gamma^{\bar{\gamma}} \not{p}_d \gamma^{\sigma},\not{p}_e,\bar{\mu},\gamma^5). 
\end{eqnarray}
After expanding out Eq.(\ref{eqn:genbgampXY}) we can write the amplitude squared as,
\begin{eqnarray}
\label{eq:genbgXYexp}
\mathcal{M}^{qn}_{OFij} \mathcal{M}^{Y\bar{n}\ast}_{SF\bar{i}\bar{j}} = 
\sum\limits_{a,b} \mathcal{C}^{ab}_{qYij\bar{i}\bar{j}} L^{ab}_{qY n\bar{n}} ,
\end{eqnarray}
where again $a,b = (\pm,\pm)$ and we have defined the Lorentz structure coefficients,
\begin{eqnarray}
\label{eqn:XYLcoeffs}
\mathcal{C}^{\pm\pm}_{qY ij\bar{i}\bar{j}} &=& 
(4D_{Y\bar{Y}i} D_{Z\bar{Z}j} D^\ast_{Y\bar{Z}\bar{i}} D^\ast_{q\bar{q}\bar{j}})^{-1} \nonumber \\
&\times&(g^i_{qR}g^j_{q R}g^{\bar{j}}_{q R} \pm g^i_{qL}g^j_{q L}g^{\bar{j}}_{q L}) \nonumber \\
&\times& (g^{i}_{\ell R}g^j_{\ell R}g^{\bar{i}2}_{\ell R}g^{\bar{j}}_{\ell R} \pm g^{i}_{\ell L}g^j_{\ell L} g^{\bar{i}2}_{\ell L}g^{\bar{j}}_{\ell L}), 
\end{eqnarray}
and Lorentz structure,
\begin{eqnarray}
\label{eqn:XYLstruc}
&&L^{\pm\pm}_{qY n\bar{n}} = 
(g_{\mu\sigma} T_{qn\pm}^{\nu\mu\bar{\gamma}} g_{\bar{\mu}\bar{\sigma}})
(g_{\nu\gamma} T_{Y\bar{n} \pm}^{\gamma\bar{\sigma}\sigma\bar{\mu}\bar{\nu}} g_{\bar{\nu}\bar{\gamma}}),
\end{eqnarray}
where the $T$ objects are the traces found in Eq.(\ref{eqn:genbgampXY}).

As in the signal case we take advantage of the fact that the Lorentz structures in Eqs.(\ref{eqn:qqLstruc}), (\ref{eqn:YYLstruc}), (\ref{eqn:YZLstruc}), and (\ref{eqn:XYLstruc}) are independent of the intermediate vector bosons to perform the sum over $i,j$ in the Lorentz coefficients defined in Eqs.(\ref{eqn:qqLcoeffs}), (\ref{eqn:YYLcoeffs}), (\ref{eqn:YZLcoeffs}), and (\ref{eqn:XYLcoeffs}) to obtain,
\begin{eqnarray}
\label{eqn:Lcoeffsum}
&&\mathcal{C}^{\pm\pm}_{qq} = \sum\limits_{ij\bar{i}\bar{j}}\mathcal{C}^{\pm\pm}_{qq ij\bar{i}\bar{j}},~
\mathcal{C}^{\pm\pm}_{YY} = \sum\limits_{ij\bar{i}\bar{j}}\mathcal{C}^{\pm\pm}_{YY ij\bar{i}\bar{j}}, \nonumber \\
&&\mathcal{C}^{\pm\pm}_{YZ} = \sum\limits_{ij\bar{i}\bar{j}}\mathcal{C}^{\pm\pm}_{YZ ij\bar{i}\bar{j}},~
\mathcal{C}^{\pm\pm}_{qY} = \sum\limits_{ij\bar{i}\bar{j}}\mathcal{C}^{\pm\pm}_{qY ij\bar{i}\bar{j}}.
\end{eqnarray}
In this way we easily take into account all possible combinations of intermediate vector bosons.

We now have all of the pieces\footnote{Expressions for the various coefficients and Lorentz structure can be obtained by emailing the authors or at~\cite{WEBSITE}.} necessary to build the total interference term ${\mathcal{M}_{OF}}_{ij} {\mathcal{M}^\ast_{SF}}_{\bar{i}\bar{j}}$ in Eq.(\ref{eqn:bgamp2}) including all contributions from the intermediate vector bosons.~Explicitly we have,
\small{
\begin{eqnarray}
\label{eqn:OFSFBGamp2}
&& {\mathcal{M}_{OF}} {\mathcal{M}^\ast_{SF}} =
\sum\limits_{ij\bar{i}\bar{j}} {\mathcal{M}_{OF}}_{ij} {\mathcal{M}^\ast_{SF}}_{\bar{i}\bar{j}} \nonumber \\
&=&\sum\limits_{a,b}\sum\limits_{n,\bar{n}} 
\Big( \mathcal{C}^{ab}_{qq} L^{ab}_{qq n\bar{n}} + \mathcal{C}^{ab}_{11} L^{ab}_{11 n\bar{n}} + \mathcal{C}^{ab}_{22} L^{ab}_{22 n\bar{n}} \Big) \nonumber \\
&+& \sum\limits_{a,b}\sum\limits_{n,\bar{n}} \Big( \mathcal{C}^{ab}_{q1} L^{ab}_{q1 n\bar{n}} + \mathcal{C}^{ab}_{1q} L^{ab}_{1q n\bar{n}} + \mathcal{C}^{ab}_{q2} L^{ab}_{q2 n\bar{n}} \Big) \nonumber \\
&+& \sum\limits_{a,b}\sum\limits_{n,\bar{n}} \Big( \mathcal{C}^{ab}_{2q} L^{ab}_{2q n\bar{n}} + \mathcal{C}^{ab}_{12} L^{ab}_{12 n\bar{n}} + \mathcal{C}^{ab}_{21} L^{ab}_{21 n\bar{n}}
\Big),
\end{eqnarray}
where the sum over intermediate vector bosons has already been implicitly performed in Eq.(\ref{eqn:Lcoeffsum}) while the sum over $n,\bar{n}$ which includes the t and u channel contributions is shown explicitly (note that this also factors from the vector boson sum).~The $\mathcal{C}^{ab}_{XY}$ coefficients are in general complex due to the factor of $i$ multiplying the decay width in the massive vector boson propagators.~The Lorentz structure is either purely real or purely imaginary depending on whether the term contains an even or odd number of traces ending in $\gamma^5$.~These traces give an overall factor of $i$ (and an epsilon tensor).~Thus if $L^{ab}_{XYn\bar{n}}$ contains an even number of these traces, then it is purely real and if it contains an odd number it is purely imaginary.~Organizing in this manner allows for easier integration when obtaining the various projections (as well as when performing convolution to include detector effects~\cite{Chen:2014pia,TechNote}). 

Plugging Eq.(\ref{eqn:OFSFBGamp2}) into Eq.(\ref{eqn:bgamp2}) and using the results from~\cite{Chen:2012jy} as well as the fact that,
\begin{eqnarray}
\label{eqn:SFfromOF}
|\mathcal{M}_{SF}|^2 = |\mathcal{M}_{OF}(p_1 \leftrightarrow p_2)|^2,
\end{eqnarray}
we can obtain the complete amplitude squared for the $q\bar{q} \rightarrow 4e/4\mu$ background process,
\begin{eqnarray}
\label{eqn:finalbg4eamp2}
|\mathcal{M}^{q\bar{q}}_{4e,4\mu}|^2 &=& 
(\frac{1}{4}) 
\Big( |\mathcal{M}_{OF}|^2 + |\mathcal{M}_{SF}|^2 \nonumber \\
&+& 2Re({\mathcal{M}_{OF}} {\mathcal{M}^\ast_{SF}}) \Big).
\end{eqnarray}
where we have included a symmetry factor of 1/4 and implicitly included a color factor of 1/3 as well as a 1/4 for averaging over initial state quark spins.

Again the fully differential cross section is found by combining with the lepton four body phase space in Eq.(\ref{eqn:phasespace}) to give,
\begin{eqnarray}
\label{eqn:bgfinal_diffcxn}
\frac{d\sigma_{q\bar{q}\rightarrow 4e/4\mu}}{dM_1^2dM_2^2d\vec{\Omega}} = 
\Pi_{4\ell} |\mathcal{M}^{q\bar{q}}_{4e/4\mu}|^2.
\end{eqnarray}
This expression can now be combined with the result for the signal differential cross section to perform detailed analysis of the golden channel.~As in the case for signal, one also finds in the interference terms a dependence on $\cos\theta_{1,2}$ in the propagator denominators, making it difficult to perform analytic integration over the angular variables to obtain the doubly differential mass spectrum as was done in the $2e2\mu$ case~\cite{Chen:2012jy}.~We thus perform this integration numerically and show in Fig.~\ref{fig:bg_doubdists} of the Appendix the doubly differential mass spectra as well as various other doubly differential distributions.~Again details of the validation procedure can be found in~\cite{Chen:2012jy,Chen:2014pia,TechNote}.

\section{Scrutinizing the Golden Channel}
\label{sec:scrutinize} 
In this section we explore the potential of the golden channel to elucidate the nature of the couplings of a spin-0 scalar to neutral electroweak gauge bosons.~We begin by examining the relative contributions of all the possible combinations of tensor structures in Eq.({\ref{eqn:sum_vertex}) to the total $\varphi \rightarrow 4\ell$ decay width.~We then perform a `toy' generator level analysis to demonstrate our parameter extraction procedure via maximization of the likelihood.~We present various parameter fits to show the flexibility of our framework and its ability to extract the effective couplings including their correlations.~We only focus on `toy' parameter extractions in this study, since a proper study of the Higgs couplings requires careful inclusion of the relevant detector effects as well as an adequate treatment of production variables.~We leave a more detailed investigation of the Higgs couplings in the golden channel including detector effects to an accompanying study~\cite{Chen:2014pia}. 

\subsection{Relative `Partial Fractions'}
\label{subsec:partial_widths}
The total decay width for $\varphi \rightarrow ZZ + Z\gamma +\gamma\gamma \rightarrow 4\ell$ can be decomposed into the various `partial widths' formed out of pairs of tensor structures in Eq.(\ref{eqn:sum_vertex}) (or operators if interpreted in terms of Eq.(\ref{eqn:siglag})).~Since each term will be quadratic in the couplings, we can label each partial width by the appropriate combination of couplings $A_{nij}A^\ast_{\bar{n}\bar{i}\bar{j}}$.~They are obtained by integrating the fully differential decay width in Eq.(\ref{eqn:final_diffcxn}) over the kinematic variables defined in Sec.~\ref{sec:events}.~We then normalize these partial widths to the standard model value to form the various `partial fractions'.

We show in Fig.~\ref{fig:Loose2e2muMatrix} a table of these partial fractions for every possible combination of $A_{nij}A^\ast_{\bar{n}\bar{i}\bar{j}}$ which can contribute to the $2e2\mu$ decay width.~For these tables we take as our phase space $4~GeV < M_{1,2}$ and $\sqrt{s} = 125~GeV$ as well as $p_{T\ell} > 2~GeV$ and $|\eta_\ell| < 2.4$ for the transverse momentum and rapidity respectively of the final state leptons. The couplings $A_{nij}$ have been separated into their real and imaginary components as $A_{nij} = A_{nijR} + iA_{nijI}$ and we have set all $A_{nijR,I} =1$.~All of the $|A_{nijR,I}|^2$ terms sit along the diagonal with the various interference terms making up the off-diagonal terms.~Note that many of the interference terms are negative indicating destructive interference between the corresponding tensor structures (or operators).

In Fig.~\ref{fig:Loose4eMatrix} we show the same plot for the $4e$ final state.~One can see the change in the partial fractions and in particular the $ZZ/\gamma\gamma$ interference terms are significantly larger than in the $2e2\mu$ channel.~The blank entries indicate terms which are identically zero after integration.~We can see that these entries are those for which CP violation in the form of interference between $A_{1,2}$ and $A_3$ tensor structures would occur.~This is indicative of the fact that after one integrates over the kinematic all information on CP violation is lost.~Of course for the fully differential decay width many of these terms are non-zero in principal allowing for sensitivity to CP violation in the golden channel.~To get a rough idea of the size of these CP violating terms, in Figs.~\ref{fig:CMS2e2muAbsMatrix} and~\ref{fig:CMS4eAbsMatrix} in the Appendix we show the integral of the absolute value of the differential decay width.

Since all couplings are set to one, these tables essentially show how much each combination of tensor structures contributes to the $\varphi \rightarrow 4\ell$ phase space relative to the standard model contribution for which we have set $A_{1ZZ} = 2$ and all other couplings to zero.~From these values of the relative partial widths, one can gain some insight into which combination of operators the golden channel might be most sensitive to.~Furthermore, for a specific model one can take the prediction for the values of the various couplings and simply multiply by the numbers given in Fig.~\ref{fig:Loose2e2muMatrix}-\ref{fig:Loose4eMatrix} to get a feel for whether those couplings might be probed in the golden channel.~For most realistic models, all couplings apart from $A_{1ZZ}$ are generated by higher dimensional operators and are expected to be small.~Of course, these rates do not contain information about the shapes in the various distributions so in principal the sensitivity is greater than might be inferred from these values.~In Sec.~\ref{subsec:CMSpartial_widths} of the Appendix we also show the same partial fractions for a `CMS-like' phase space as well as show the same tables for the standard model prediction.~Of course for a scalar resonance with a mass much larger or smaller than $125~GeV$ these numbers can change significantly. 
\begin{figure*}
\includegraphics[width=0.85\textwidth]{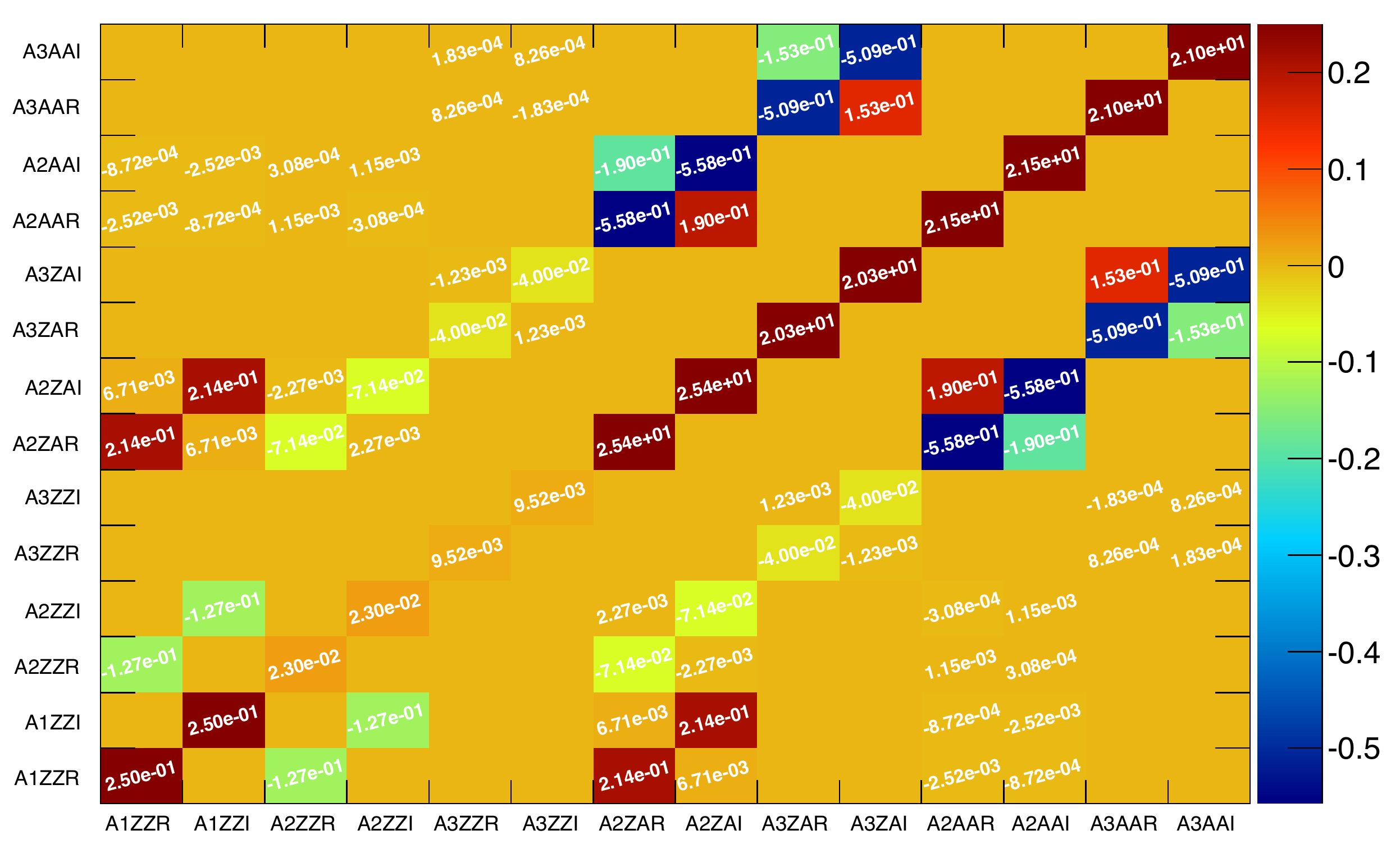}
\caption{Here we plot the relative partial fractions for each possible combination of tensor structures in Eq.(\ref{eqn:vertex}) for the $2e2\mu$ final state.~For these partial fractions we take a `loose' phase space of $4~GeV < M_{1,2}$, and $\sqrt{s} = 125~GeV$ as well as $p_{T\ell} > 2~GeV$ and $|\eta_\ell| < 2.4$.~They have been normalized to the standard model value for which we take $A_{1ZZ} = 2$ and all other couplings zero.}
\label{fig:Loose2e2muMatrix}
\end{figure*}
\begin{figure*}
\includegraphics[width=0.85\textwidth]{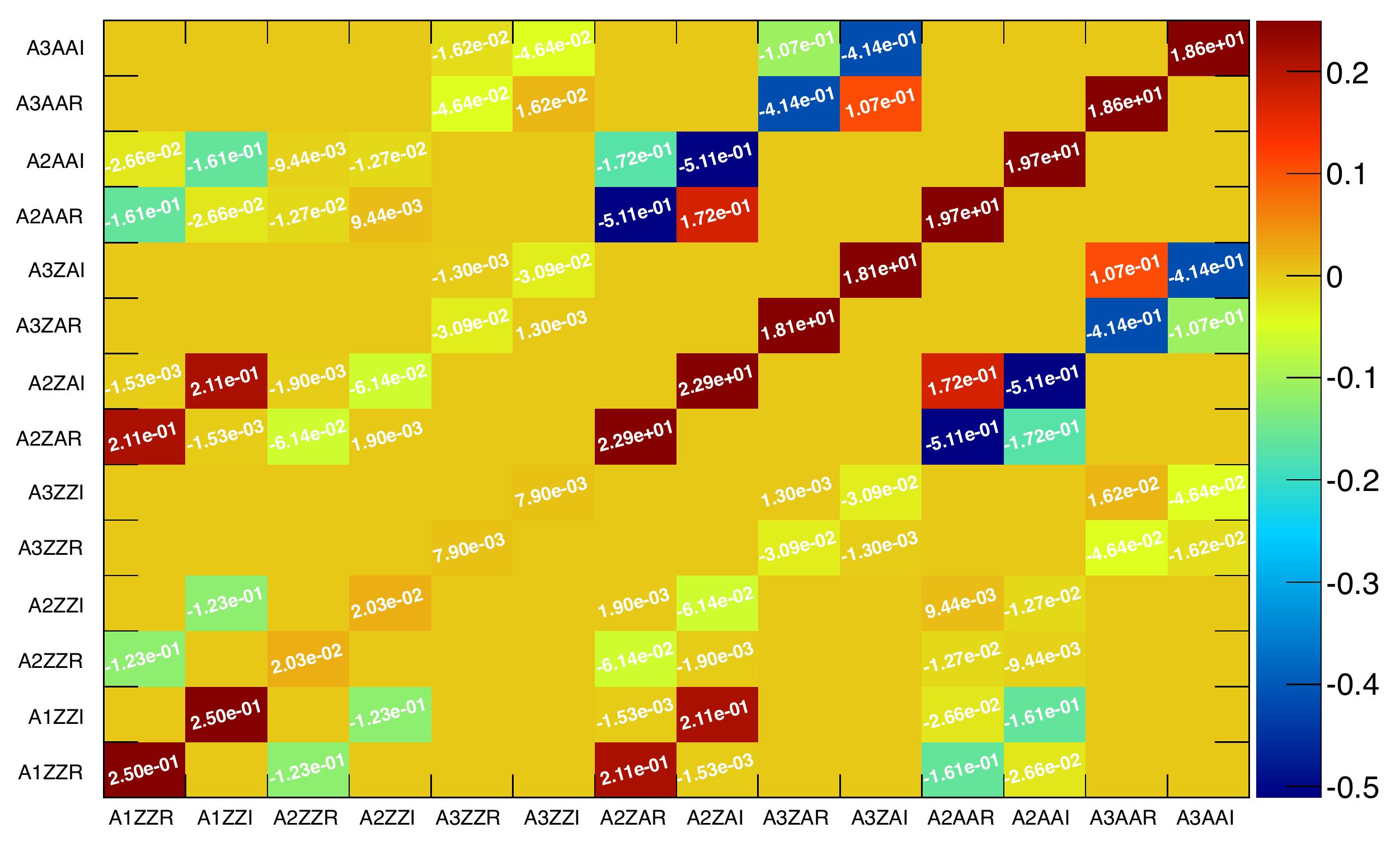}
\caption{Here we plot the relative partial fractions for each possible combination of tensor structures in Eq.(\ref{eqn:vertex}) for the $4e/4\mu$ final state.~For these partial fractions we take a `loose' phase space of $4~GeV < M_{1,2}$, and $\sqrt{s} = 125~GeV$ as well as $p_{T\ell} > 2~GeV$ and $|\eta_\ell| < 2.4$.~They have been normalized to the standard model value for which we take $A_{1ZZ} = 2$ and all other couplings zero.}
\label{fig:Loose4eMatrix}
\end{figure*}

\subsection{Simplified Analysis}
\label{subsec:analysis}
In order to demonstrate the flexibility and potential of our framework, we perform a simplified generator level analysis neglecting any detector effects and at a fixed center of mass energy of $\sqrt{s} = m_\varphi = 125~GeV$.~To do this we construct a maximum likelihood analysis using the fully differential cross sections in Eqs.(\ref{eqn:final_diffcxn}) and (\ref{eqn:bgfinal_diffcxn}) to build the signal plus background \emph{pdf} from which the total likelihood will be constructed.~Thus we have,
\begin{eqnarray}
\label{eqn:sigPbg_norm_pdf}
&& \mathcal{P}_{S+B}(\mathcal{O} | f, \vec{\lambda}) =   f \times \mathcal{P}_{B}(s,M_1,M_2,\vec{\Omega}) \\
&& + (1-f) \times \mathcal{P}_{S}(m_h^2,M_1,M_2,\vec{\Omega} | \vec{\lambda}). \nonumber
\end{eqnarray}
where $\mathcal{O} = (s,M_1,M_2,\vec{\Omega})$ is our final set of observables and $f$ is the background fraction, which we must also extract. The signal and background \emph{pdfs} are given by,
\begin{eqnarray}
\label{eqn:SandB_pdf}
\mathcal{P}_{S}(m_h^2,M_1,M_2,\vec{\Omega} | \vec{\lambda}) &=& \frac{d\sigma_{h\rightarrow 4\ell}}{dM_1^2dM_2^2d\vec{\Omega}}\nonumber\\
\mathcal{P}_{B}(s,M_1,M_2,\vec{\Omega}) &=& \frac{d\sigma_{q\bar{q}\rightarrow 4\ell}}{dM_1^2dM_2^2d\vec{\Omega}},
\end{eqnarray}
where they have been normalized over $\mathcal{O}$ (at fixed $\sqrt{s}$). With the \emph{pdfs} in hand we can now write the likelihood of obtaining a particular data set containing $N$ events as,
\begin{eqnarray}
\label{eqn:likelihood}
&& L(f, \vec{\lambda}) = \prod_{\mathcal{O}}^N \mathcal{P}_{S+B}(\mathcal{O} | f, \vec{\lambda}).
\end{eqnarray}
After constructing $L(f, \vec{\lambda})$ we then maximize with respect to $f$ and $\vec{\lambda}$ to extract the values which maximize the likelihood $\hat{\lambda}$ and $\hat{f}$ for a given data set.~To asses the error we then repeat this for a large number of pseudo experiments to obtain distributions for $\hat{\lambda}$ and $\hat{f}$ with a corresponding spread. Below we show the results for an example parameter point.~More details on this procedure can be found in~\cite{Chen:2014pia} and~\cite{TechNote}.


\subsection{Fit Definition}
\label{subsec:fit_def}

To examine the Higgs couplings to neutral gauge bosons, we take as our hypothesis the vertex in Eq.(\ref{eqn:vertex}).~We can use an overall phase rotation to make one of the parameters real.~Furthermore, we can avoid the need for the absolute normalization if we instead fit to ratios of couplings.~Which parameter to make real and which ratios to construct explicitly is a matter of choice the most convenient of which depends on the fit being performed.~Thus, in terms of the vertex as defined in Eqs.(\ref{eqn:sum_vertex}), we are explicitly fitting to,
\begin{eqnarray}
\label{eqn:fit_lagrangian}
\Gamma^{\mu\nu}_{ij}(k,k^\prime) &\propto& 
\mathcal{R}^{ij}_1 V_1^{\mu\nu} + \mathcal{R}^{ij}_2  V_2^{\mu\nu} + \mathcal{R}^{ij}_3 V_3^{\mu\nu}
\end{eqnarray}
where $\mathcal{R}^{ij}_n$ are complex ratios defined as $\mathcal{R}^{ij}_n = A_{nij}/|A|$ where $|A|$ is some normalization to be chosen for each fit. Since one of the $\mathcal{R}^{ij}_n$ can always be made real there are in principal twelve undetermined parameters to fit for when neglecting the overall normalization (note $\mathcal{R}^{Z\gamma}_1 = \mathcal{R}^{\gamma\gamma}_1 = 0$).~Fitting to ratios also makes any dependence on the production variables, $\vec{p}_T$ and $Y$ minimal since they mainly only affect selection efficiencies when detector effects are eventually included~\cite{Chen:2014pia}. 

\subsection{Example Parameter Extraction}
\label{toyfit}
As a demonstration of our ability to perform parameter extraction, we analyze the following example parameter point:
\begin{itemize}
\item $\vec{\lambda} \equiv (A_{1ZZ} = 1, A_{2ZZ} = 0, A_{3ZZ} = 5.1, A_{2Z\gamma} = 0.05, A_{3Z\gamma} = -0.1, A_{2\gamma\gamma} = 0.07, A_{3\gamma\gamma} = -0.08$).
\end{itemize}
Note that even though $A_{2ZZ}$ is zero we still fit for it and therefore it is floated when performing the maximization.~Thus we allow for all operators in Eq.(\ref{eqn:siglag}) to be `turned on' simultaneously, but we assume all coefficients to be real.~Our framework can easily also allow for non-zero phases, but we do not consider them here for simplicity.~The pseudo-data set to which we fit is obtained by generating large samples from the analytic expressions using a simply constructed event generator\footnote{The event generator can be obtained from~\cite{WEBSITE}.}.~We generate both signal and background events at fixed energy $\sqrt{s} = 125~GeV$ and $M_{1,2} > 4~GeV$.~Since we seek only to demonstrate the validity of our parameter extraction framework, we focus on the $2e2\mu$ final state for simplicity.~It would be interesting, however, to perform a dedicated study and examine how the sensitivity of the $2e2\mu$ final state compares to the $4e/4\mu$ final state for different choices of phase space, but we leave this for future work.~The parameter extraction is performed by maximizing the likelihood function as described above.

We first perform a simultaneous extraction of all parameters including the correlations assuming a pure signal sample.~We show in Fig.~\ref{fig:fits1} one dimensional results for a large set of pseudo experiments containing 1000 events each.~We have explicitly fit to the ratios of couplings $R^{ij}_n = A_{nij}/|A|$ where here we take $|A| = |A_{1ZZ}|$ (thus fixing $R^{ZZ}_1 = 1$).~The distribution for the extracted parameters obtained for the set of pseudo experiments is shown in blue with the true value indicated by the red vertical line.~One can see that the true value sits near the center of the distribution, an indication that the maximization procedure is working properly and that the global maximum of the likelihood function is in fact being obtained in each pseudo experiment. The efficiency of convergence in our maximization is $\gtrsim 99 \%$ and takes on the order of a few minutes to complete~\cite{Chen:2014pia,TechNote}. 

Of course there are also correlations between the parameters.~To see this we can examine the different parameters in pairs as shown in Fig.~\ref{fig:fits3} again for 1000 events for each pseudo experiment and assuming a pure signal sample.~The true value is indicated by the intersection of the two solid black lines which as can be seen falls near the center of the distribution.~The colors indicate the density of pseudo experiments returning a particular value for the extracted parameters as indicated on the x and y axis and we have fit to $R^{ij}_n = A_{nij}/|A|$ where $|A| = |A_{1ZZ}|$.~We can see that in the majority of pseudo experiments the fit returns values close to the true ones.~One can also see in these plots some of the potential correlations between the various parameters though of course the full set of correlations between the six parameters which are contained in the fit can not be displayed easily.~This also demonstrates the importance of including all possible couplings in the differential cross section. 

We can also examine how the results change when the $q\bar{q} \rightarrow 4\ell$ background is included.~As discussed above, when including both signal and background we must also extract the background fraction $f$.~In Fig.~\ref{fig:fits4} we show our results including background in the likelihood.~In the top left plot we show the distribution of the extracted background fraction for 1000 signal plus 250 background events for a large set of pseudo experiments.~In the additional plots we compare the results assuming a pure signal sample shown in black to those which include both signal and background which are shown in red. We fit to $R^{ij}_n = A_{nij}/|A|$, but now take the overall normalization to be $|A| = \sqrt{\sum\limits_{n,ij}|A_{nij}|^2}$.~We can see that the couplings which are affected the most by the inclusion of background are the $Z\gamma$ couplings.~This can be understood by the fact that near the signal region of $125~GeV$, the background is primarily composed of the $Z\gamma$ intermediate state~\cite{Chen:2012jy}.~In general, however, one can see that the effect of including background is small, an indication that there is strong discrimination between signal and background as implied by the differences in the various doubly differential spectra shown in~\cite{Chen:2012jy} and in Figs.~\ref{fig:sig_doubdists}-\ref{fig:bg_doubdists}.

It is also important to study how the spread of the distribution changes as a function of the numbers of events.~In Fig.~\ref{fig:fits2} we have plotted the results for various number of signal events ranging from 30 to 3000 per pseudo experiment with a $20\%$ background fraction.~The color indicates the density of pseudo experiments which return a value of the parameter as indicated on the y-axis.~Here we again fit to $R^{ij}_n = A_{nij}/|A|$, and take the overall normalization to be $|A| = \sqrt{\sum\limits_{n,ij}|A_{nij}|^2}$. We can easily see by the color that the spread roughly decreases with the expected $1/\sqrt{N}$ scaling as the number of events is increased.~The true value is now indicated by the solid black line, which as can be seen sits within the red region indicating that in most of the pseudo experiments the fit procedure returns a value of the parameter close to the true value.~Quantifying more precisely how the spread, or more accurately the error, changes as a function of number of events requires the inclusion of detector effects and is beyond the scope of this study, but a more detailed analysis (using CMS criteria) is left to ongoing/future work.

One of the interesting questions to ask, is whether the golden channel is sensitive to the $Z\gamma$ and $\gamma\gamma$ couplings of $\varphi$ assuming it is the recently discovered resonance at $\sim 125$~GeV.~Since it has been firmly established that this resonance couples to $ZZ$ through the $Z_\mu Z^\mu$ operator with a strength consistent with the SM prediction~\cite{Falkowski:2013dza} it may perhaps be difficult to extract the $Z\gamma$ and $\gamma\gamma$ couplings since they only occur through higher dimensional operators and will have couplings $\sim \mathcal{O}(10^{-2}-10^{-3})$, thus suppressing the partial widths corresponding to those operators in Figs.~\ref{fig:Loose2e2muMatrix}-\ref{fig:Loose4eMatrix}.~Determining whether this is in fact impossible requires a detailed analysis including detector effects which is beyond the scope of this paper and we leave it to a future study.  
\begin{figure*}
\includegraphics[scale=.37]{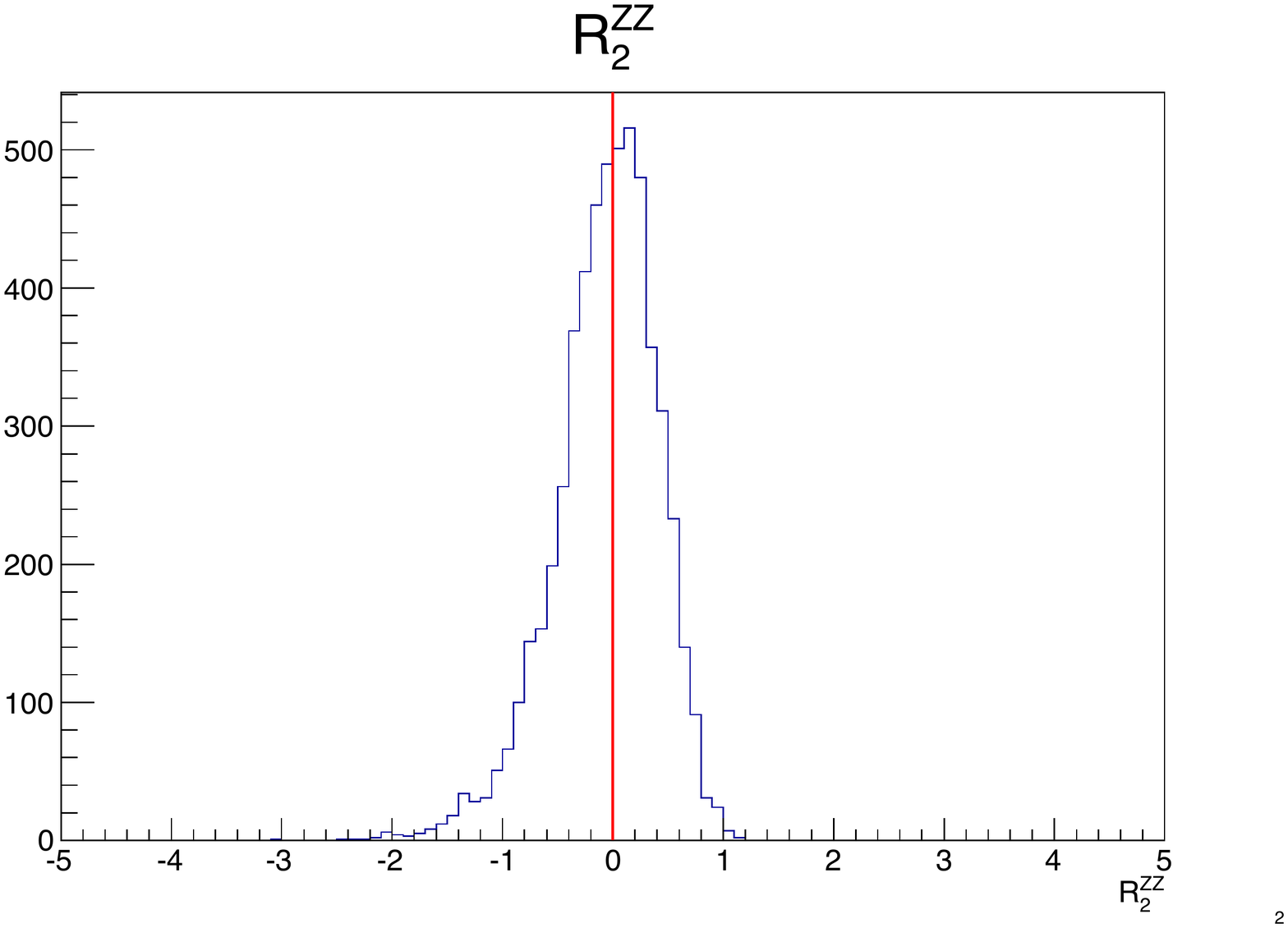}
\includegraphics[scale=.37]{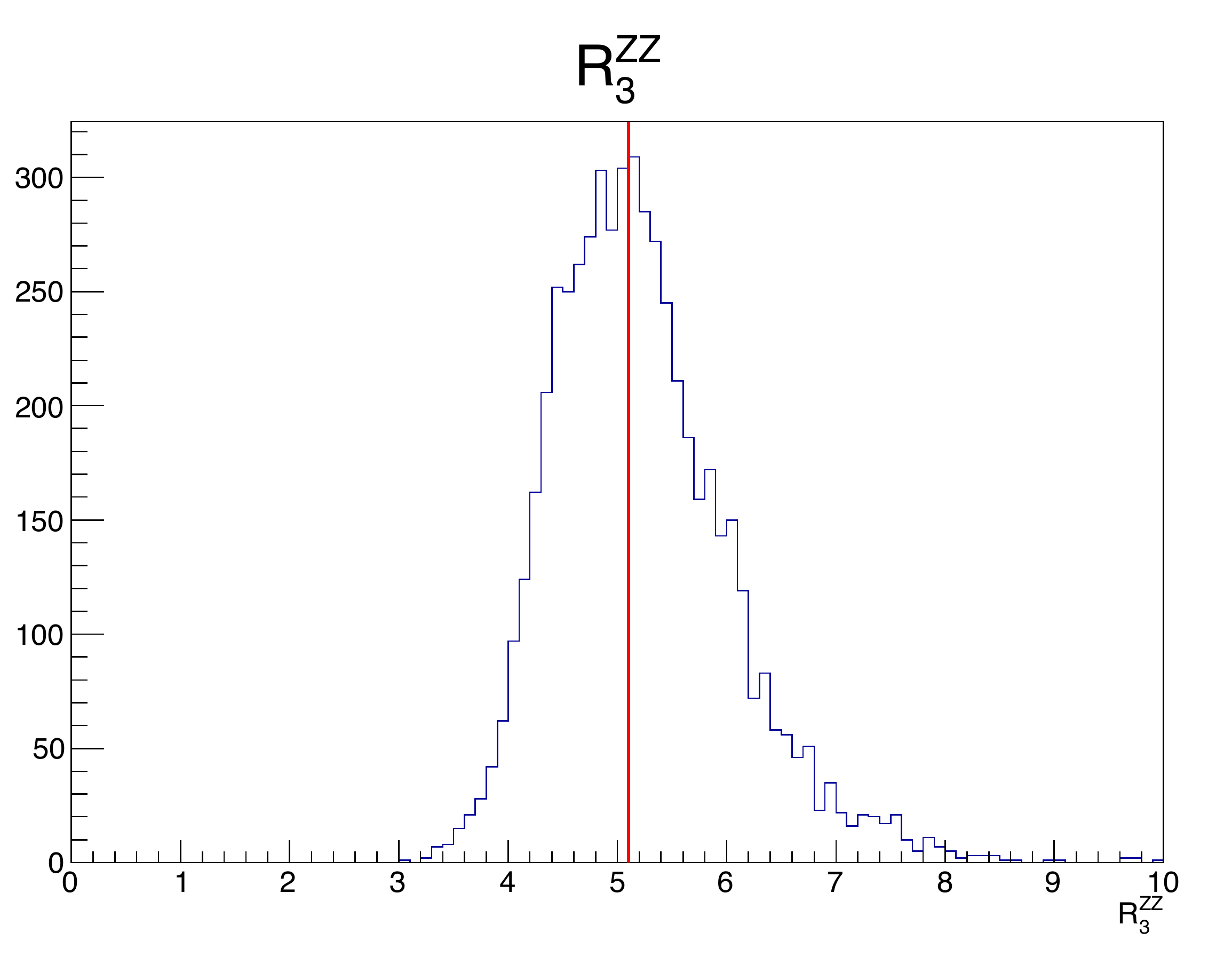}
\includegraphics[scale=.37]{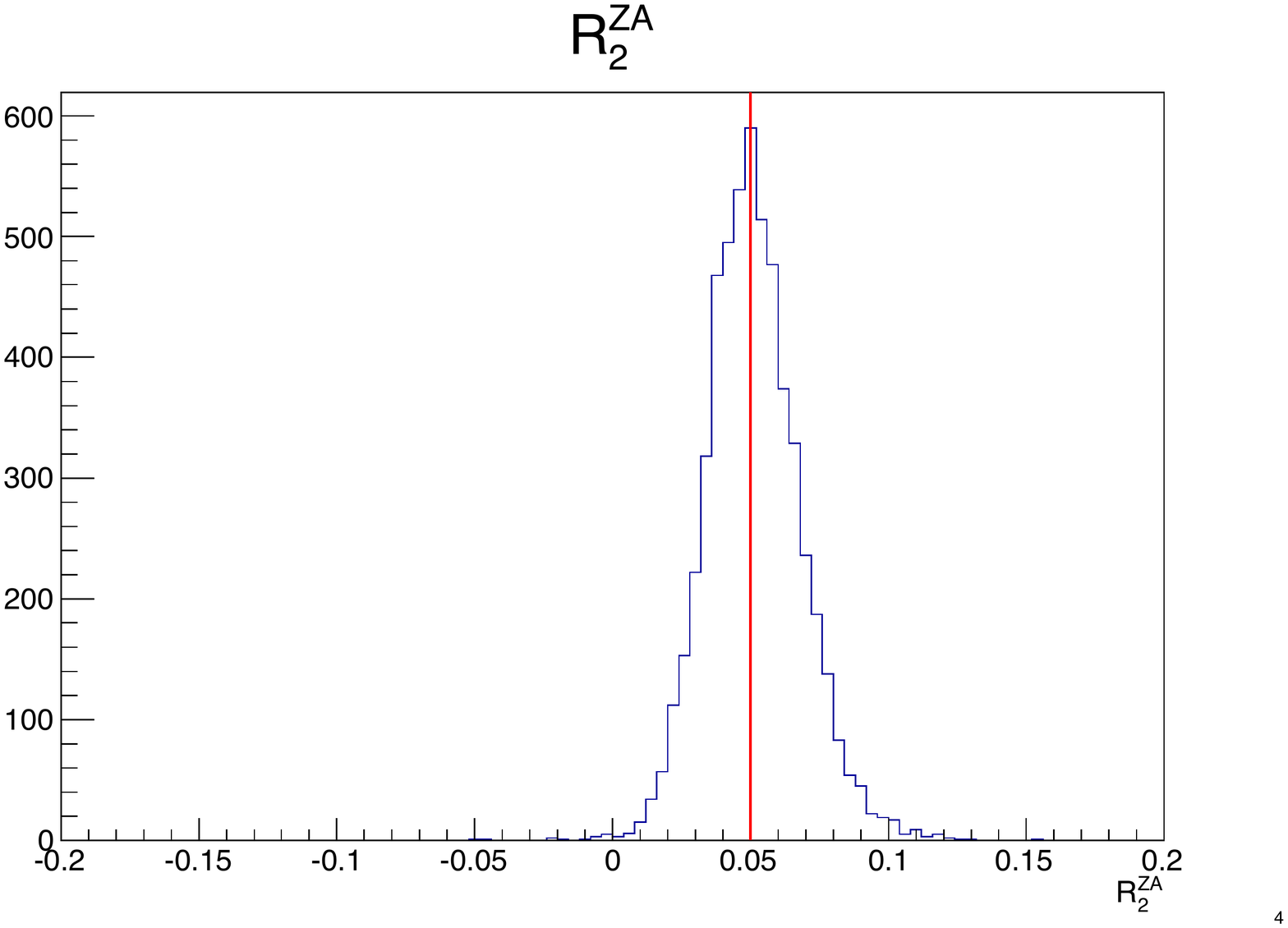} 
\includegraphics[scale=.37]{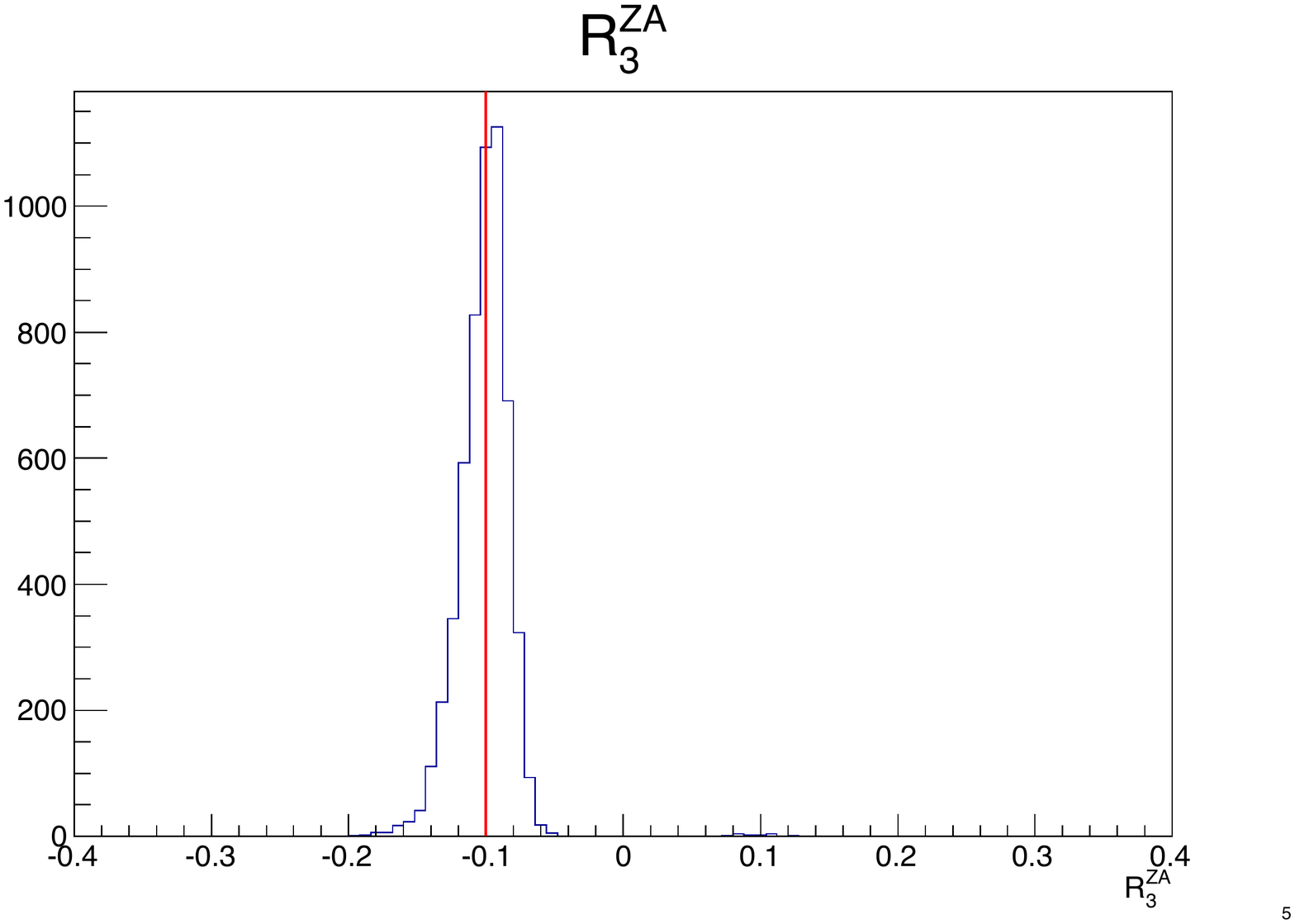}
\includegraphics[scale=.37]{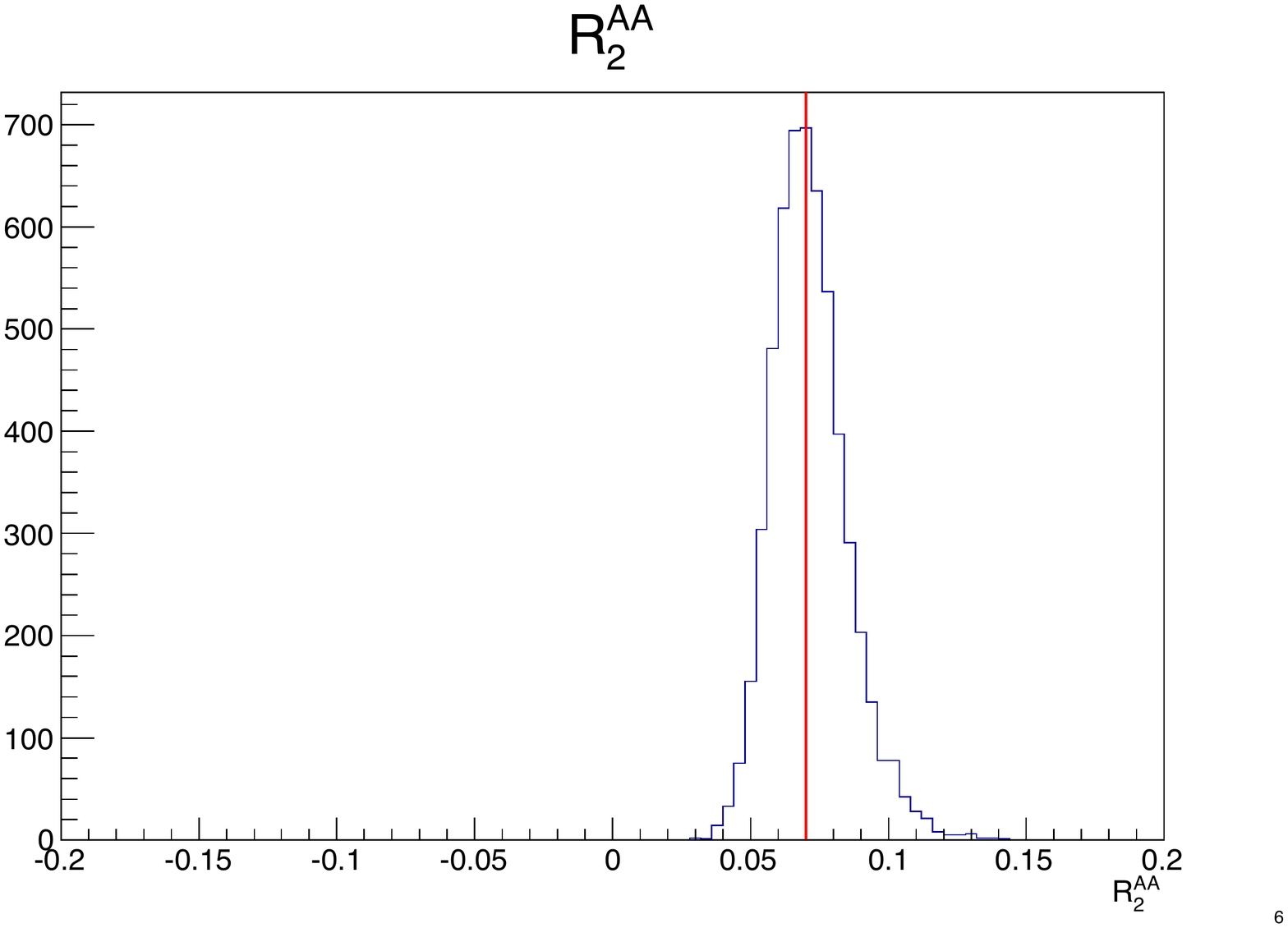}
\includegraphics[scale=.37]{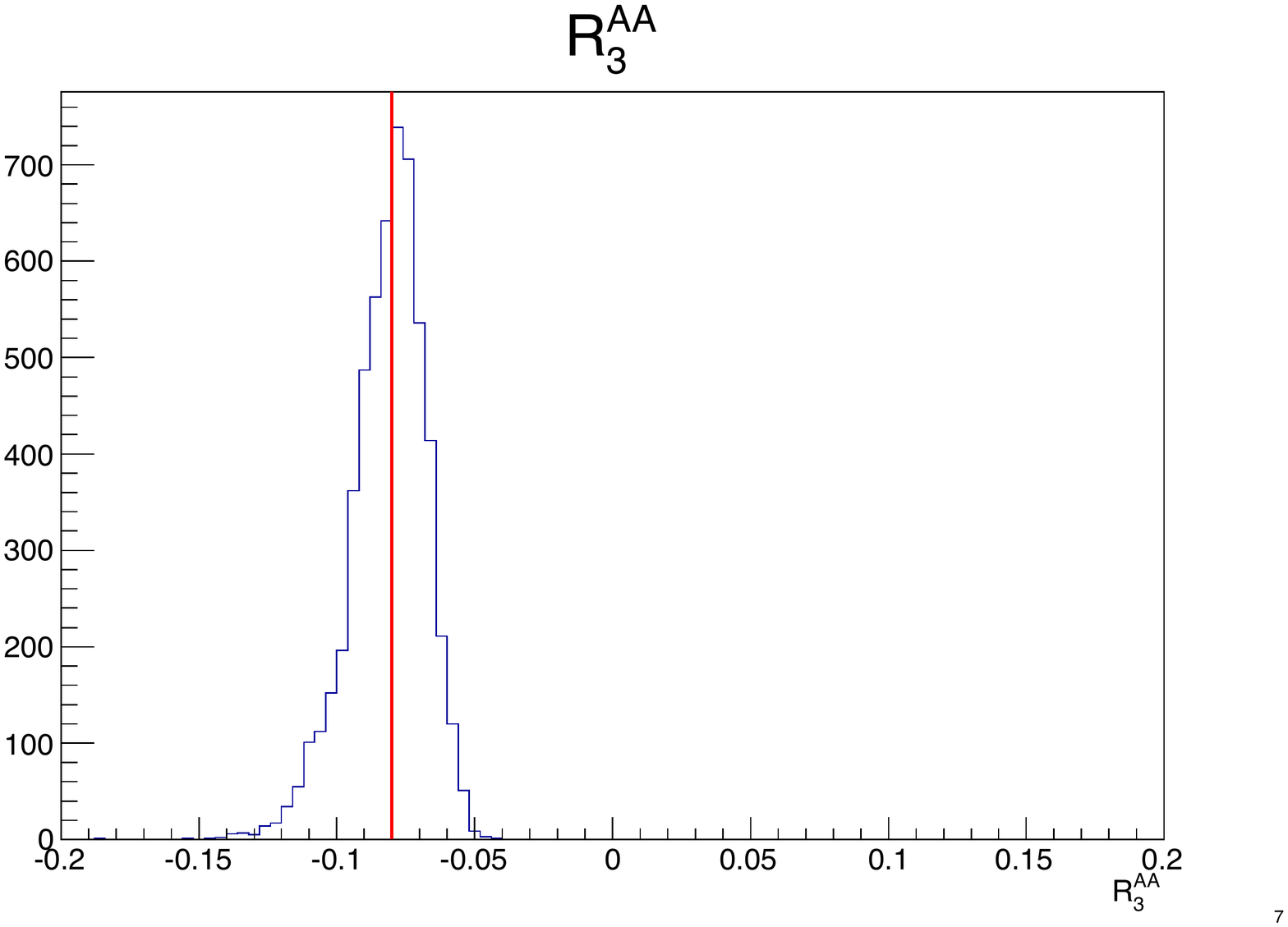} 
\caption{One dimensional results for the extraction of the six parameters for a large set of pseudo experiments containing 1000 events each of a pure signal sample.~We have explicitly fit to the ratios of couplings $R^{ij}_n = A_{nij}/|A|$. Here we choose the normalization to be $|A| = |A_{1ZZ}|$ which fixes $R^{ZZ}_1 = 1$.~The distribution for the extracted parameters obtained for the set of pseudo experiments is shown in blue with the true value indicated by the red vertical line.}
\label{fig:fits1}
\end{figure*}
\begin{figure*}
\includegraphics[scale=.32]{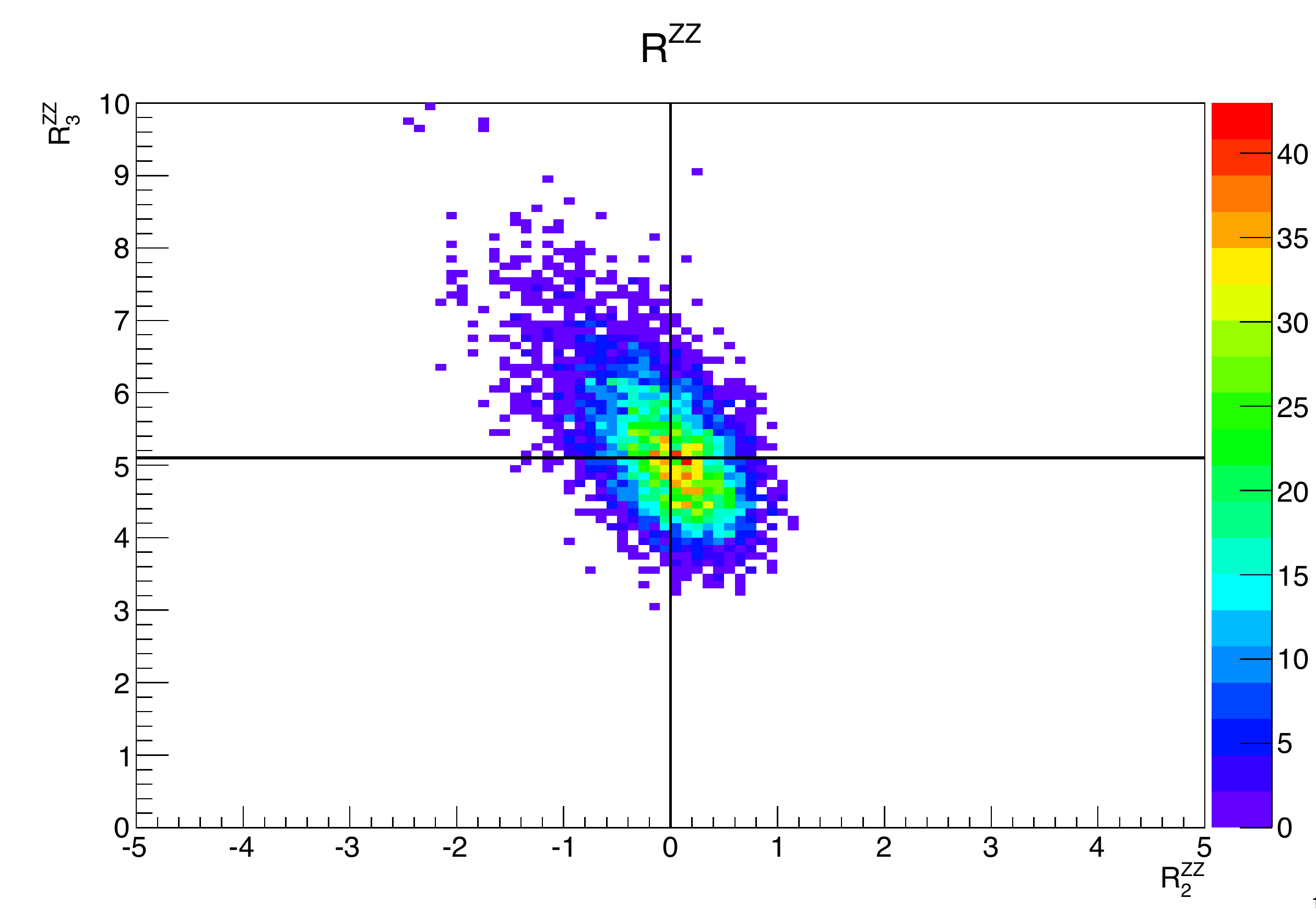}
\includegraphics[scale=.32]{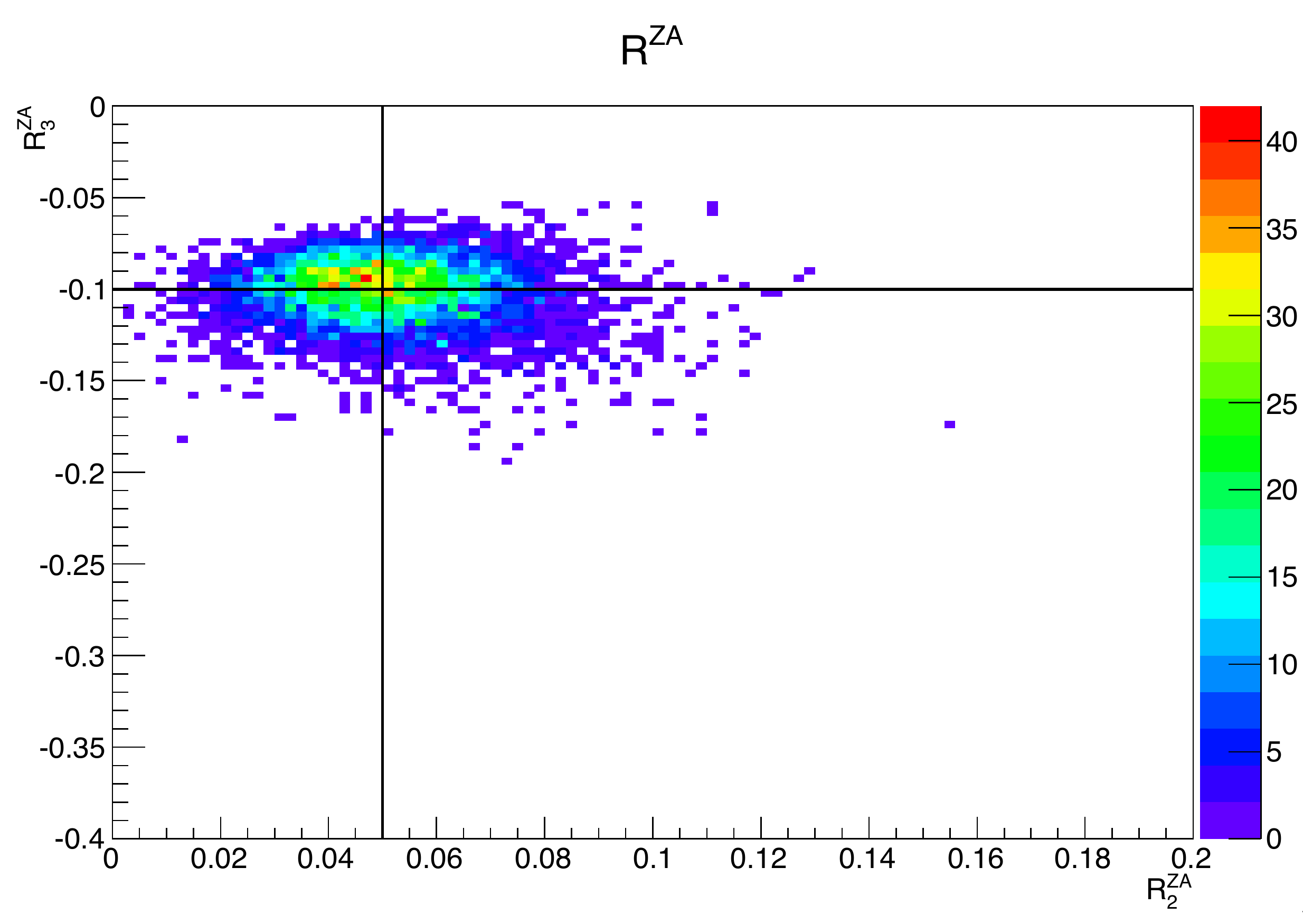}
\includegraphics[scale=.32]{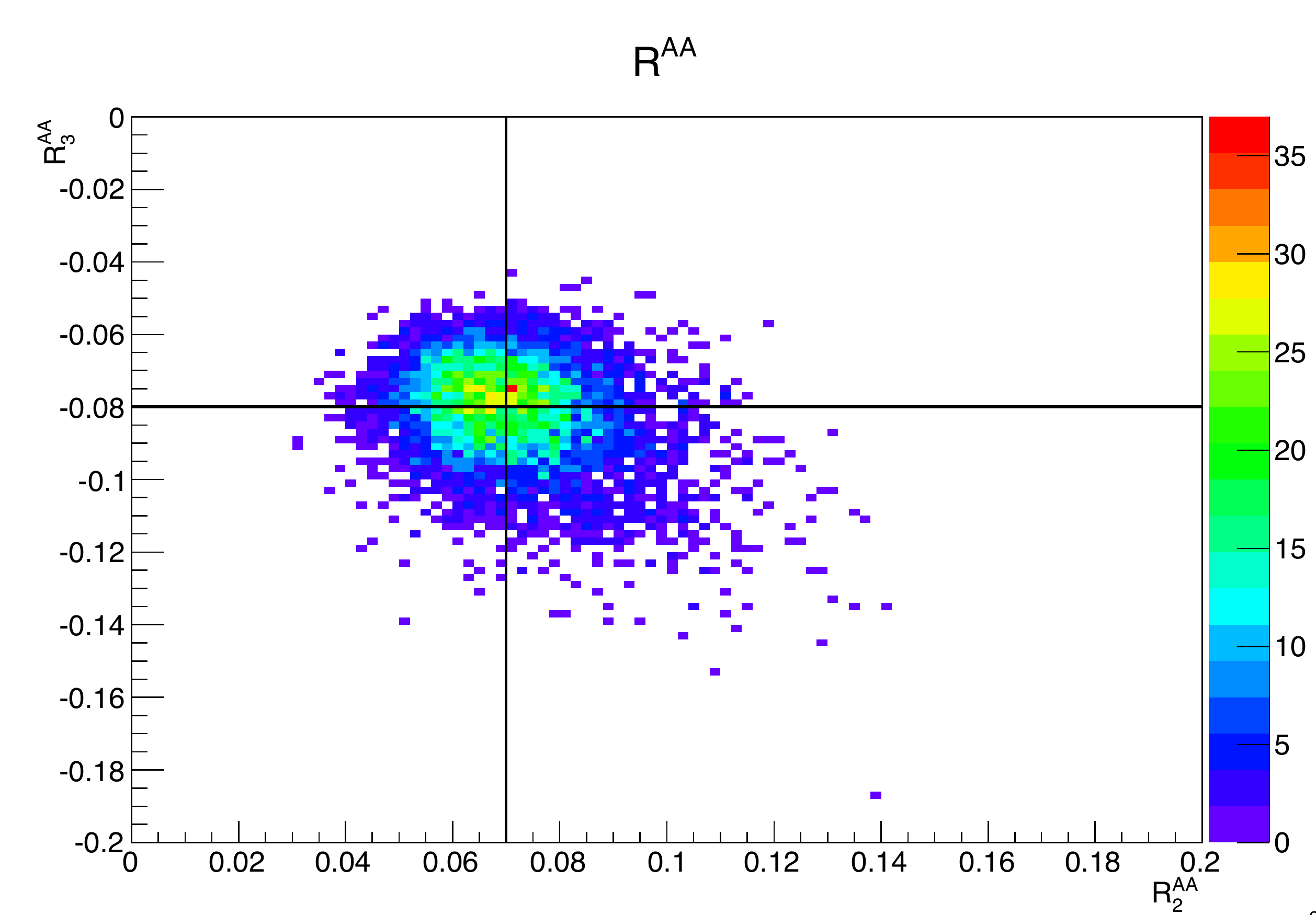} 
\includegraphics[scale=.32]{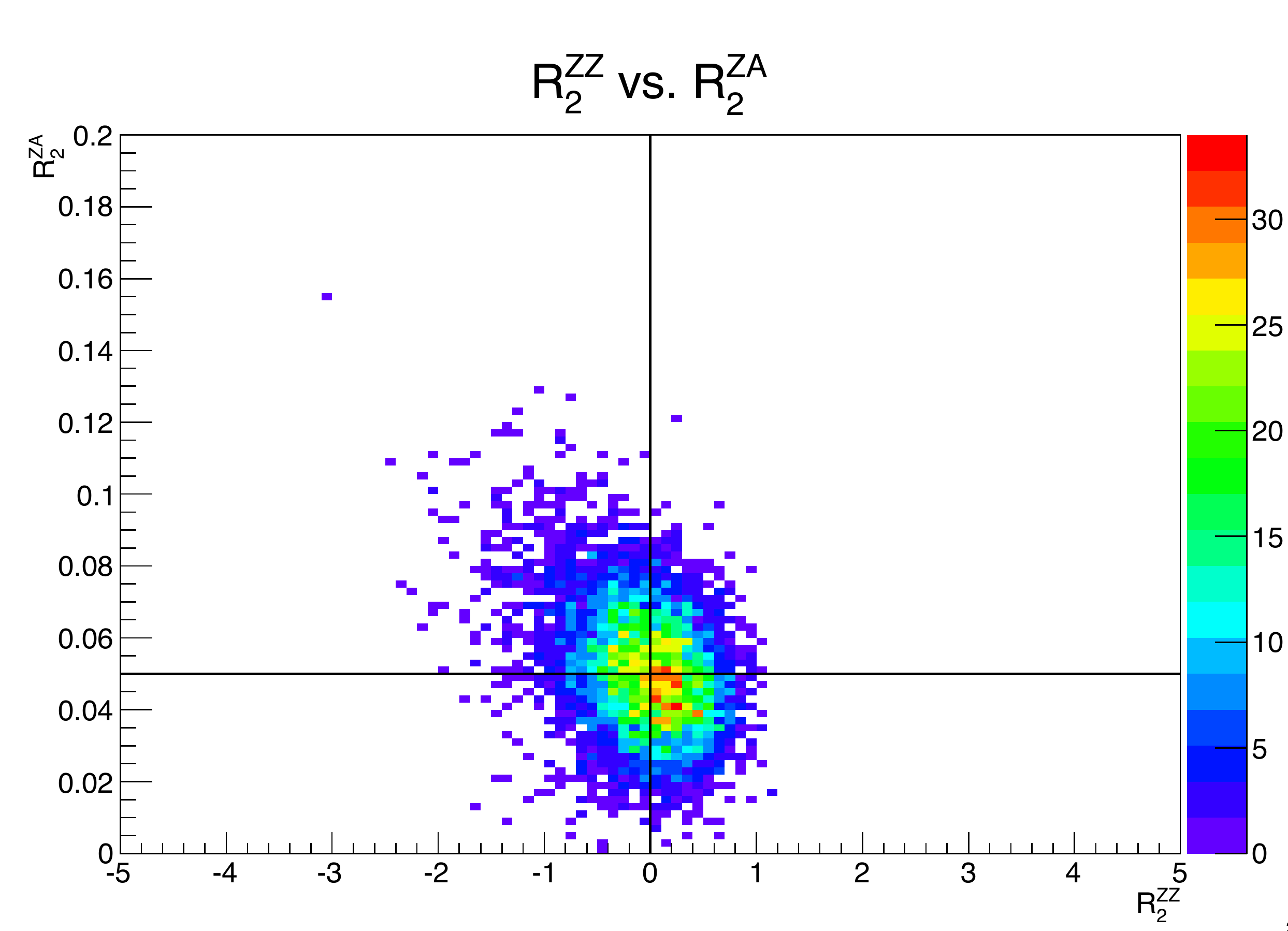}
\includegraphics[scale=.32]{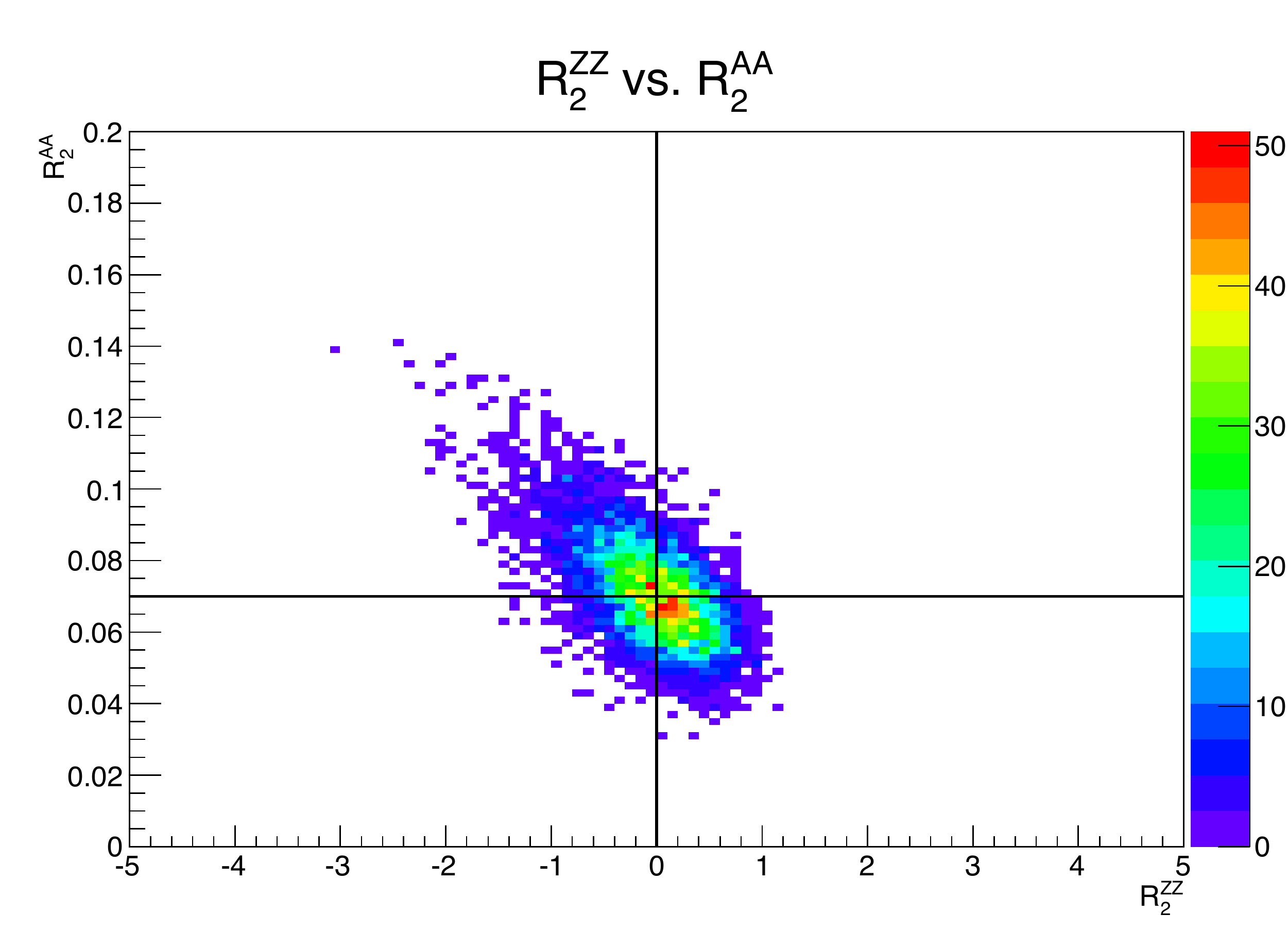}
\includegraphics[scale=.32]{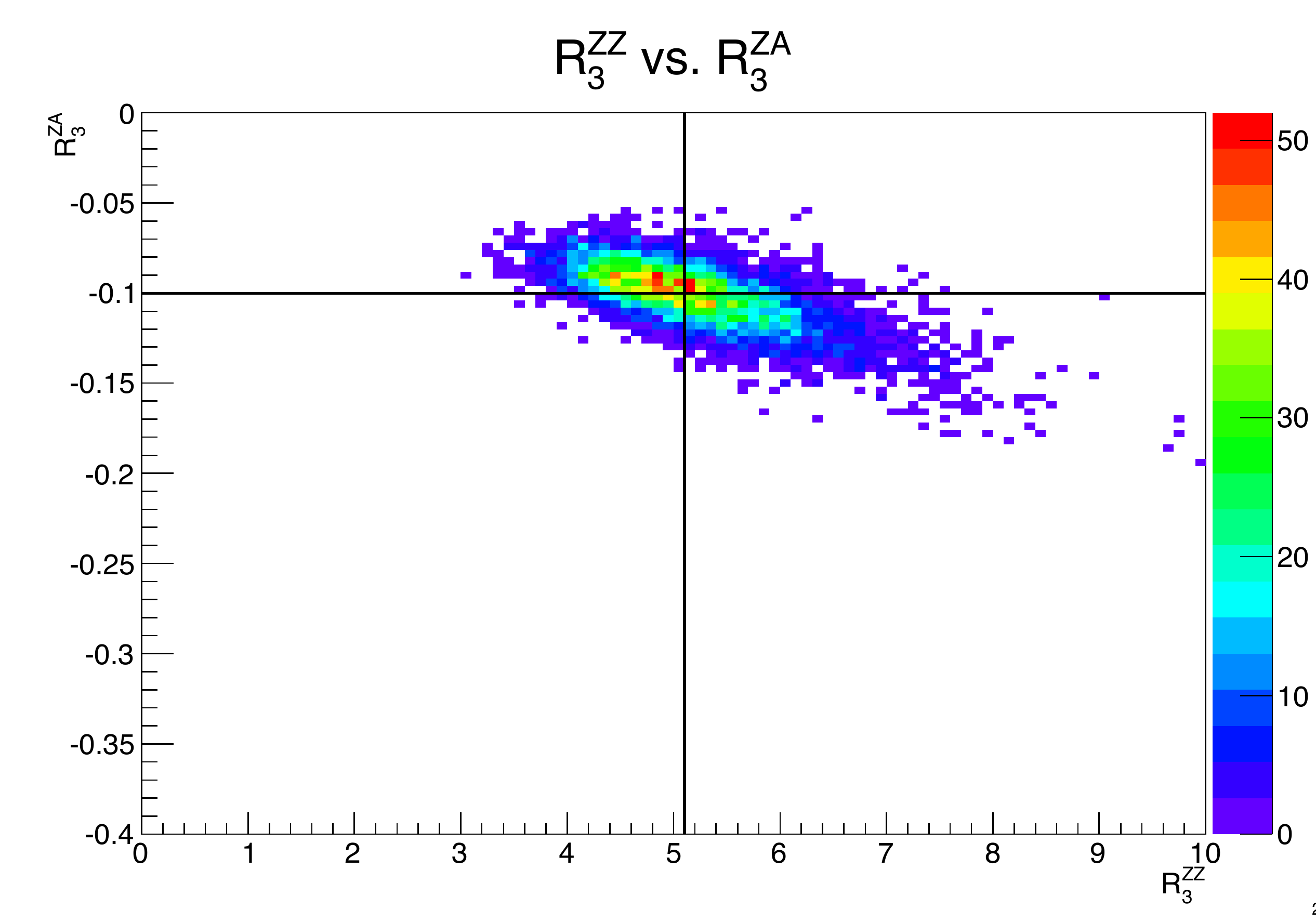} 
\includegraphics[scale=.32]{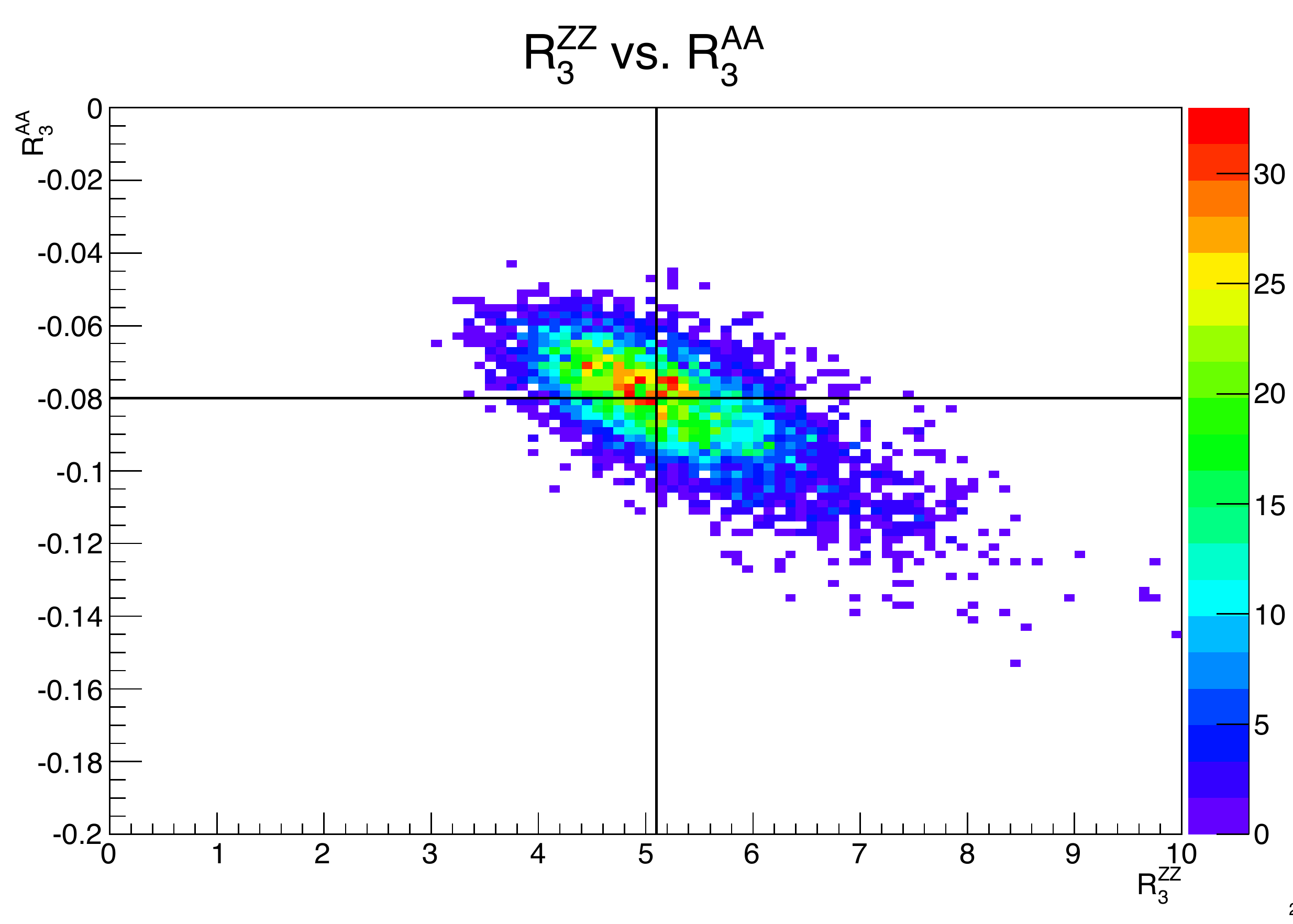}
\includegraphics[scale=.32]{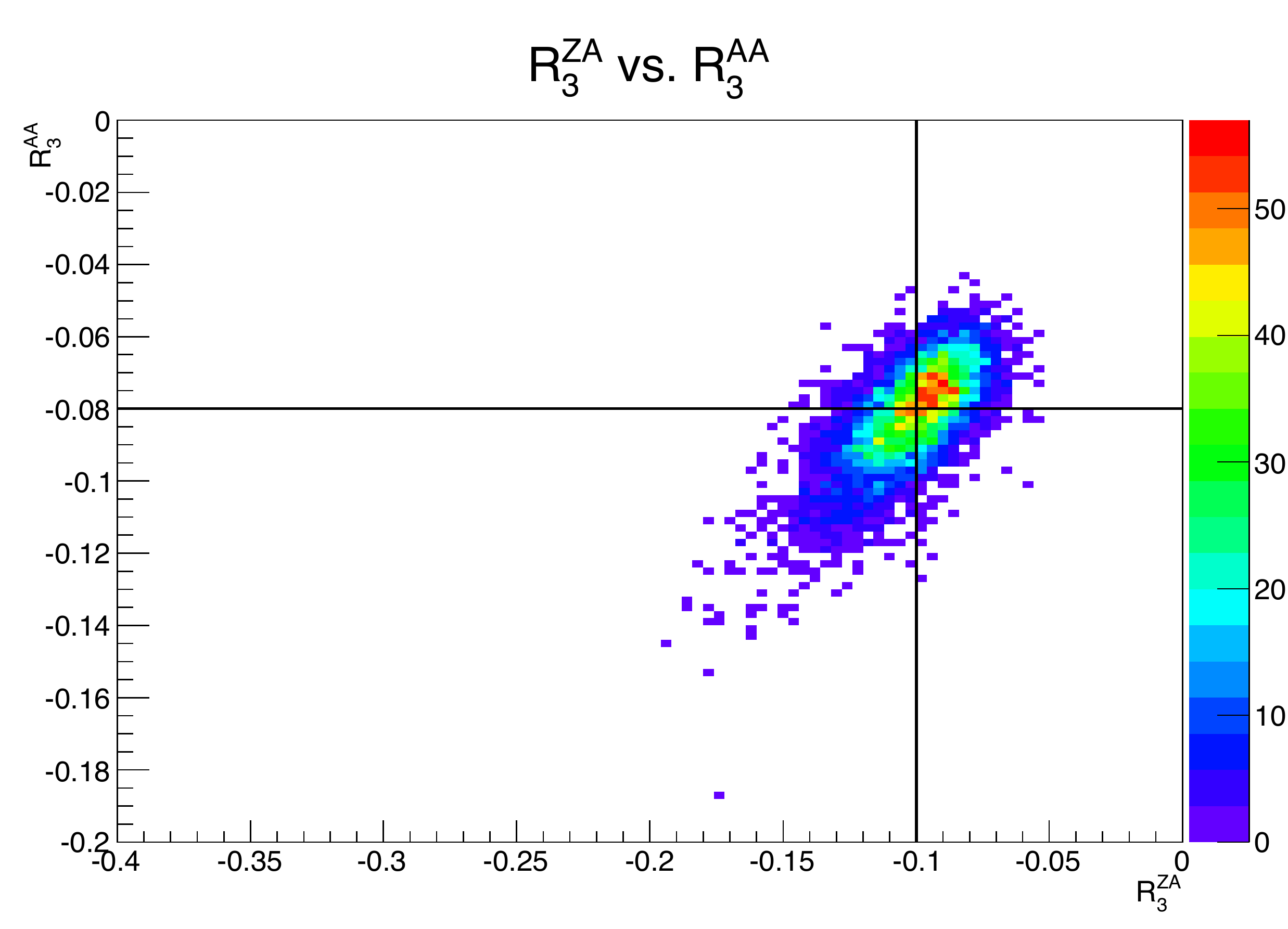} 
\caption{Here we examine the correlations between pairs of parameters.~We conduct a large set of pseudo experiments with 1000 events for each and assuming a pure signal sample.~The true value is indicated by the intersection of the two solid black lines and again we have fit to the ratios $R^{ij}_n = A_{nij}/|A|$ and take the normalization to be $|A| = |A_{1ZZ}|$.}
\label{fig:fits3}
\end{figure*}
\begin{figure*}
\includegraphics[scale=.33]{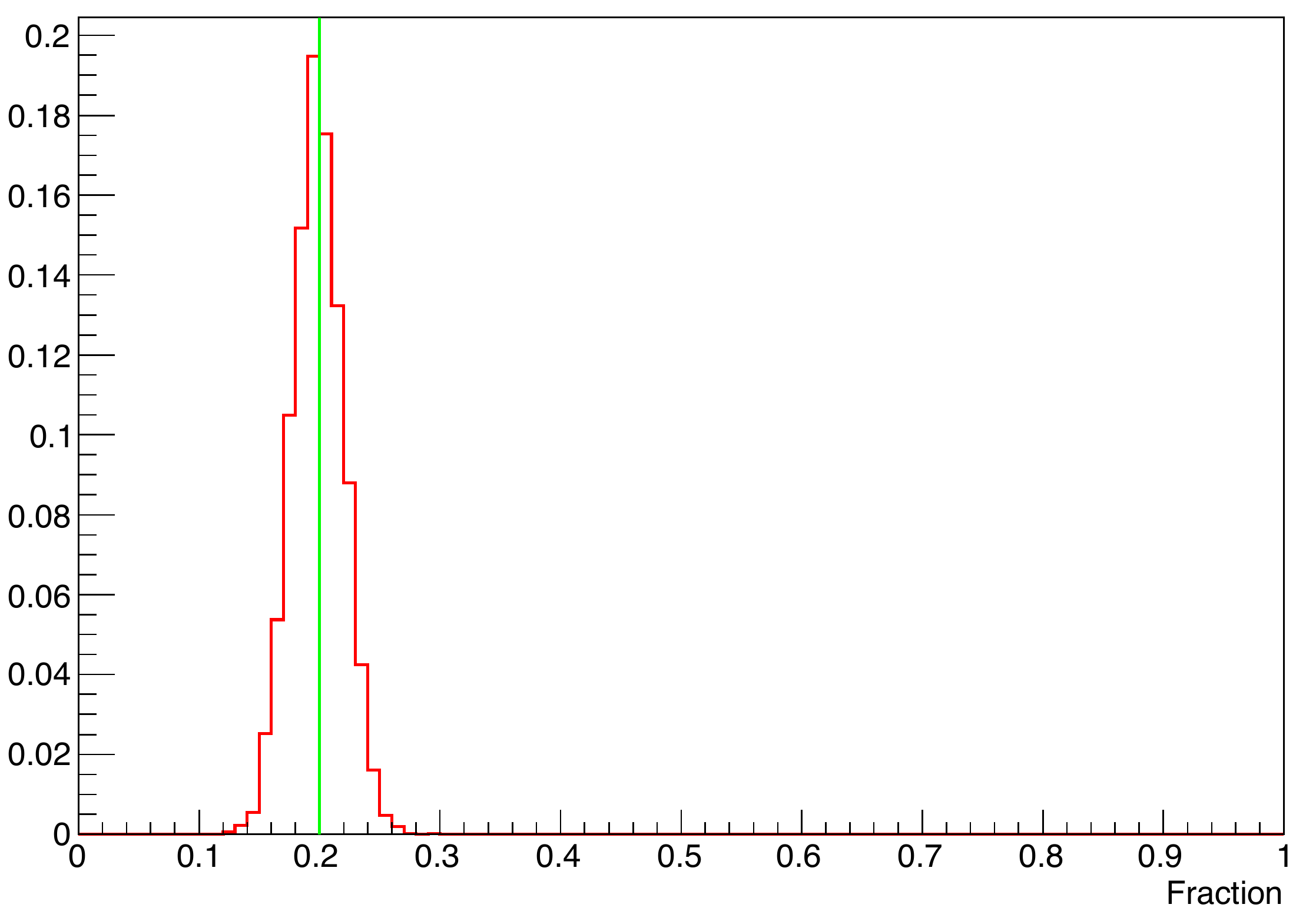}
\includegraphics[scale=.33]{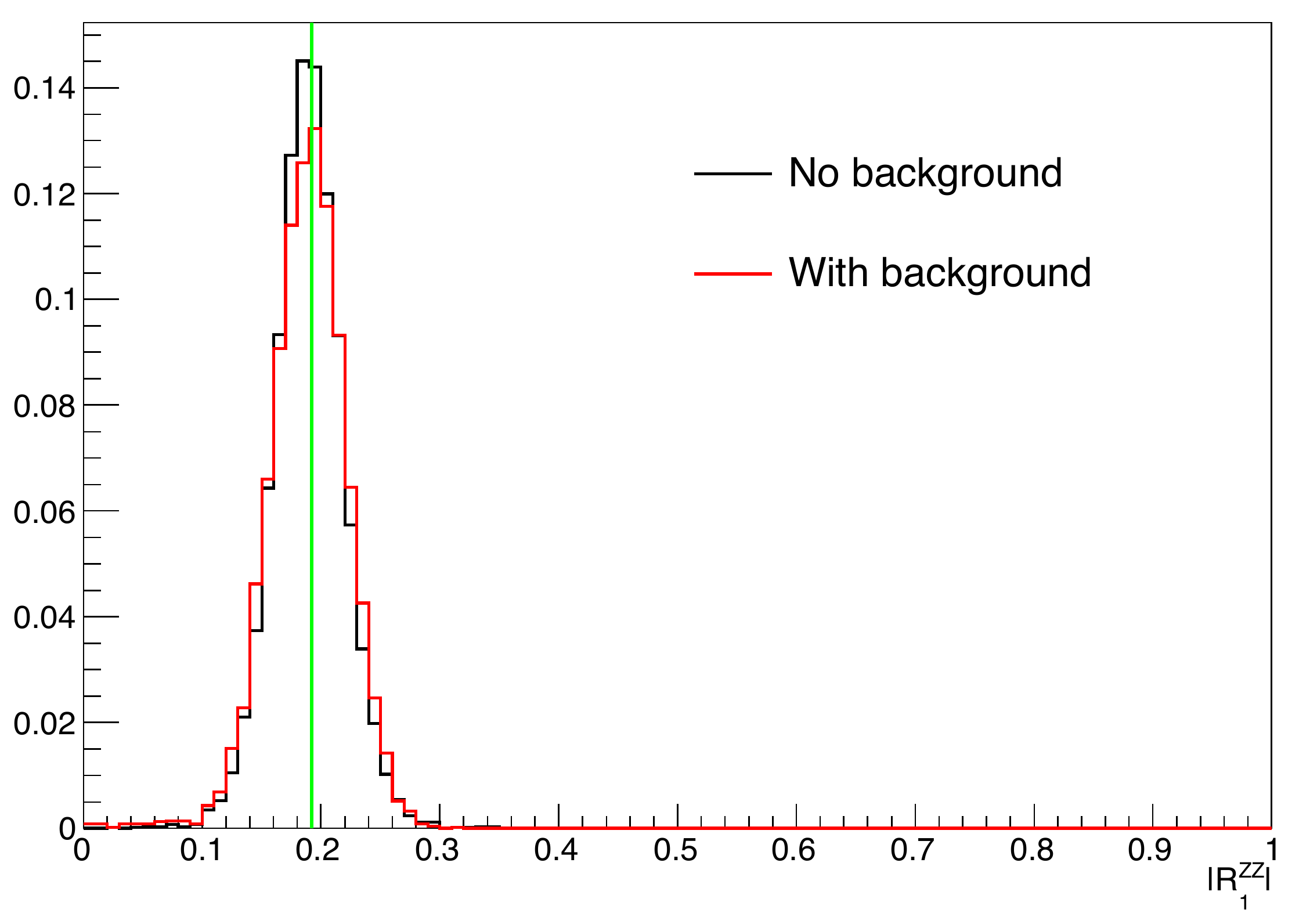}
\includegraphics[scale=.33]{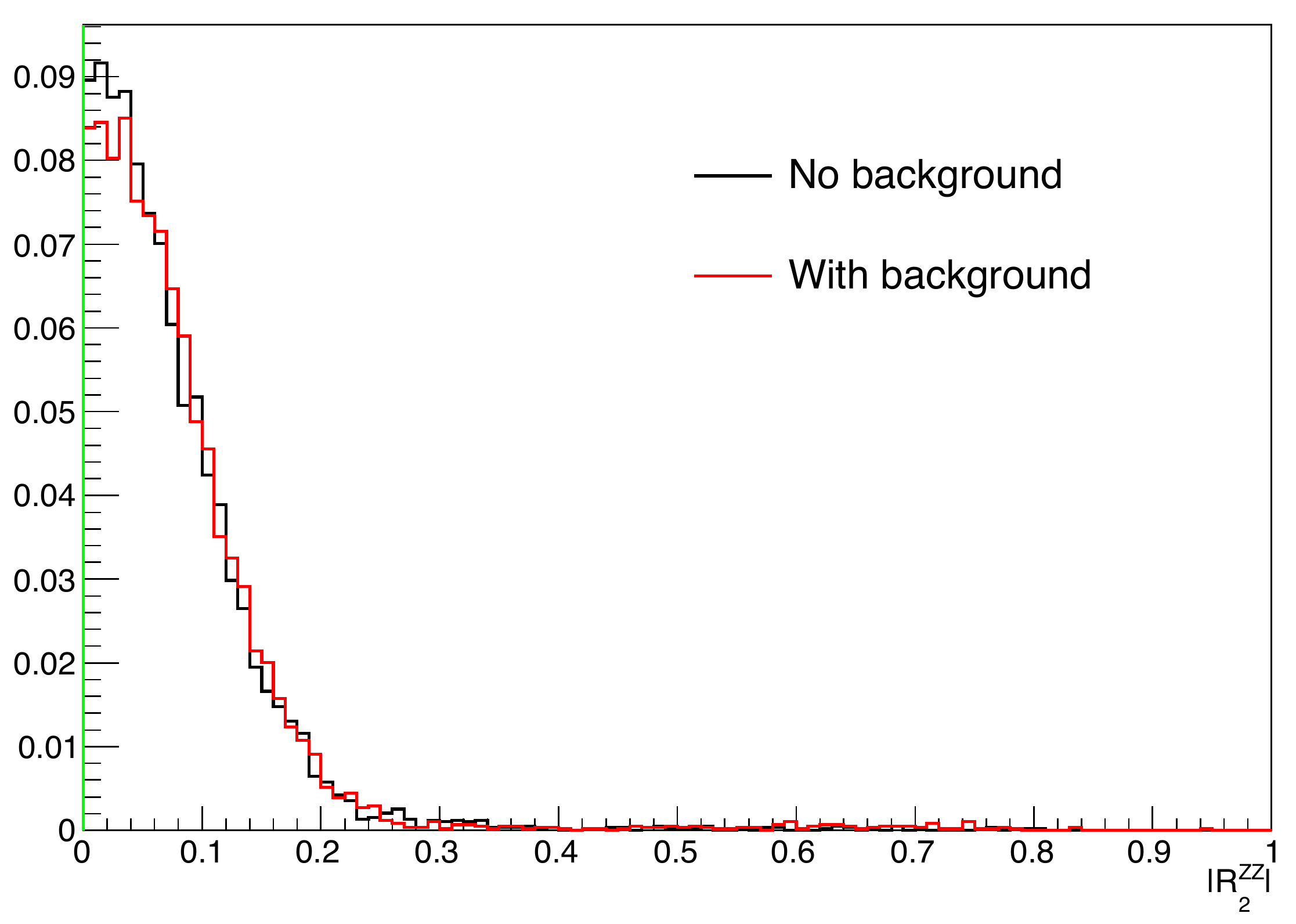}
\includegraphics[scale=.33]{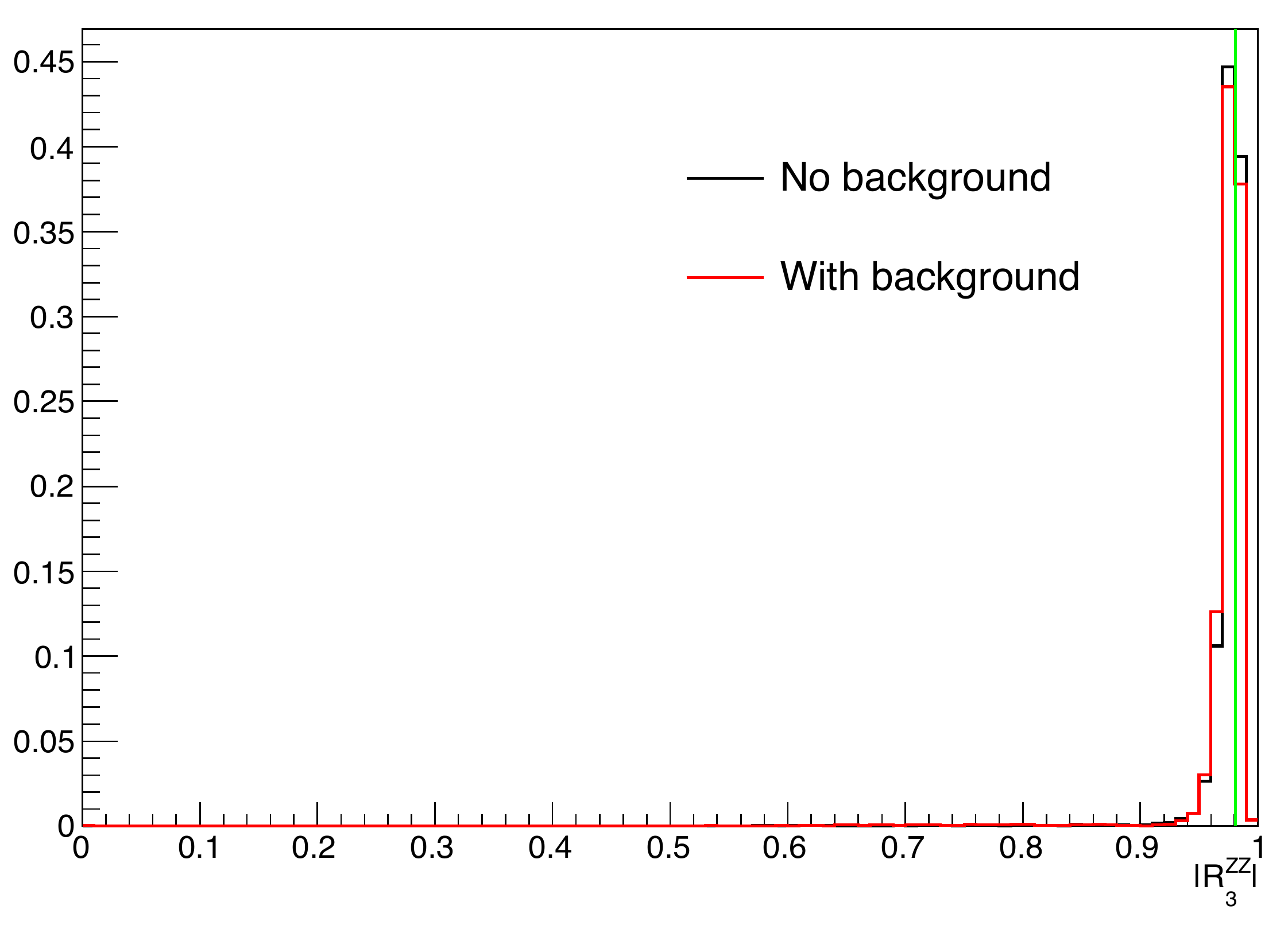}
\includegraphics[scale=.33]{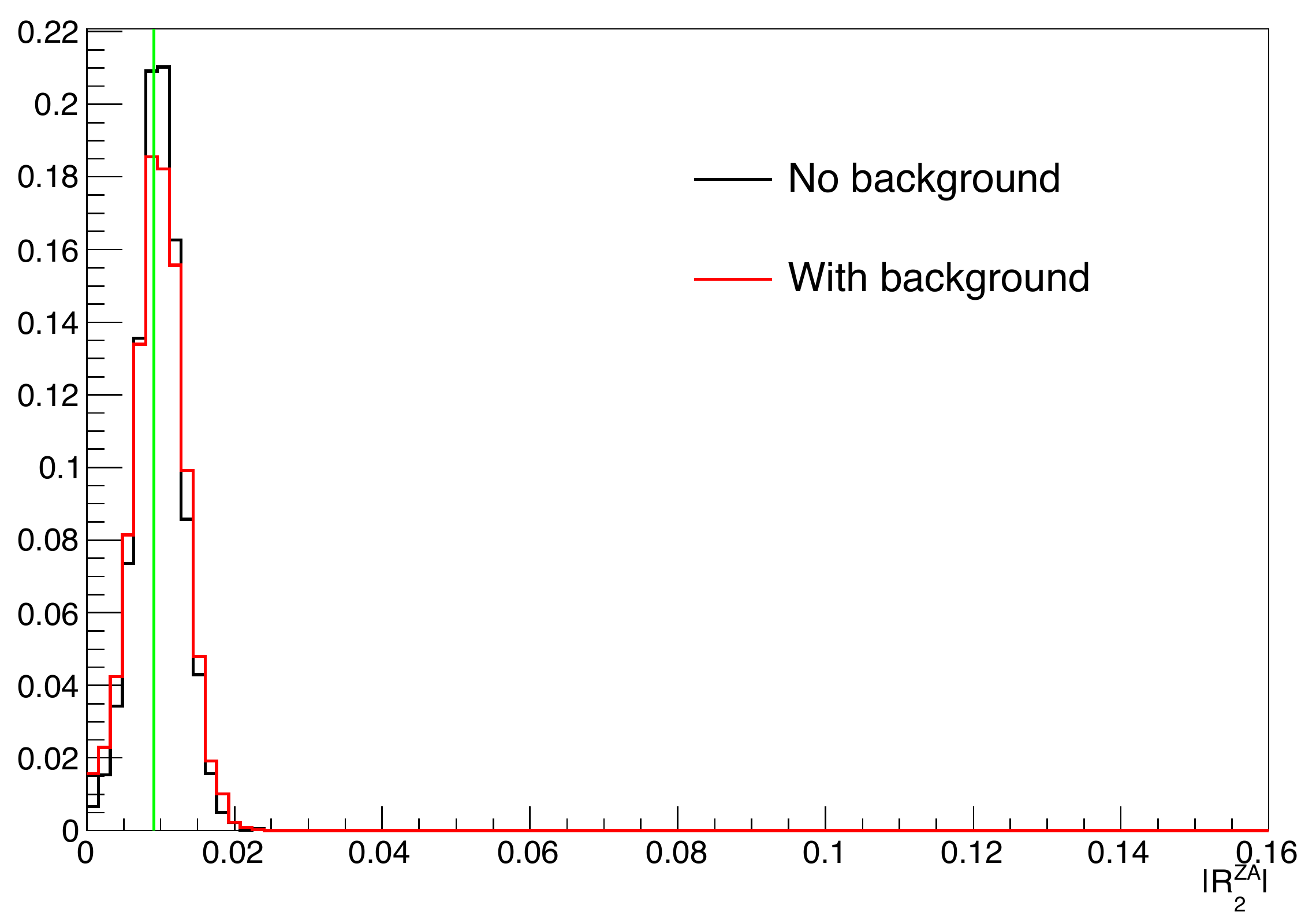}
\includegraphics[scale=.33]{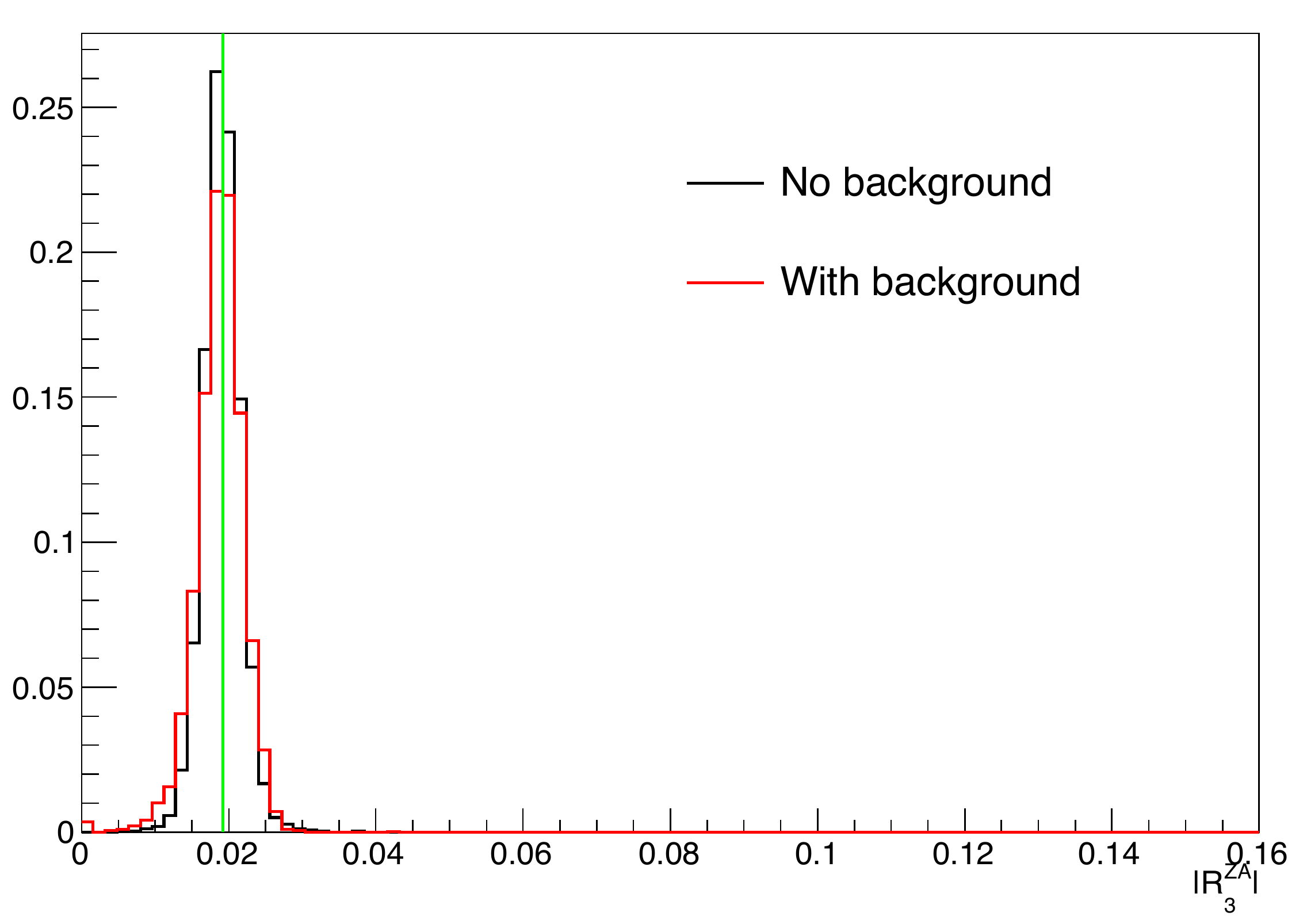}
\includegraphics[scale=.33]{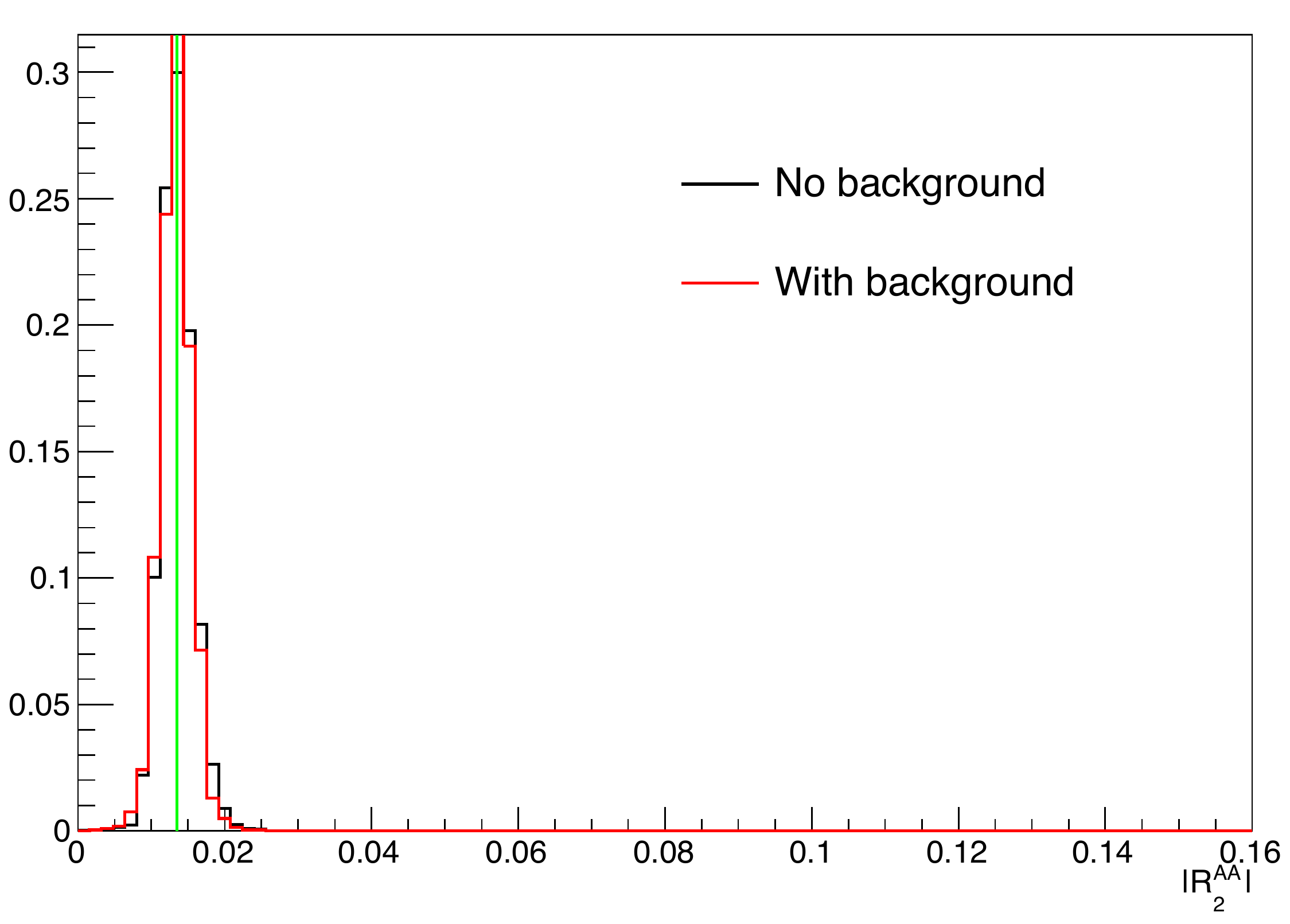}
\includegraphics[scale=.33]{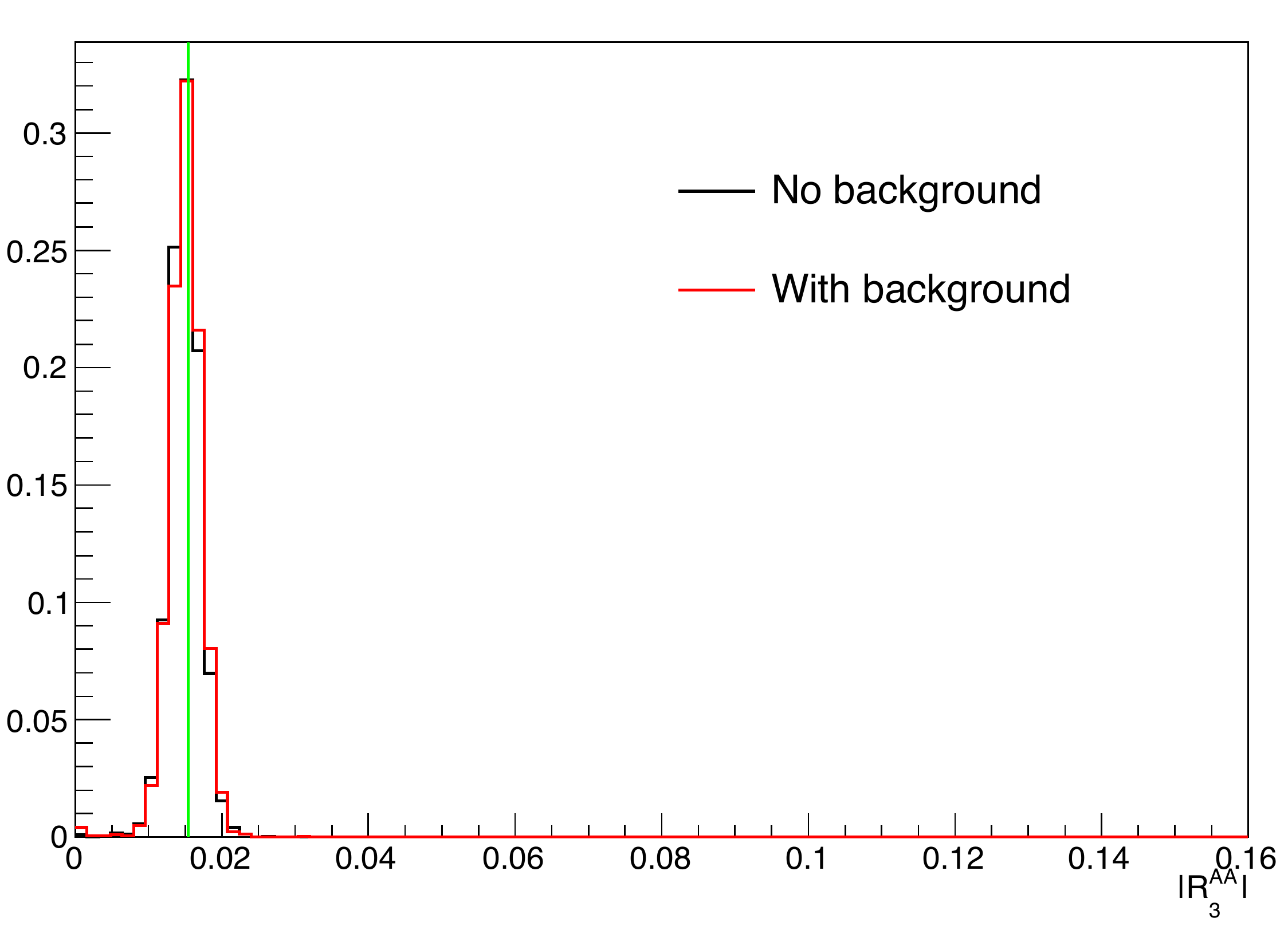} 
\caption{Results including both signal and background in the likelihood. In the top left plot we show the distribution of the extracted background fraction for 1000 signal plus 250 background events for a large set of pseudo experiments.~In the additional plots we compare the results of the parameter extraction assuming a pure signal sample (black) to those which include both signal and background (red).~We fit to the ratios $R^{ij}_n = A_{nij}/|A|$ and take the normalization to be $|A| = \sqrt{\sum\limits_{n,ij}|A_{nij}|^2}$.}
\label{fig:fits4}
\end{figure*}
\begin{figure*}
\includegraphics[scale=.32]{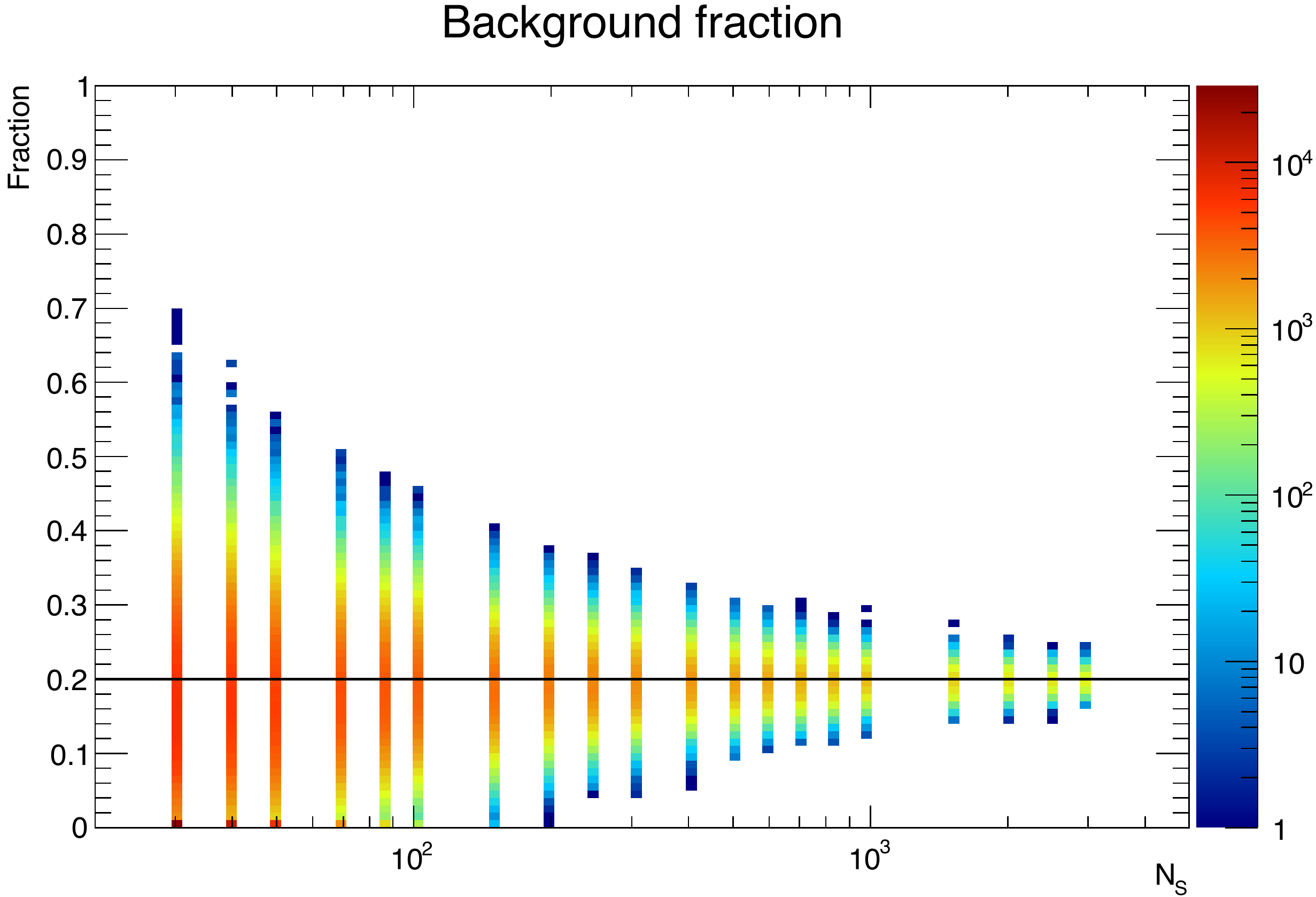}
\includegraphics[scale=.32]{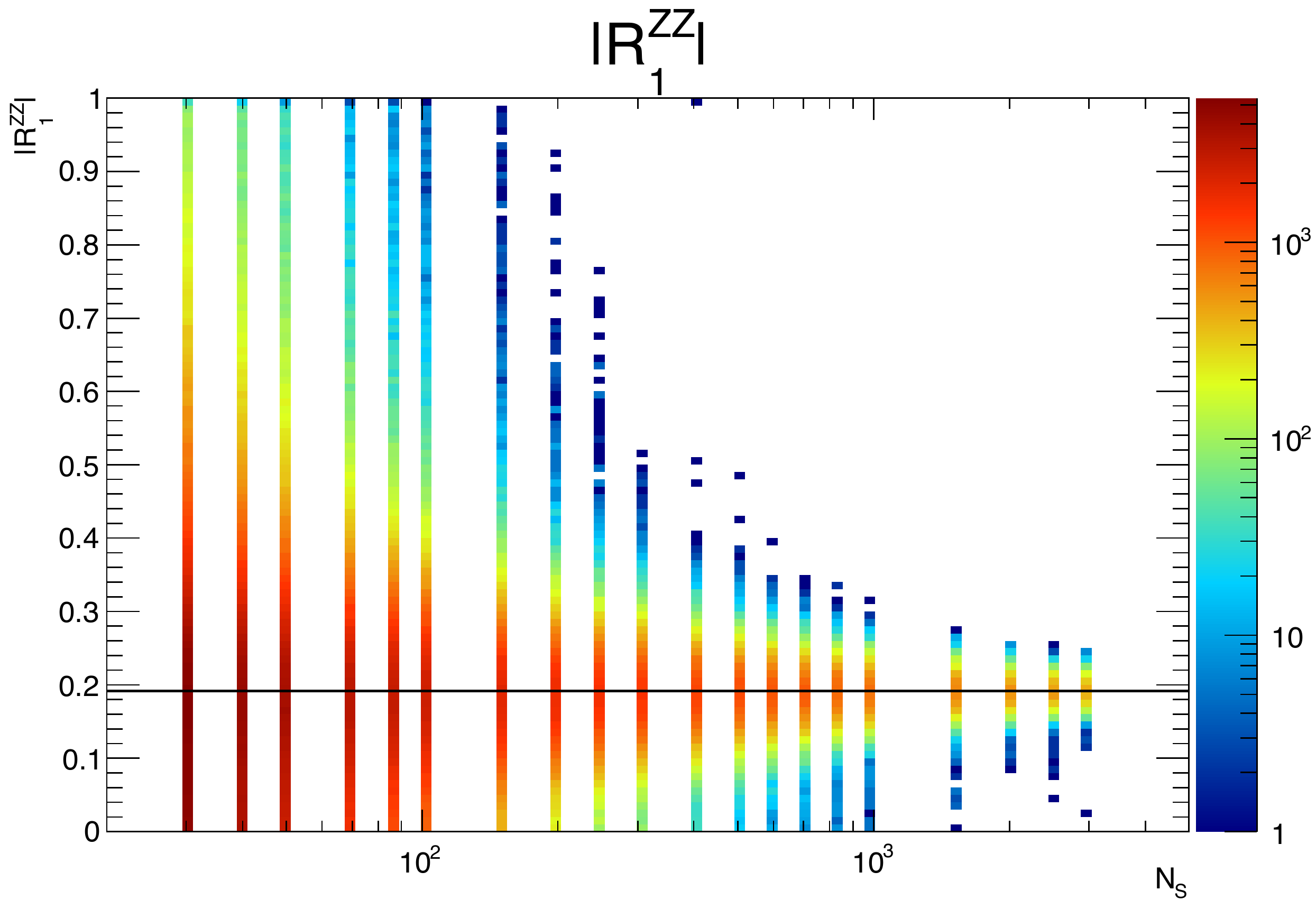}
\includegraphics[scale=.32]{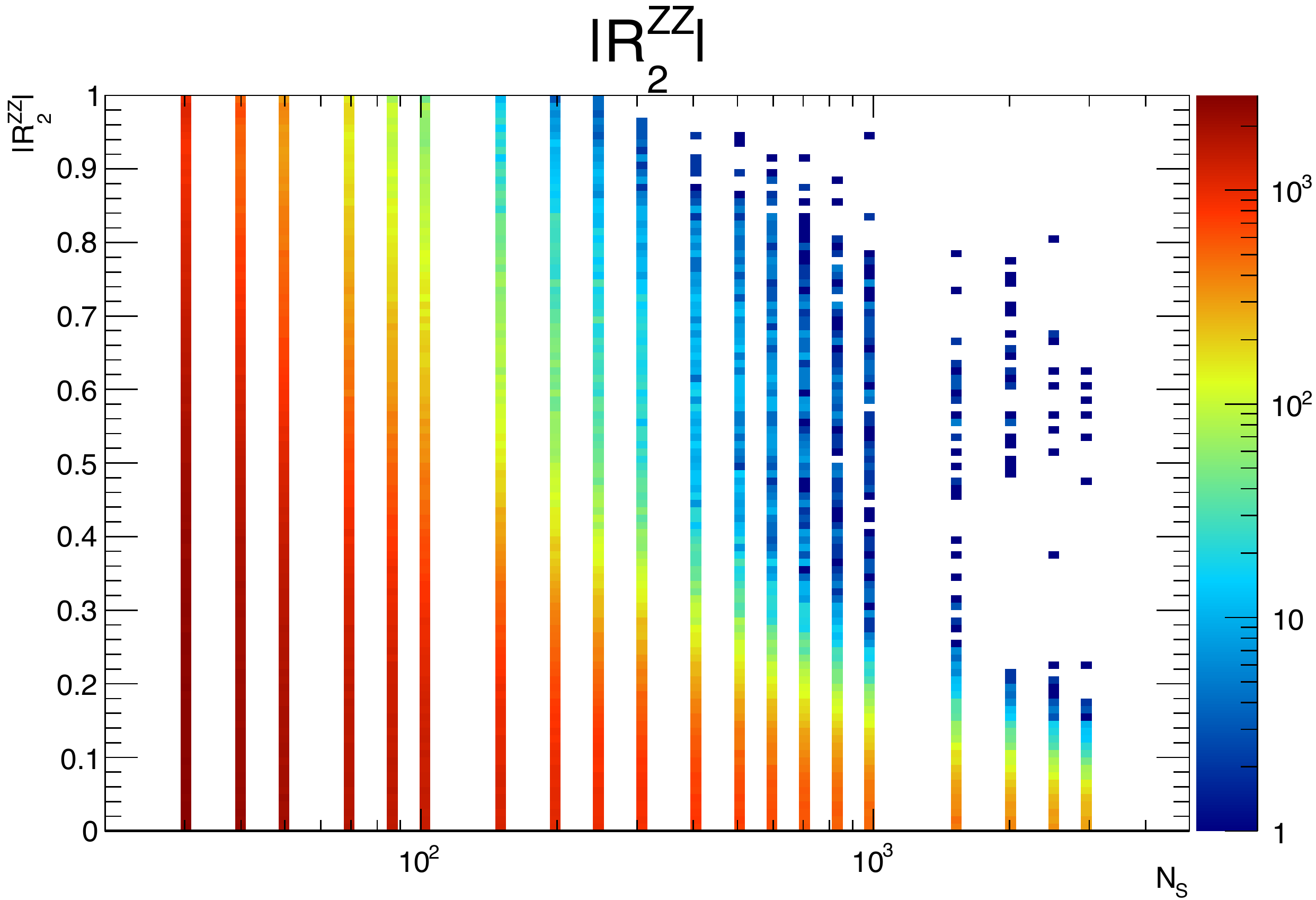}
\includegraphics[scale=.32]{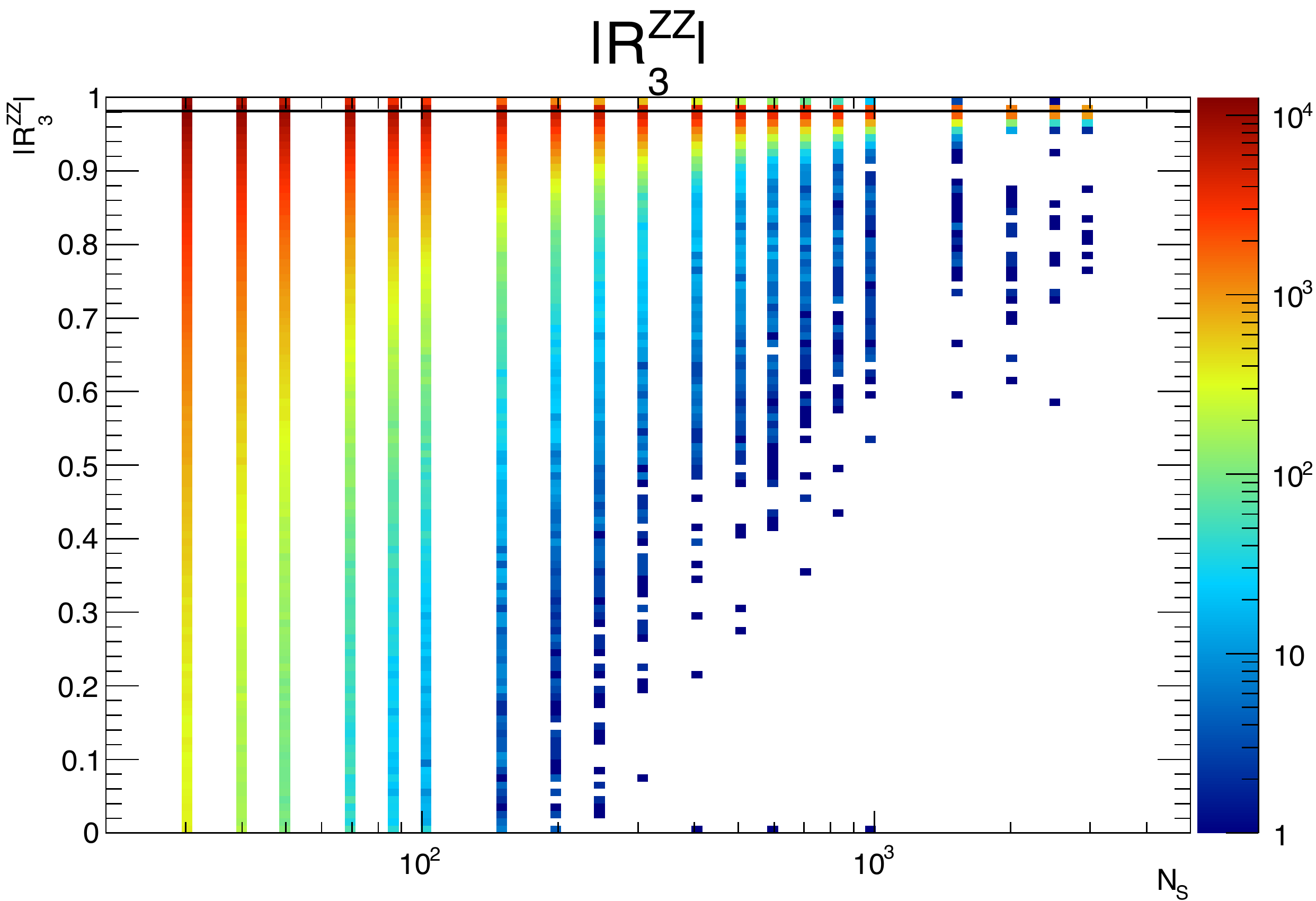}
\includegraphics[scale=.32]{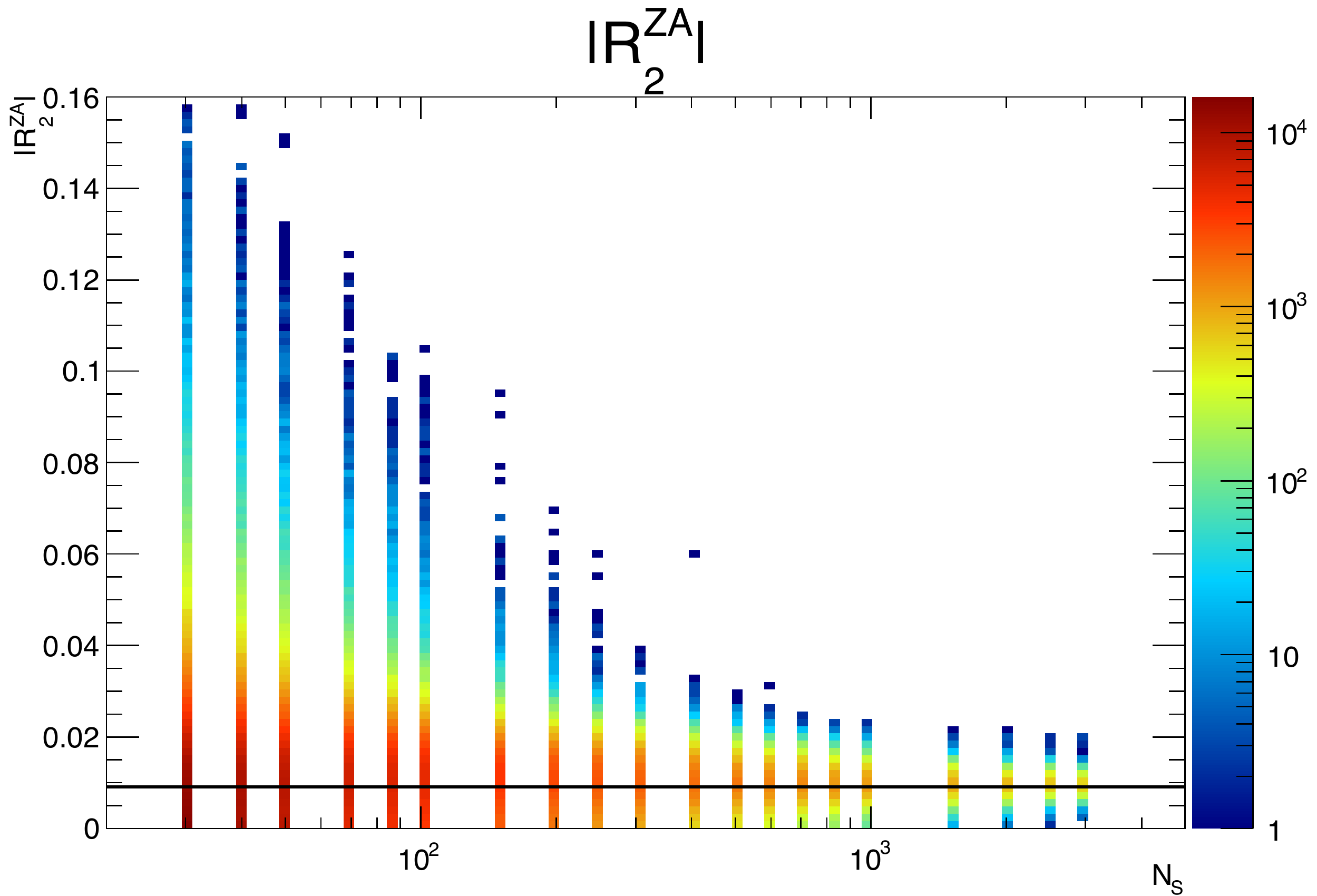}
\includegraphics[scale=.32]{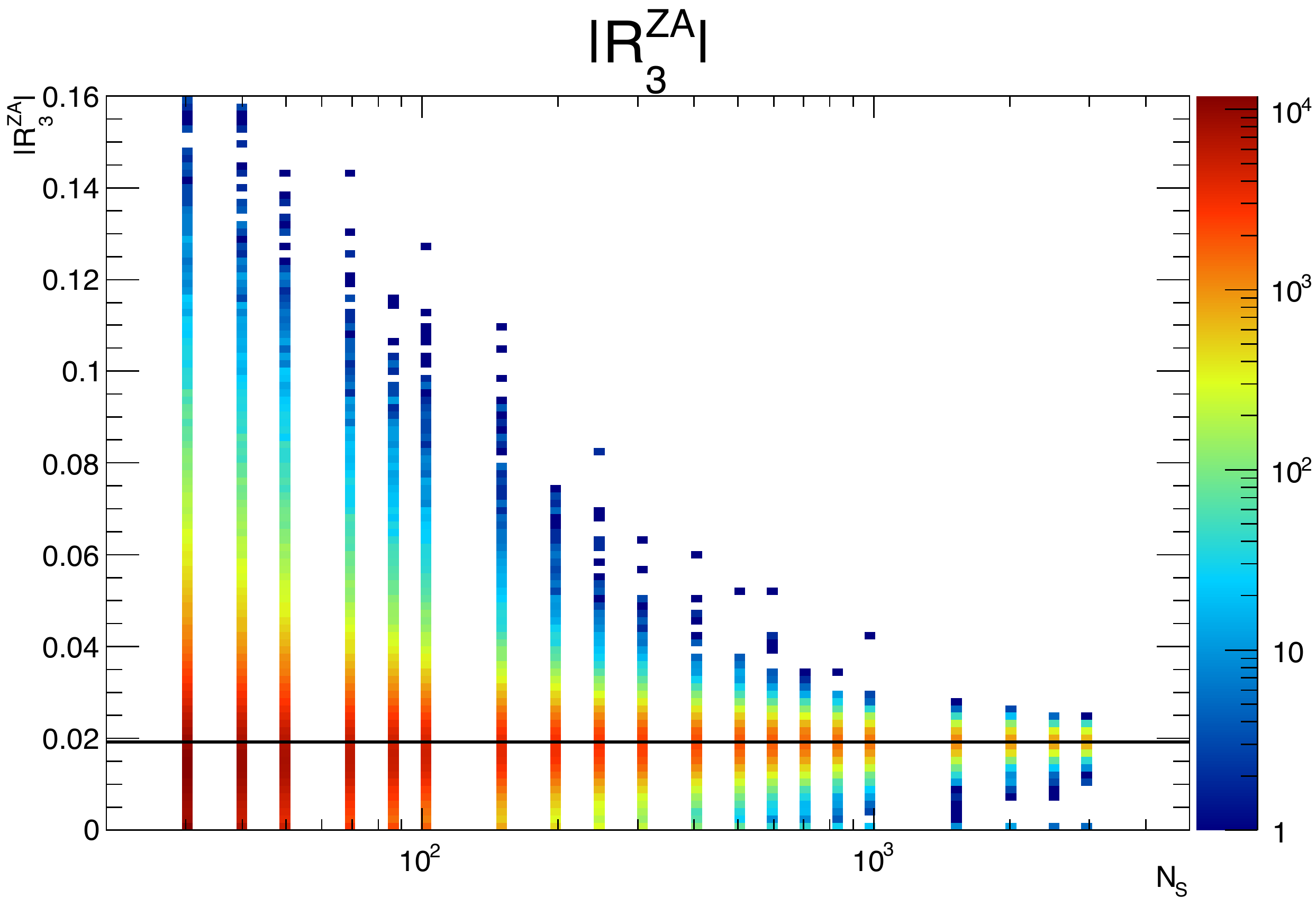}
\includegraphics[scale=.32]{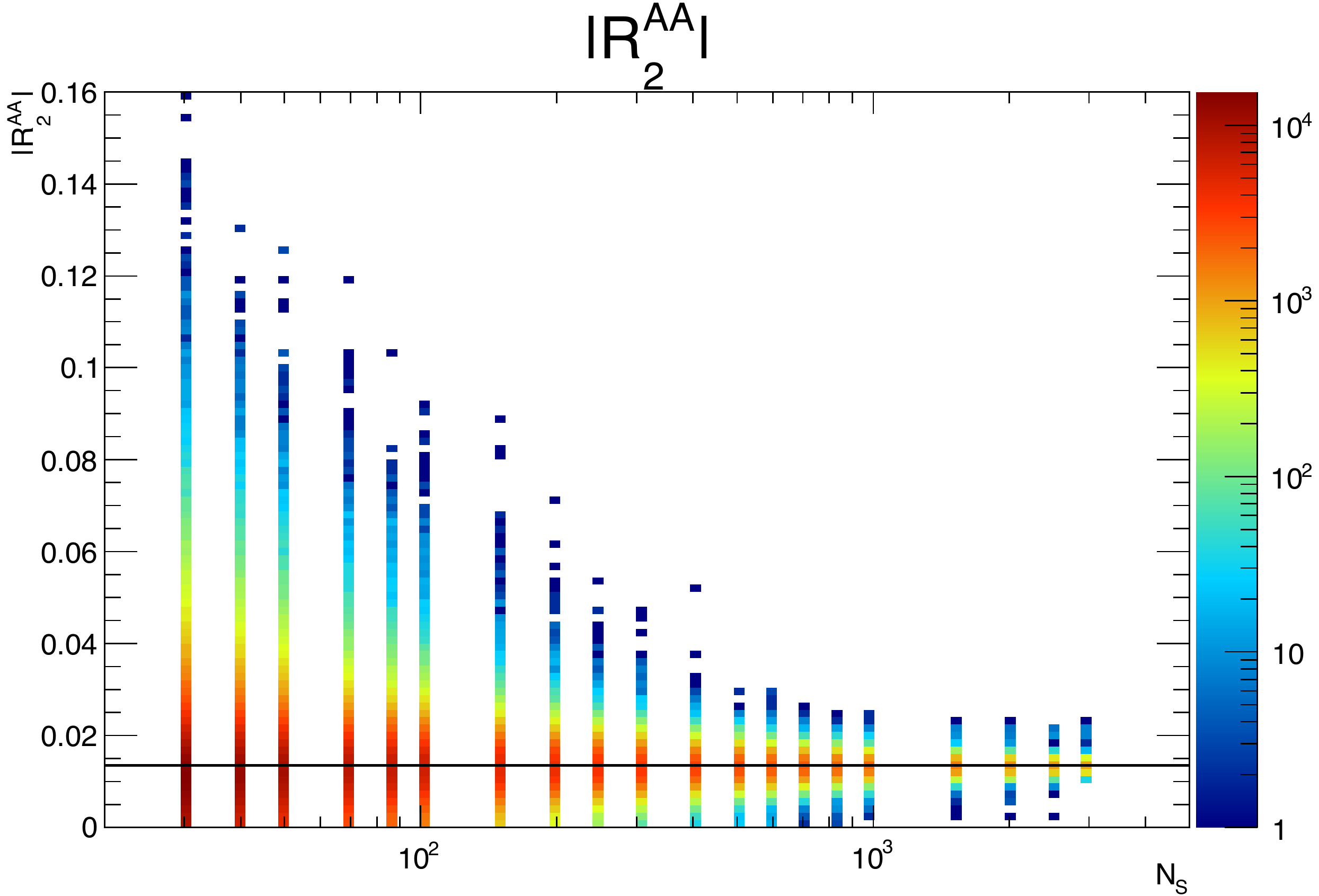}
\includegraphics[scale=.32]{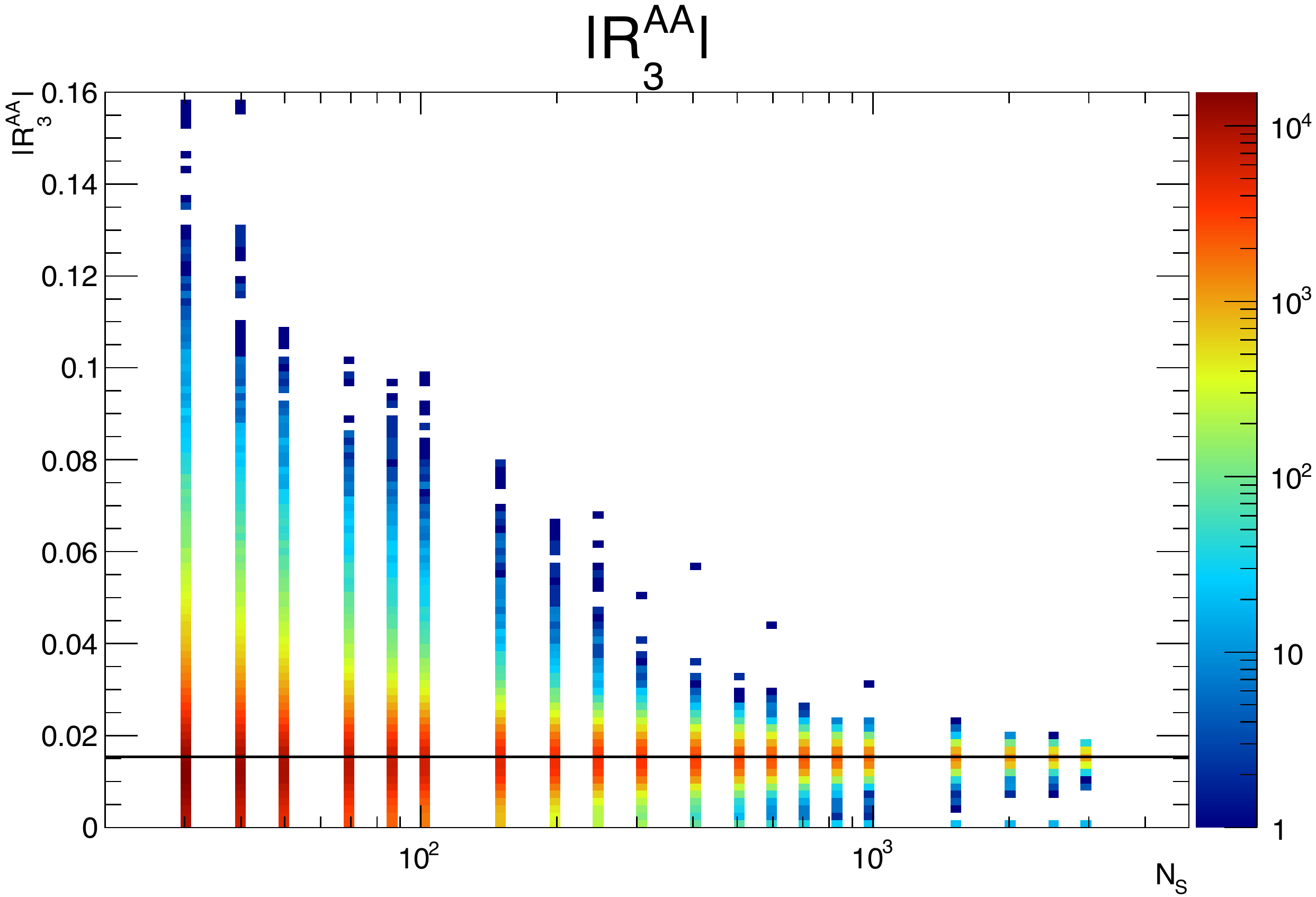} 
\caption{Here we show the distribution of extracted parameters as a function of the numbers of events for 30 to 3000 signal events per pseudo experiment with a $20\%$ background fraction.~In the top left plot we show the distribution of the extracted background fraction while in the remaining plots we show the various couplings.~The color indicates the density of pseudo experiments which return a value of the parameter as indicated on the y-axis while the true value is now indicated by the solid black line.~We fit to the ratios $R^{ij}_n = A_{nij}/|A|$ and take the normalization to be $|A| = \sqrt{\sum\limits_{n,ij}|A_{nij}|^2}$.}
\label{fig:fits2}
\end{figure*}


\section{Conclusions and Outlook}
\label{sec:conc} 
In this study we have completed the first stage in the construction of a comprehensive analysis framework which builds upon earlier studies~\cite{Gao:2010qx,Bolognesi:2012mm,Anderson:2013fba} and is aimed at extracting as much information as possible from the Higgs golden channel.~First we extended previous analytic calculations for both signal and background in the $2e2\mu$ Higgs `golden channel' to include the $4e/4\mu$ final states with the interference between identical final states.~We have presented an overview of the calculations of the expressions as well as showing various doubly differential projections and relative `partial fractions' for every combination of tensor structures.

We have also shown the potential of using these analytic expressions to perform parameter extraction of the various couplings of a spin-0 scalar to neutral electroweak gauge bosons including any correlations between parameters by implementing them into a maximum likelihood analysis.~In order to show the validity of our maximization procedure we have focused on a simplified generator level analysis which includes both signal and background at fixed $\sqrt{s}$.~As our example parameter point, we have performed a simultaneous extraction of all parameters assuming real couplings (and overall normalization) of our scalar to $ZZ$, $Z\gamma$, and $\gamma\gamma$ pairs as well as the background fraction.~We have shown that our maximum likelihood analysis gives accurate extraction of the parameters as well as the background fraction.

A more accurate analysis of course requires the inclusion of detector and systematic effects.~We have not addressed these issues here and instead have left them for a series of accompanying studies of the golden channel~\cite{Chen:2014pia,TechNote} where we also demonstrate the advantage of analytic expressions when including detector effects.~We have also neglected the use of `production variables' into our set of observables since this requires careful treatment of the production mechanism which is beyond the scope of this study and furthermore would introduce additional systematic uncertainties.~Since we fit to ratios of couplings and do not attempt to extract the overall normalization however, our results and analysis procedure are not overly sensitive to the production mechanism.~We hope to include a detailed description of the production mechanism in future studies.~In addition, we hope to conduct a detailed comparison between the sensitivity of the $2e2\mu$ and $4e/4\mu$ final states for different choices of the phase space cuts in order to determine the optimal phase space for extracting particular couplings to neutral electroweak gauge bosons. 

In summary, we have demonstrated the potential of using analytic expressions in the golden channel to extract the couplings of a spin-0 scalar to neutral electroweak gauge bosons and have completed the first stage in the construction of a comprehensive analysis framework aimed at maximizing the power of this channel.~This framework can now readily be adapted to include the relevant detector effects as well as any systematic uncertainties.

\vskip 0.2 cm

\noindent
{\bf Acknowledgments:} We thank Artur Apresyan, Emanuele DiMarco, Adam Falkowski, Andrei Gritsan, Roni Harnik, Alex Mott, Nhan Tran, Roberto Vega, Si Xie, and Felix Yu for helpful discussions and Kunal Kumar and Shashank Shalgar for help in validation with Madgraph.~We especially thank Ian Low, Joe Lykken and Maria Spiropulu for providing us with the resources and guidance necessary to complete this study.~R.V.M. is supported by the Fermilab Graduate Student Fellowship in Theoretical Physics and the ERC Advanced Grant Higg@LHC.~Fermilab is operated by Fermi Research Alliance, LLC, under Contract
No.~DE-AC02-07CH11359 with the United States Department of Energy. Y.C. is supported by the Weston Havens Foundation and DOE grant No.~DE-FG02-92-ER-40701.~This work is also sponsored in part by the DOE grant No.~DE-FG02-91ER40684.


\section{Appendix}
\label{sec:Appendix}
In this Appendix we examine the `partial fractions' of the various pairs of tensor structures which are found in Eq.(\ref{eqn:vertex}).~We also display a number of doubly differential spectra for a standard model signal as well as the $q\bar{q} \rightarrow 4\ell$ background.~Finally, we also show our validation of the signal and background calculations for the matrix element squared.

\subsection{Relative `Partial Fractions' for CMS Cuts}
\label{subsec:CMSpartial_widths}
The total decay width for $\varphi \rightarrow ZZ + Z\gamma +\gamma\gamma \rightarrow 4\ell$ can be decomposed into the various `partial widths' formed out of pairs of tensor structures in Eq.(\ref{eqn:sum_vertex}) (or operators if interpreted in terms of Eq.(\ref{eqn:siglag})).~Since each term will be quadratic in the couplings, we can label each partial width by the appropriate combination of couplings $A_{nij}A^\ast_{\bar{n}\bar{i}\bar{j}}$.~They are obtained by integrating the fully differential decay width in Eq.(\ref{eqn:final_diffcxn}) over the kinematic variables defined in Sec.~\ref{sec:events}.~We then normalize these partial widths to the standard model value to form the various `partial fractions'.

We show in Fig.~\ref{fig:CMS2e2muMatrix} a table of these partial fractions for every possible combination of $A_{nij}A^\ast_{\bar{n}\bar{i}\bar{j}}$ which can contribute to the $2e2\mu$ decay width.~For these partial fractions we now take a `CMS-like' phase space of $40~GeV\lesssim M_{1}, 12~GeV\lesssim M_{2}$ and $\sqrt{s} = 125~GeV$ as well as $p_{T\ell} > 20, 10, 7, 7~GeV$ for the ordering of final state lepton $p_T$ and $|\eta_\ell| < 2.4$ for their rapidity.~The couplings $A_{nij}$ have been separated into their real and imaginary components as $A_{nij} = A_{nijR} + iA_{nijI}$ and we have set all $A_{nijR,I} =1$.~All of the $|A_{nijR,I}|^2$ terms sit along the diagonal with the various interference terms making up the off-diagonal terms.~Note that many of the interference terms are negative indicating destructive interference between the corresponding tensor structures (or operators).

In Fig.~\ref{fig:CMS4eMatrix} we show the same plot for the $4e$ final state.~One can see the change in the partial fractions and in particular the $ZZ/\gamma\gamma$ interference terms are significantly larger than in the $2e2\mu$ channel.~The blank entries indicate terms which are identically zero after integration.~We can see that these entries are those for which CP violation in the form of interference between $A_{1,2}$ and $A_3$ tensor structures would occur.~This is indicative of the fact that after one integrates over the kinematic all information on CP violation is lost.~Of course for the fully differential decay width many of these terms are non-zero in principal allowing for sensitivity to CP violation in the golden channel.

Since all couplings are set to one, these tables essentially show how much each combination of tensor structures contributes to the $\varphi \rightarrow 4\ell$ phase space relative to the contribution from the partial width for which we have set $A_{1ZZ} = 2$ and all other couplings to zero.~From these values of the relative partial fractions, one can gain some insight into which combination of operators the golden channel might be most sensitive to.~Furthermore, for a specific model one can take the prediction for the values of the various couplings and simply multiply by the numbers given in Fig.~\ref{fig:CMS2e2muMatrix}-\ref{fig:CMS4eMatrix} to get a feel for whether those couplings might be probed in the golden channel.

For most realistic models, all couplings apart from $A_{1ZZ}$ are generated by higher dimensional operators and are expected to be small.~In Figs.~\ref{fig:CMS2e2muMatrixSM}-\ref{fig:CMS4eMatrixSM} we also show the same tables for the standard model prediction including the $Z\gamma$ and $\gamma\gamma$ couplings for which we have $A_{1ZZ} = 2, A_{2Z\gamma} \simeq 0.007, A_{2Z\gamma} \simeq -0.008$~\cite{Low:2012rj}\footnote{Note that there is a conversion done here from the parametrization in~\cite{Low:2012rj} to ours.} while all other couplings zero.~These values are normalized the same as in Figs.~\ref{fig:CMS2e2muMatrix}-\ref{fig:CMS4eMatrix}.~Of course, these rates do not contain information about the shapes in the various distributions so in principal the sensitivity is greater than might be inferred from these values.~Whether or not the golden channel has sensitivity to these couplings in the standard model requires careful study, which we leave for ongoing work.

In Figs.~\ref{fig:CMS2e2muAbsMatrix} and~\ref{fig:CMS4eAbsMatrix} we show the integral of the absolute value of the differential decay width.~This gives a better indication of the shape differences in the different combinations of operators since some of them can integrate to zero when the absolute value is not taken.~Furthermore we can see in this table some of the potential sensitivity in the golden channel to CP violation.~Note that there are two sources of CP violation which occur.~One is due to the interference between the $A_{1,2}$ and $A_3$ tensor structures, while the other occurs in the interference between the real and imaginary components of the couplings from different tensor structures.

One could also imagine attempting to find different sets of cuts in order to maximize the contribution of a particular combination of operators.~In addition, the sensitivity between the $2e2\mu$ and $4e/4\mu$ final states may differ depending on the phase space that is chosen. We leave a detailed investigation of this issue to future work.~These tables, however, obviously only give a partial picture of the sensitivity to the different operator combinations and are meant to be used only as a guide.~Of course when performing parameter extraction the full kinematic information of the differential decay width is used.~Obviously, for a scalar resonance with a mass much larger or smaller than $125~GeV$ these numbers can change significantly. 

\begin{figure*}
\includegraphics[width=0.8\textwidth]{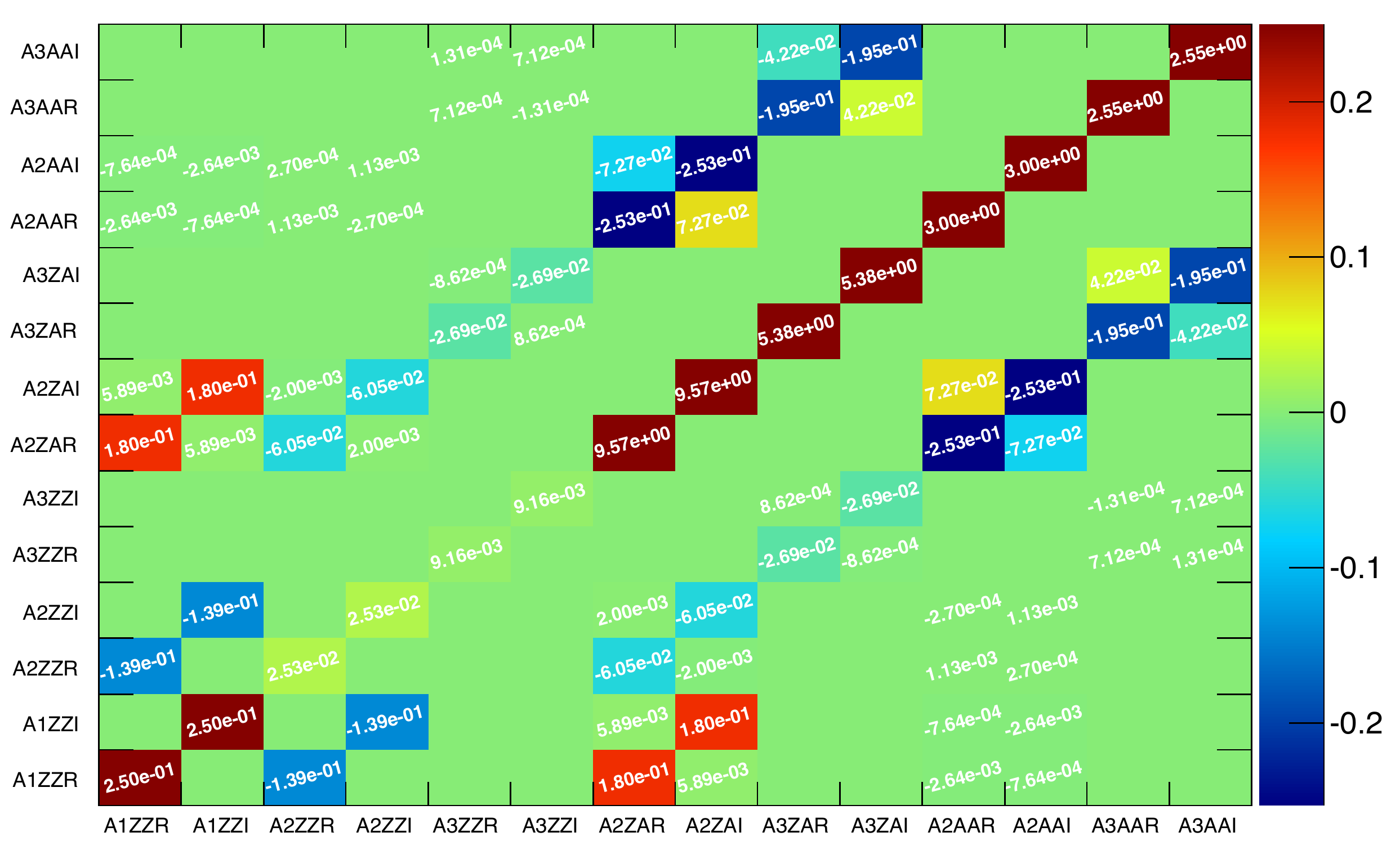}
\caption{Here we plot the relative partial fractions for each possible combination of tensor structures in Eq.(\ref{eqn:vertex}) for the $2e2\mu$ final state.~For these partial fractions we take a `CMS-like' phase space of $40~GeV\lesssim M_{1}, 12~GeV\lesssim M_{2}$, and $\sqrt{s} = 125~GeV$ as well as $p_{T\ell} > 20, 10, 7, 7~GeV$ for the ordering of final state lepton $p_T$ and $|\eta_\ell| < 2.4$ for their rapidity.~They have been normalized to the standard model value for which we take $A_{1ZZ} = 2$ and all other couplings zero.}
\label{fig:CMS2e2muMatrix}
\end{figure*}
\begin{figure*}
\includegraphics[width=0.8\textwidth]{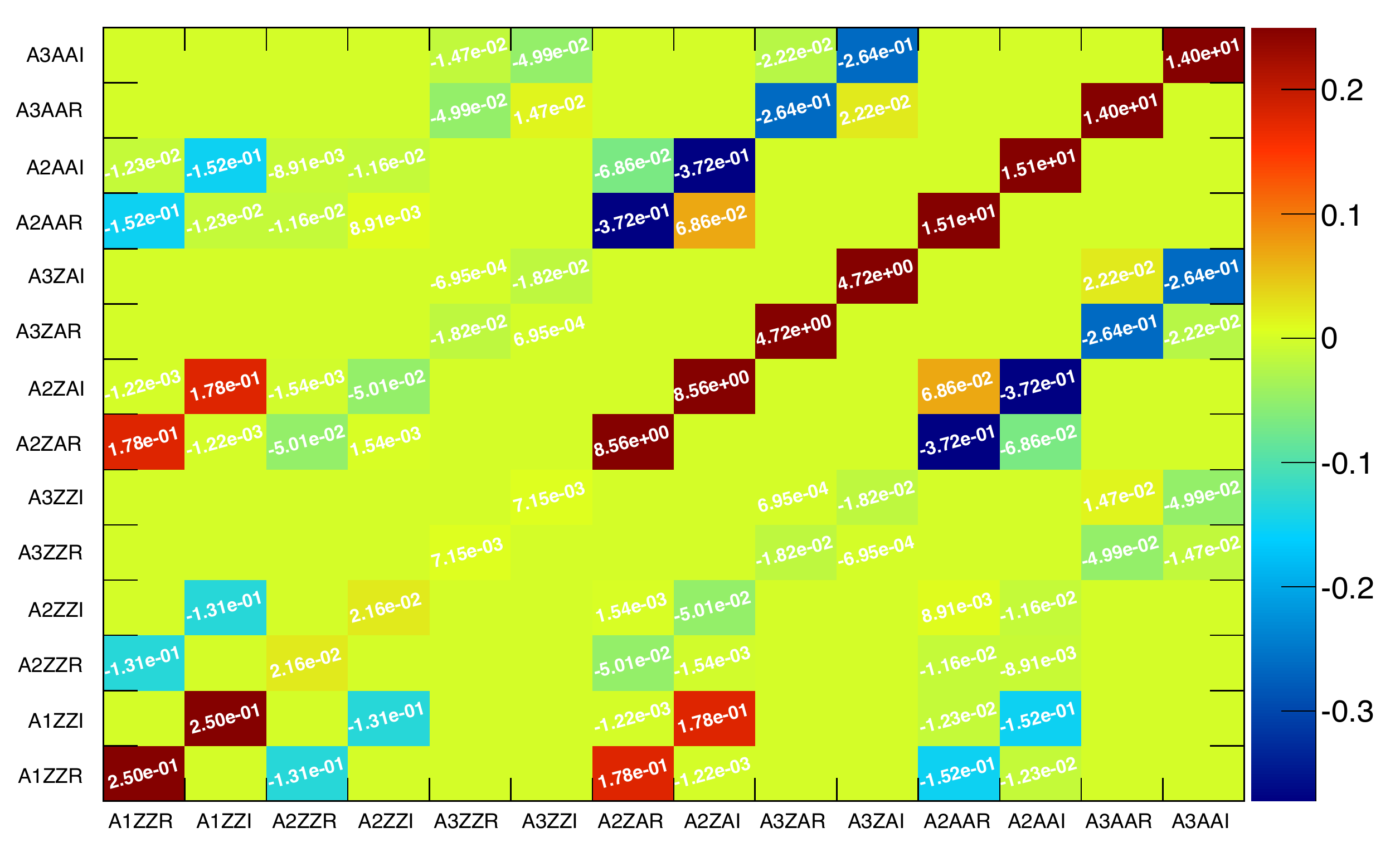}
\caption{Here we plot the relative partial fractions for each possible combination of tensor structures in Eq.(\ref{eqn:vertex}) for the $4e/4\mu$ final state.~For these partial fractions we take a `CMS-like' phase space of $40~GeV\lesssim M_{1}, 12~GeV\lesssim M_{2}$, and $\sqrt{s} = 125~GeV$ as well as $p_{T\ell} > 20, 10, 7, 7~GeV$ for the ordering of final state lepton $p_T$ and $|\eta_\ell| < 2.4$ for their rapidity.~They have been normalized to the standard model value for which we take $A_{1ZZ} = 2$ and all other couplings zero.}
\label{fig:CMS4eMatrix}
\end{figure*}
\begin{figure*}
\includegraphics[width=0.8\textwidth]{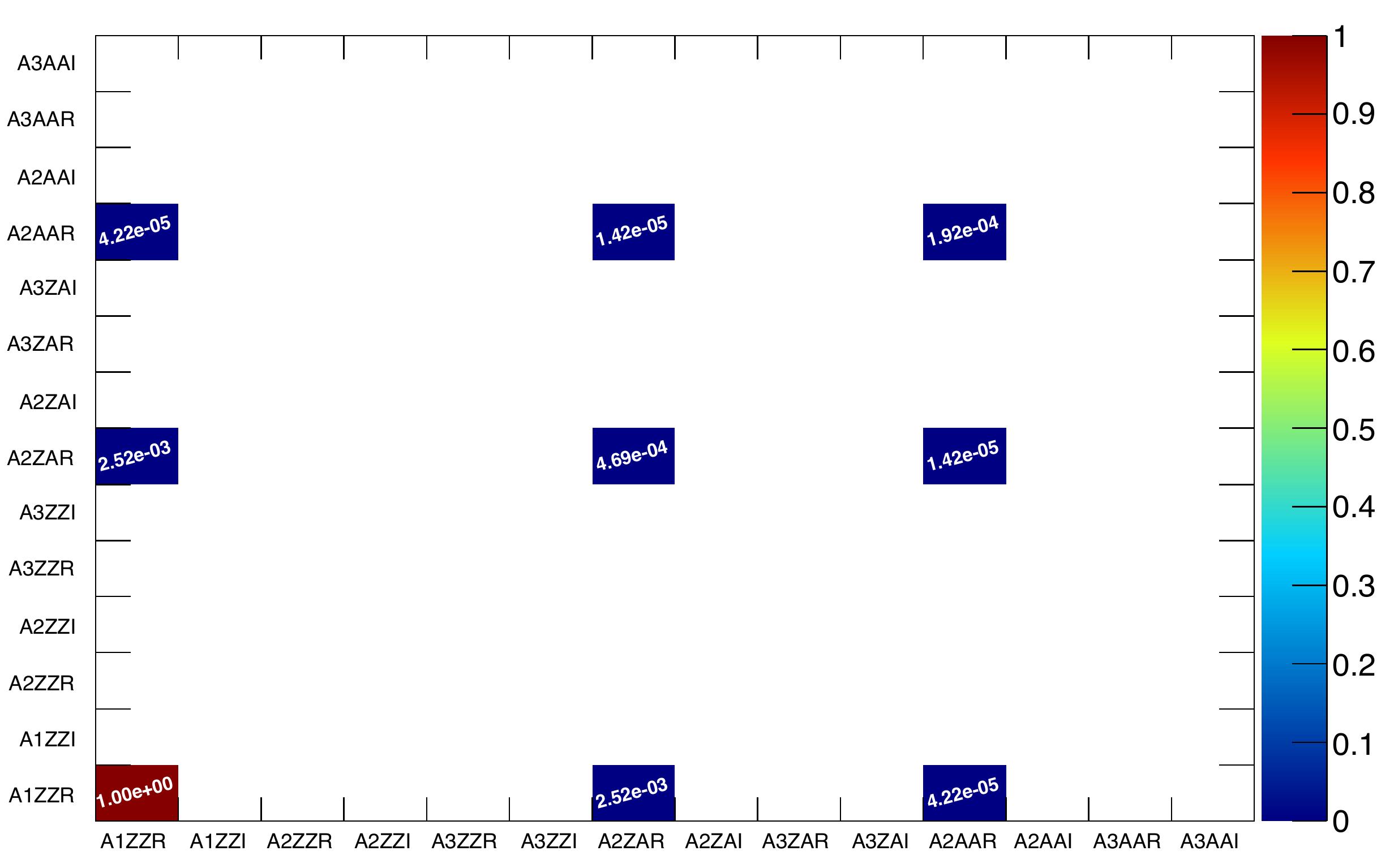}
\caption{Here we plot the relative partial fractions for the standard model values including the $Z\gamma$ and $\gamma\gamma$ couplings ($A_{1ZZ} = 2, A_{2Z\gamma} \simeq 0.007, A_{2\gamma\gamma} \simeq -0.008$)~\cite{Low:2012rj} in the $2e2\mu$ final state.~For these partial fractions we take a `CMS-like' phase space of $40~GeV\lesssim M_{1}, 12~GeV\lesssim M_{2}$ and $\sqrt{s} = 125~GeV$ as well as $p_{T\ell} > 20, 10, 7, 7~GeV$ for the ordering of final state lepton $p_T$ and $|\eta_\ell| < 2.4$ for their rapidity.~They have been normalized to the partial width where we take $A_{1ZZ} = 2$ and all other couplings zero.}
\label{fig:CMS2e2muMatrixSM}
\end{figure*}
\begin{figure*}
\includegraphics[width=0.8\textwidth]{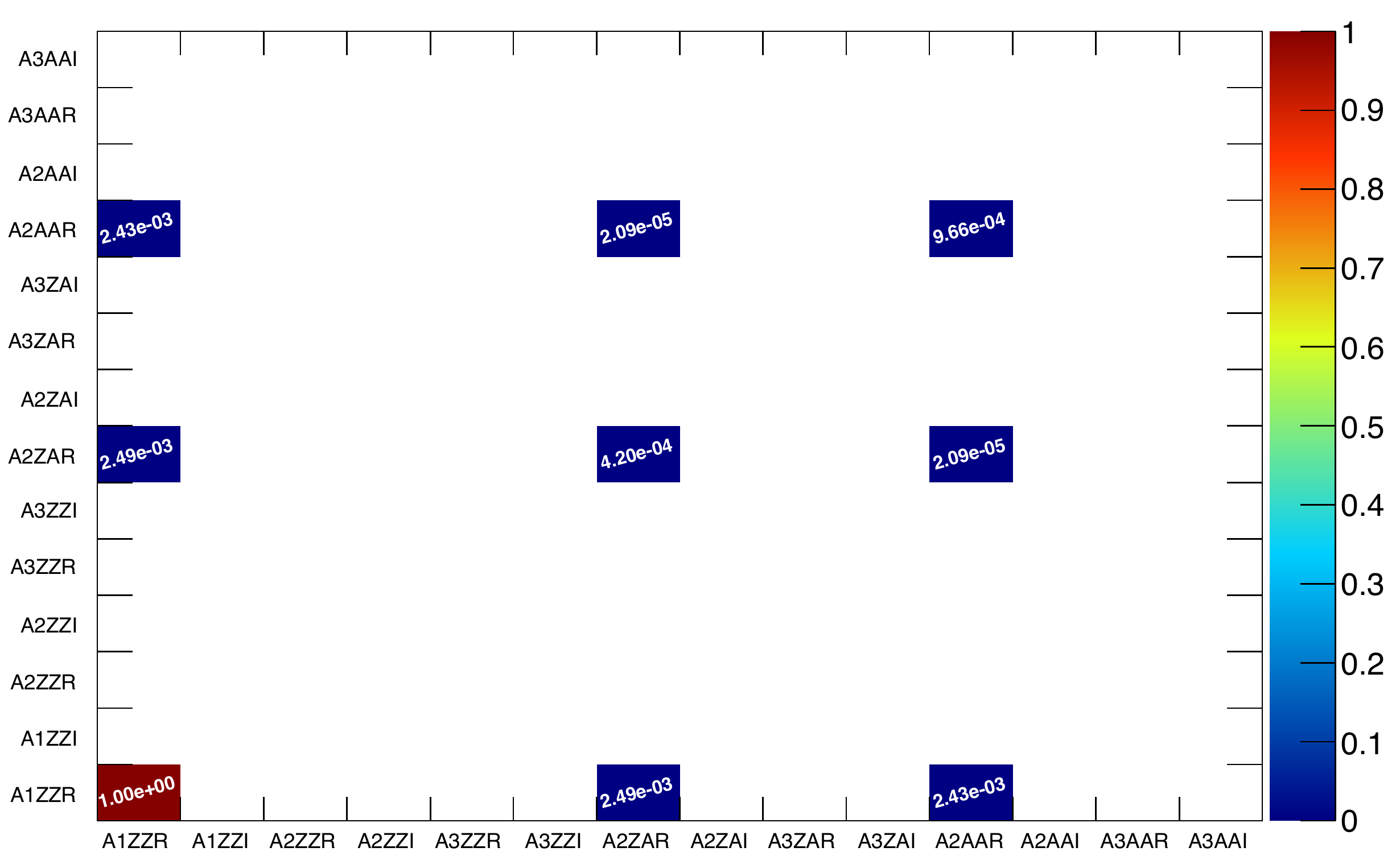}
\caption{Here we plot the relative partial fractions for the standard model values including the $Z\gamma$ and $\gamma\gamma$ couplings ($A_{1ZZ} = 2, A_{2Z\gamma} \simeq 0.007, A_{2\gamma\gamma} \simeq -0.008$)~\cite{Low:2012rj} in the $4e/4\mu$ final state.~For these partial fractions we take a `CMS-like' phase space of $40~GeV\lesssim M_{1}, 12~GeV\lesssim M_{2}$ and $\sqrt{s} = 125~GeV$ as well as $p_{T\ell} > 20, 10, 7, 7~GeV$ for the ordering of final state lepton $p_T$ and $|\eta_\ell| < 2.4$ for their rapidity.~They have been normalized to the partial width where we take $A_{1ZZ} = 2$ and all other couplings zero.}
\label{fig:CMS4eMatrixSM}
\end{figure*}
\begin{figure*}
\includegraphics[width=0.8\textwidth]{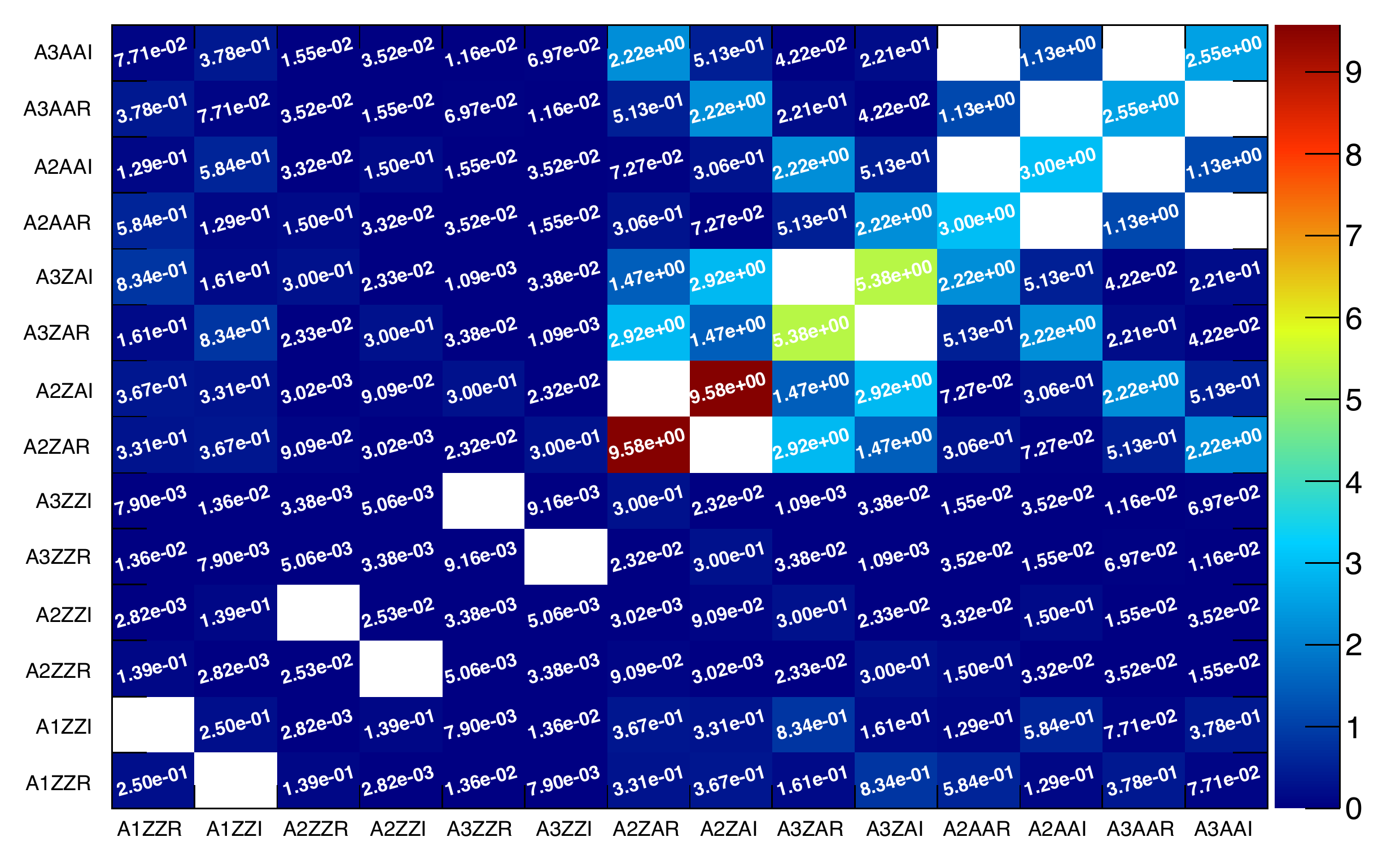}
\caption{Here we plot the relative integrated absolute value of the fully differential decay rate for each possible combination of vertex structures in Eq.(\ref{eqn:vertex}) for the $2e2\mu$ final state.~For these partial fractions we take a `CMS-like' phase space of $40~GeV\lesssim M_{1}, 12~GeV\lesssim M_{2}$ and $\sqrt{s} = 125~GeV$ as well as $p_{T\ell} > 20, 10, 7, 7~GeV$ for the ordering of final state lepton $p_T$ and $|\eta_\ell| < 2.4$ for their rapidity.~They have been normalized to the partial width where we take $A_{1ZZ} = 2$ and all other couplings zero.}
\label{fig:CMS2e2muAbsMatrix}
\end{figure*}
\begin{figure*}
\includegraphics[width=0.8\textwidth]{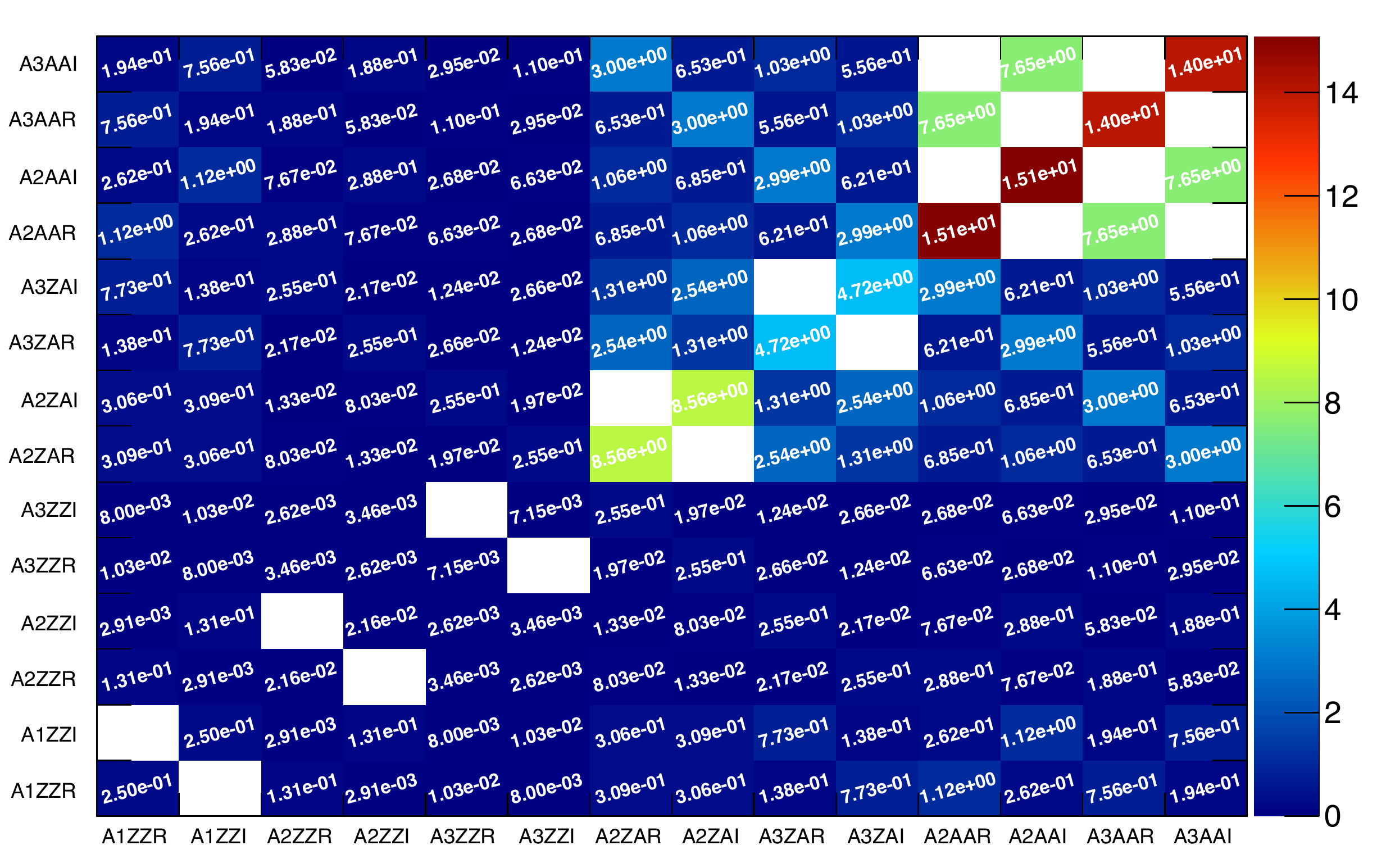}
\caption{Here we plot the relative integrated absolute value of the fully differential decay rate for each possible combination of vertex structures in Eq.(\ref{eqn:vertex}) for the $4e/4\mu$ final state.~For these partial fractions we take a `CMS-like' phase space of $40~GeV\lesssim M_{1}, 12~GeV\lesssim M_{2}$ and $\sqrt{s} = 125~GeV$ as well as $p_{T\ell} > 20, 10, 7, 7~GeV$ for the ordering of final state lepton $p_T$ and $|\eta_\ell| < 2.4$ for their rapidity.~They have been normalized to the partial width where we take $A_{1ZZ} = 2$ and all other couplings zero.}
\label{fig:CMS4eAbsMatrix}
\end{figure*}

\subsection{Doubly Differential spectra}
\label{sec:bgdists}
In Fig.~\ref{fig:sig_doubdists}-\ref{fig:bg_doubdists} we show various combinations of the doubly differential spectra for both the signal and background in the $4e/4\mu$ final state.~These are primarily for illustration purposes, but from these one can get an idea of the correlations between the different kinematic variables.~One can also see from these spectra the strong discriminating power between the signal and background in the golden channel. For the signal plots in Fig.~\ref{fig:sig_doubdists} we only show the standard model result for which only $A_{1ZZ}$ is non-zero\footnote{Doubly differential spectra for a variety of other signal hypothesis and phase space cuts as well as other pairs of variables not shown here can be found on~\cite{Chen:2012jy,WEBSITE}.}.~The background spectra are shown in Fig.~\ref{fig:bg_doubdists}.~For all distributions the phase space is defined as 4~GeV $< M_1 < 120$~GeV and 4~GeV $< M_2 <$ 120~GeV with $\sqrt{s}=125$~GeV for signal and background. We also take $|\eta_\ell | < 2.4$ and $p_{T\ell} > 2~GeV$ for the lepton rapidity and transverse momentum.~For these distributions we show the $(M_1, M_2)$, $(M_1, \Phi_1)$, $(M_1, \Phi)$, $(M_2, \Phi_1)$, $(M_2, \Phi)$, $(\Phi_1, \Phi)$ doubly differential spectra.
\begin{figure*}
\includegraphics[width=0.32\textwidth]{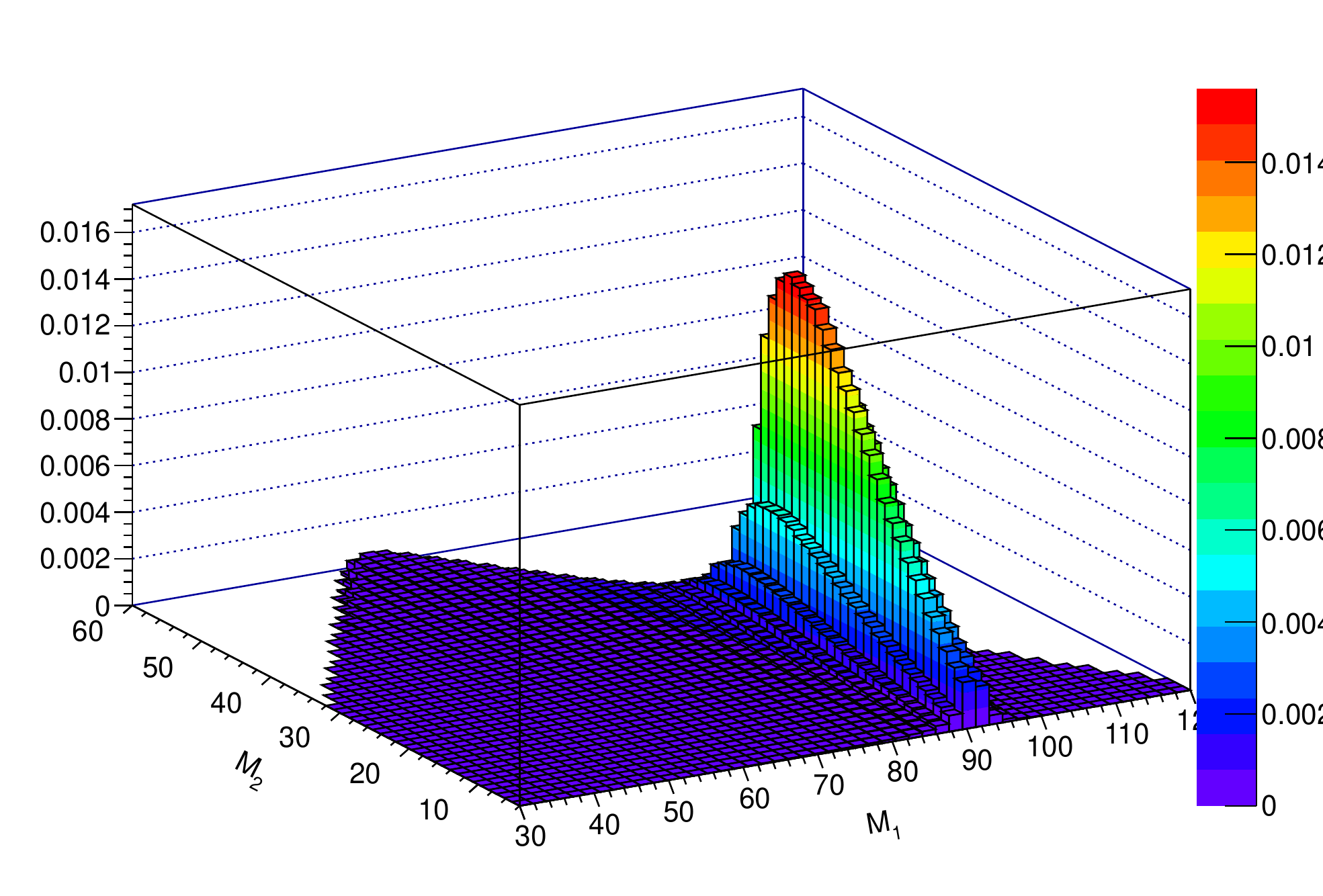}
\includegraphics[width=0.32\textwidth]{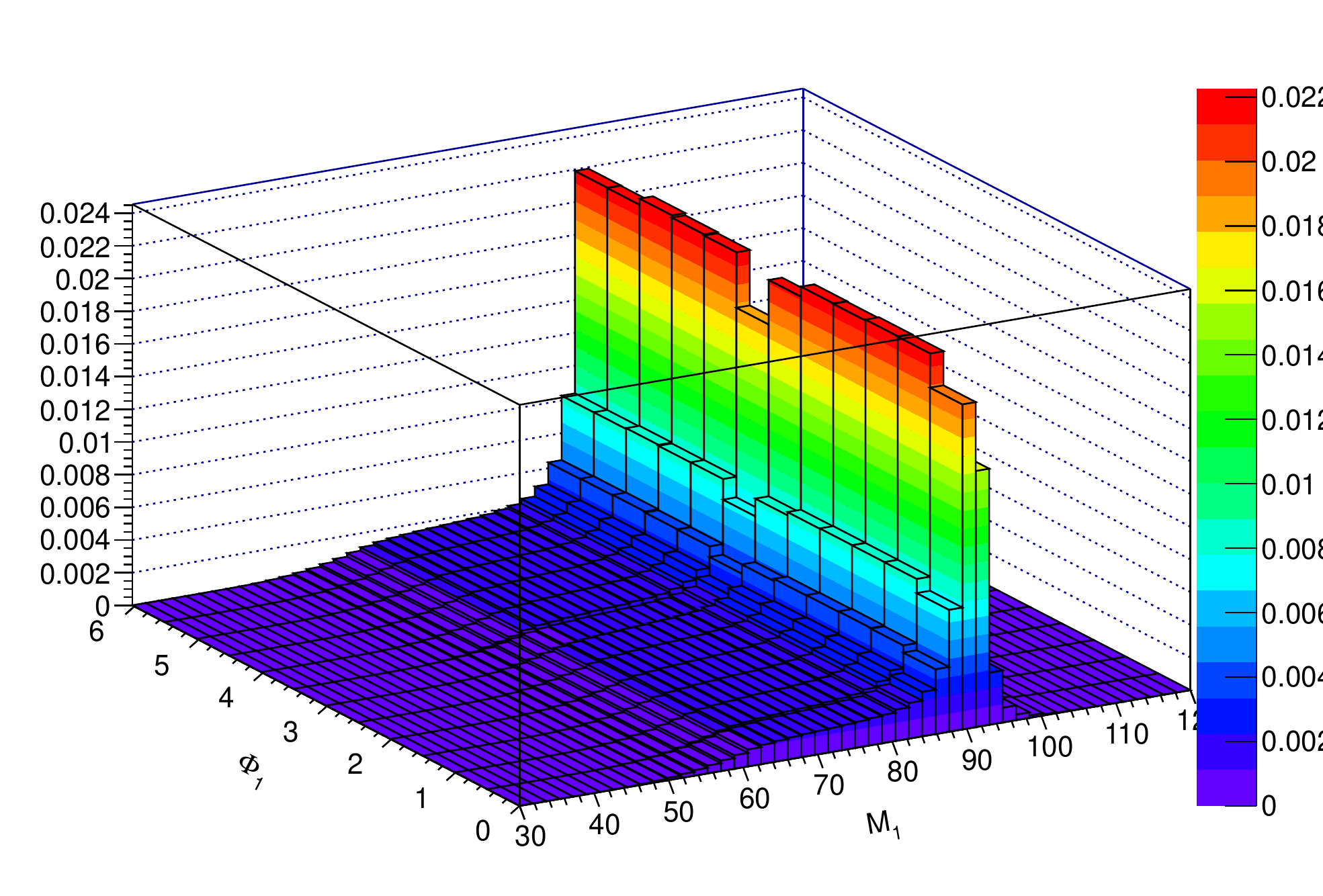}
\includegraphics[width=0.32\textwidth]{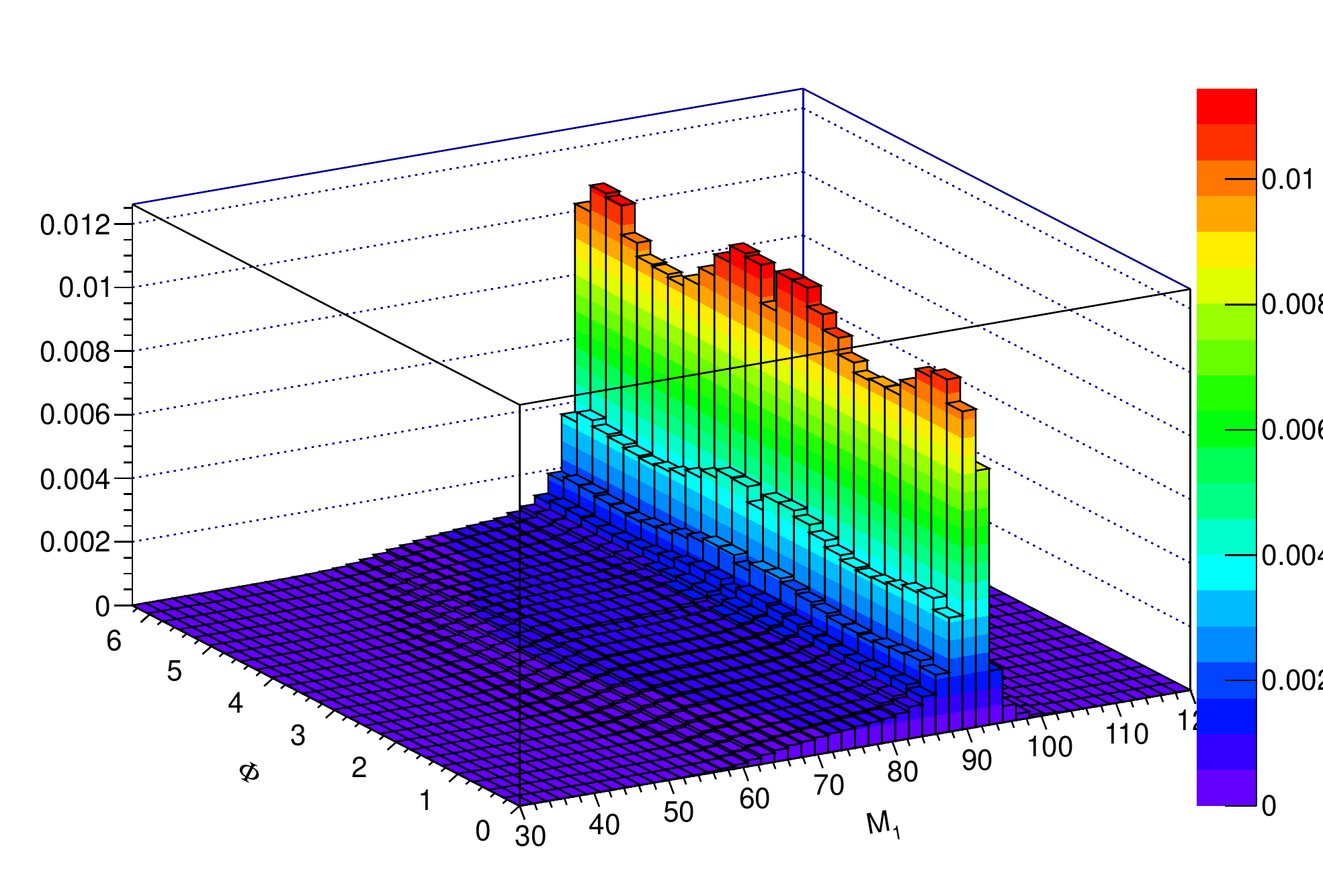}
\includegraphics[width=0.32\textwidth]{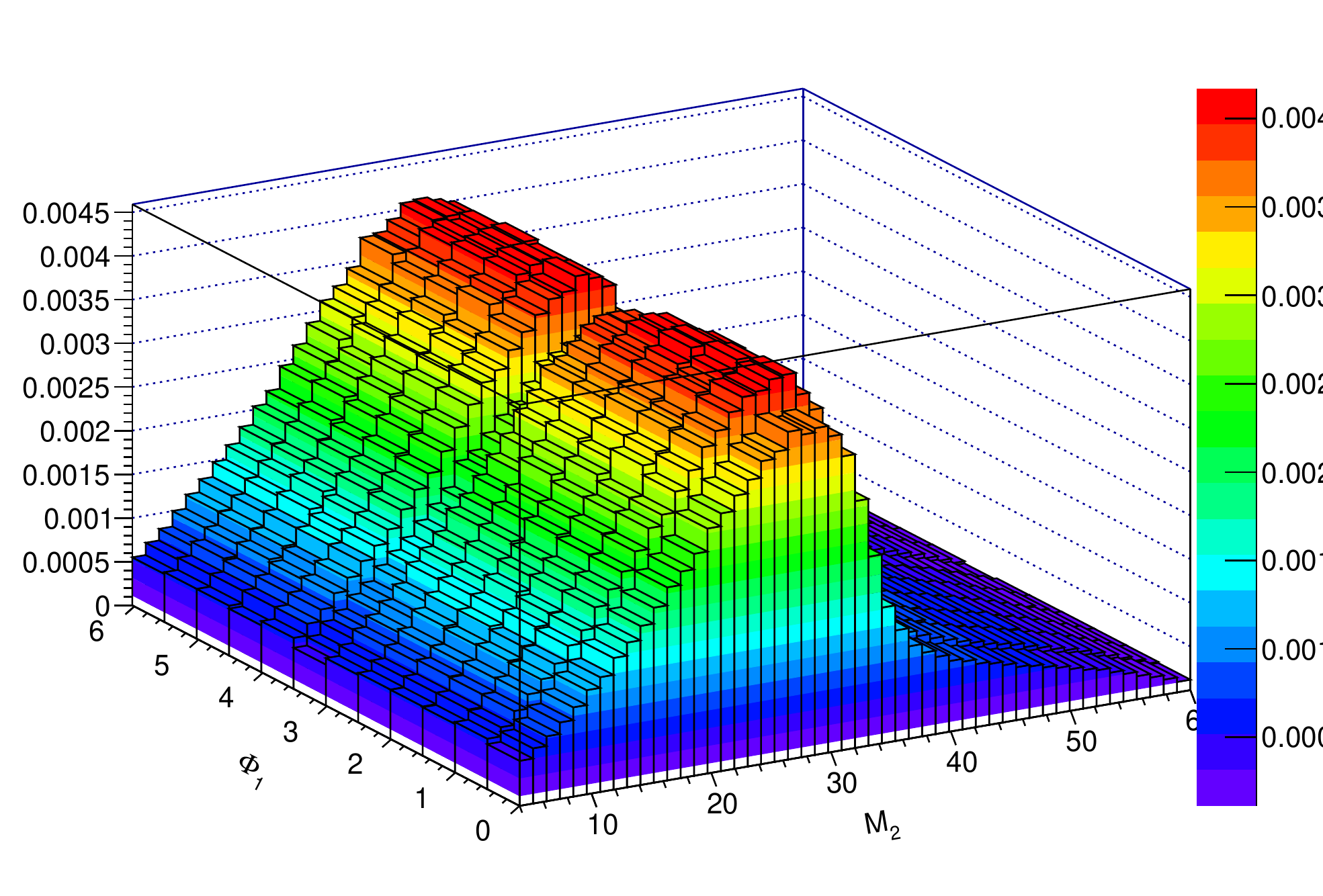}
\includegraphics[width=0.32\textwidth]{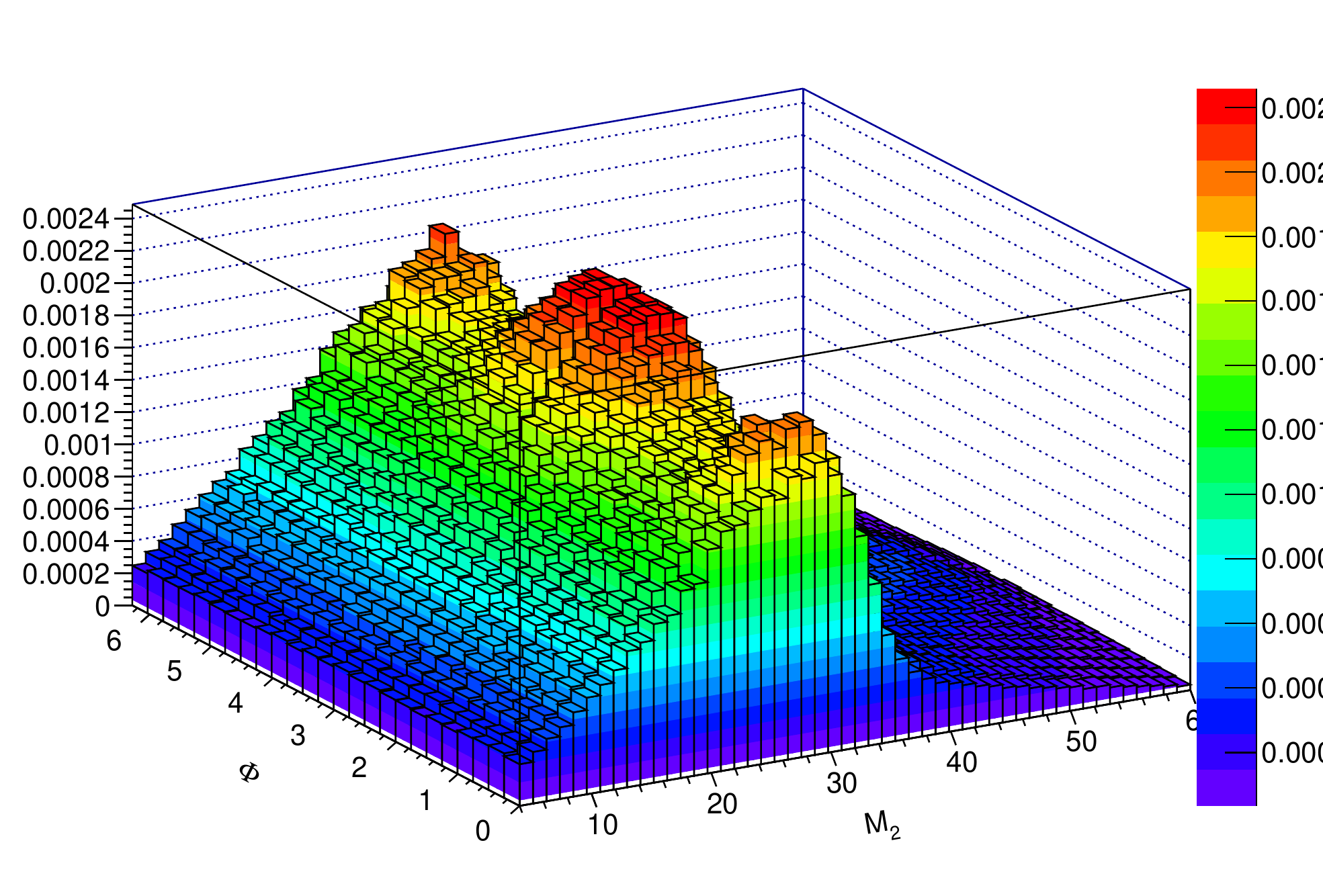}
\includegraphics[width=0.32\textwidth]{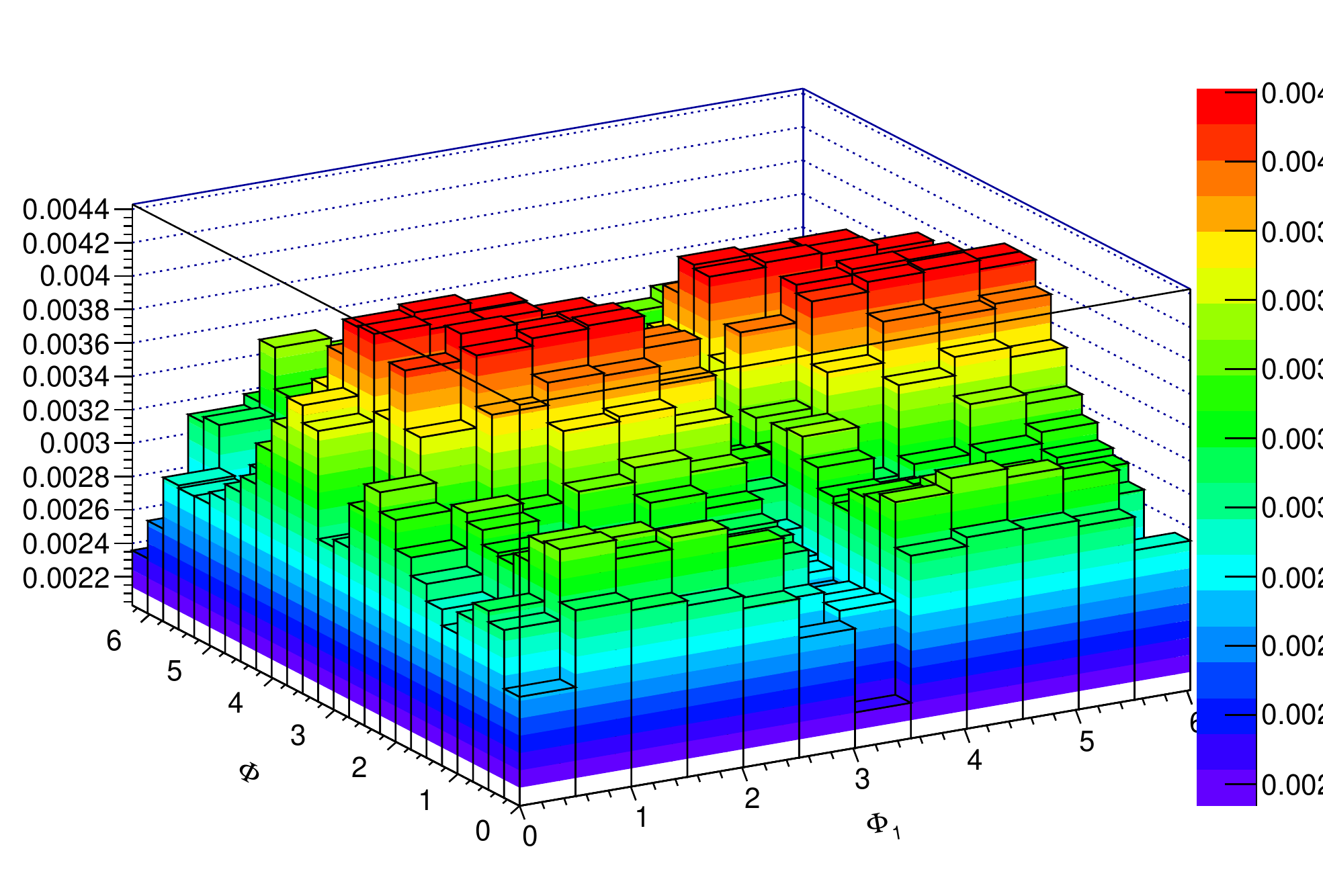}
\caption{The doubly differential spectra for a SM signal assuming only $A_{1ZZ}$ is non-zero (i.e. the SM point).~The phase space is defined as 4~GeV $< M_1 < 120$~GeV and 4~GeV $< M_2 <$ 120~GeV with $\sqrt{s}=125$~GeV and interpreting $M_1 > M_2$ when reconstructing the $4e/4\mu$ final state.~We also take $|\eta_\ell | < 2.4$ and $p_{T\ell} > 2~GeV$.~From left to right and top to bottom we show the $(M_1, M_2)$, $(M_1, \Phi_1)$, $(M_1, \Phi)$, $(M_2, \Phi_1)$, $(M_2, \Phi)$, $(\Phi_1, \Phi)$ doubly differential spectra.}
\label{fig:sig_doubdists}
\end{figure*}
\begin{figure*}
\includegraphics[width=0.32\textwidth]{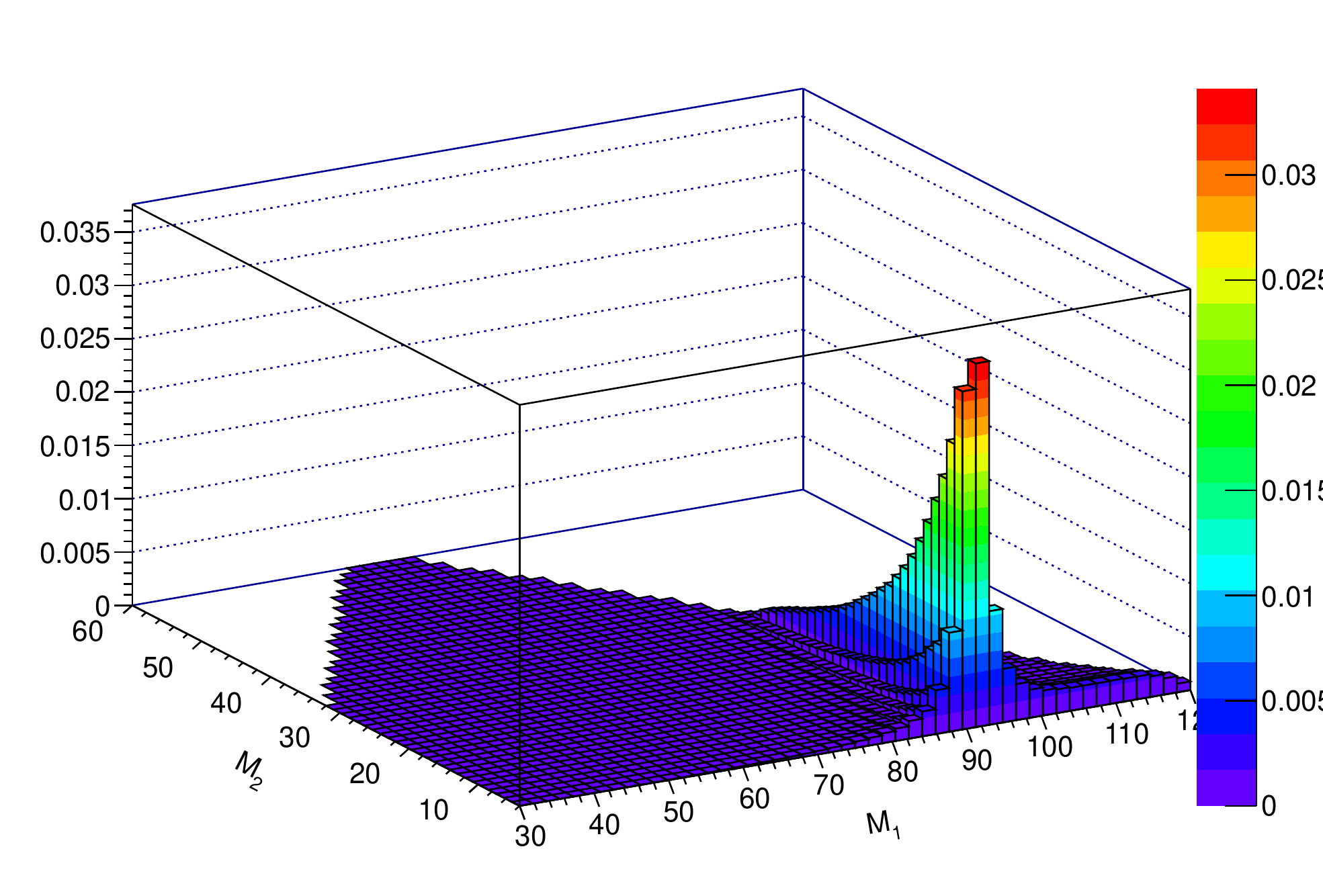}
\includegraphics[width=0.32\textwidth]{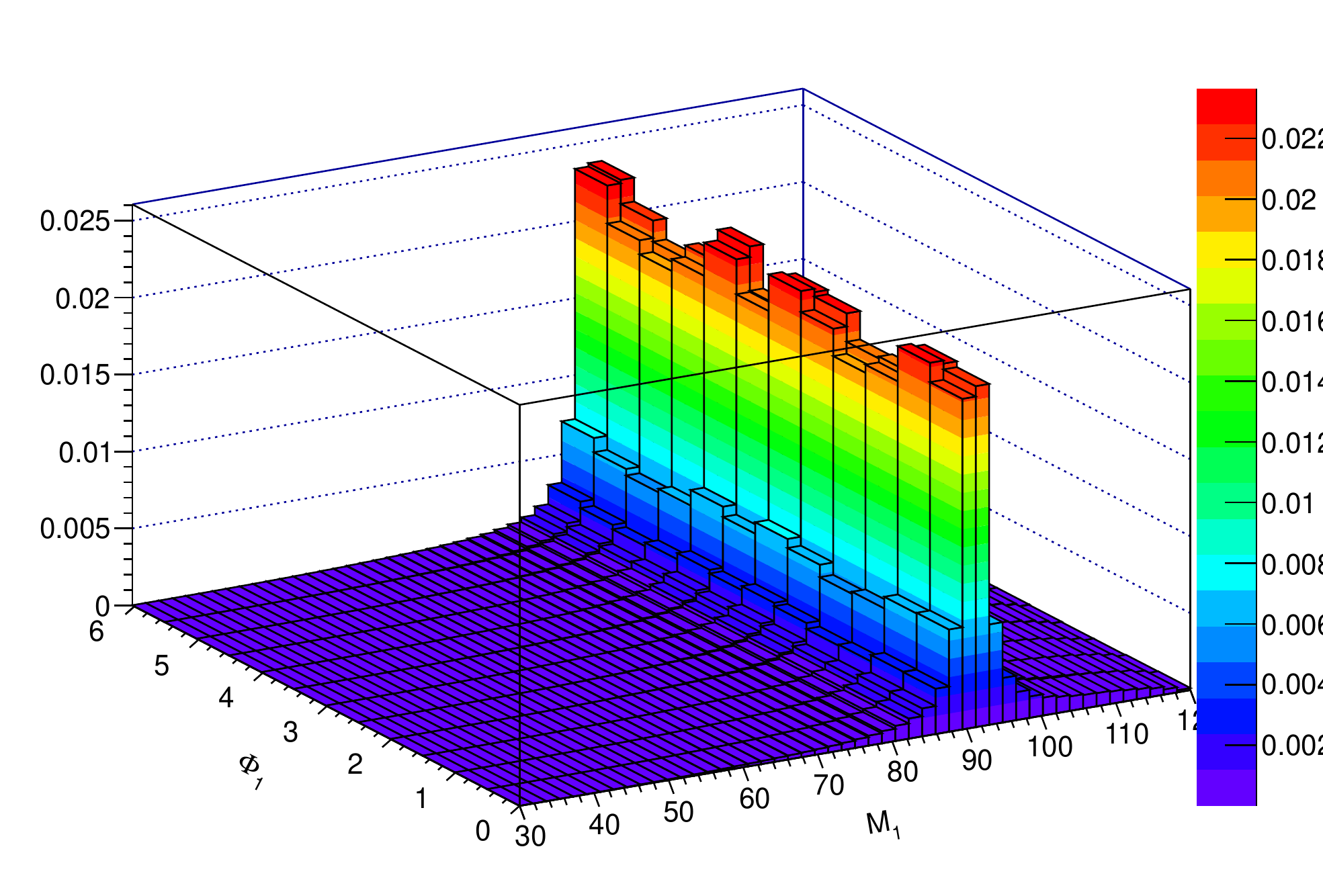}
\includegraphics[width=0.32\textwidth]{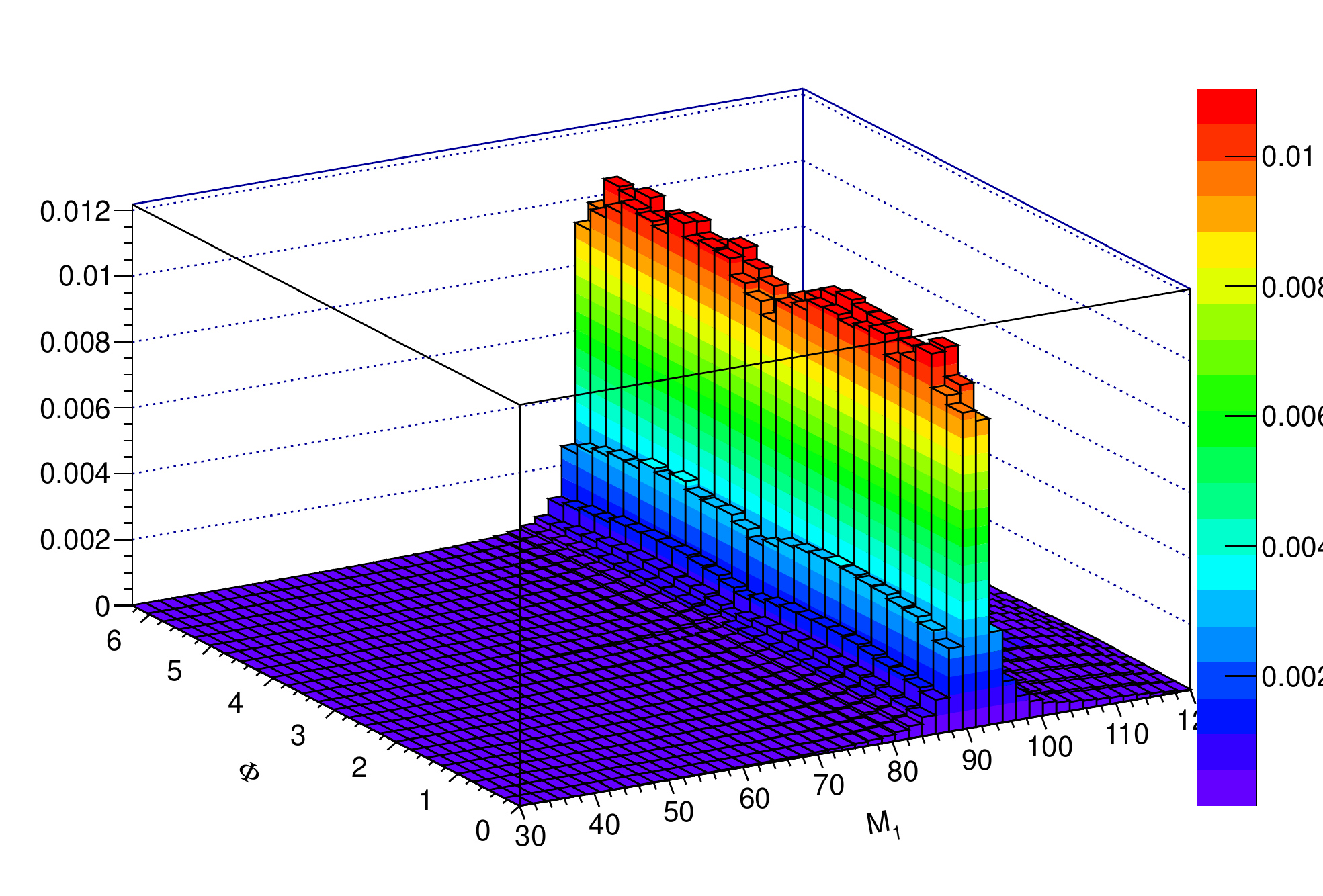}
\includegraphics[width=0.32\textwidth]{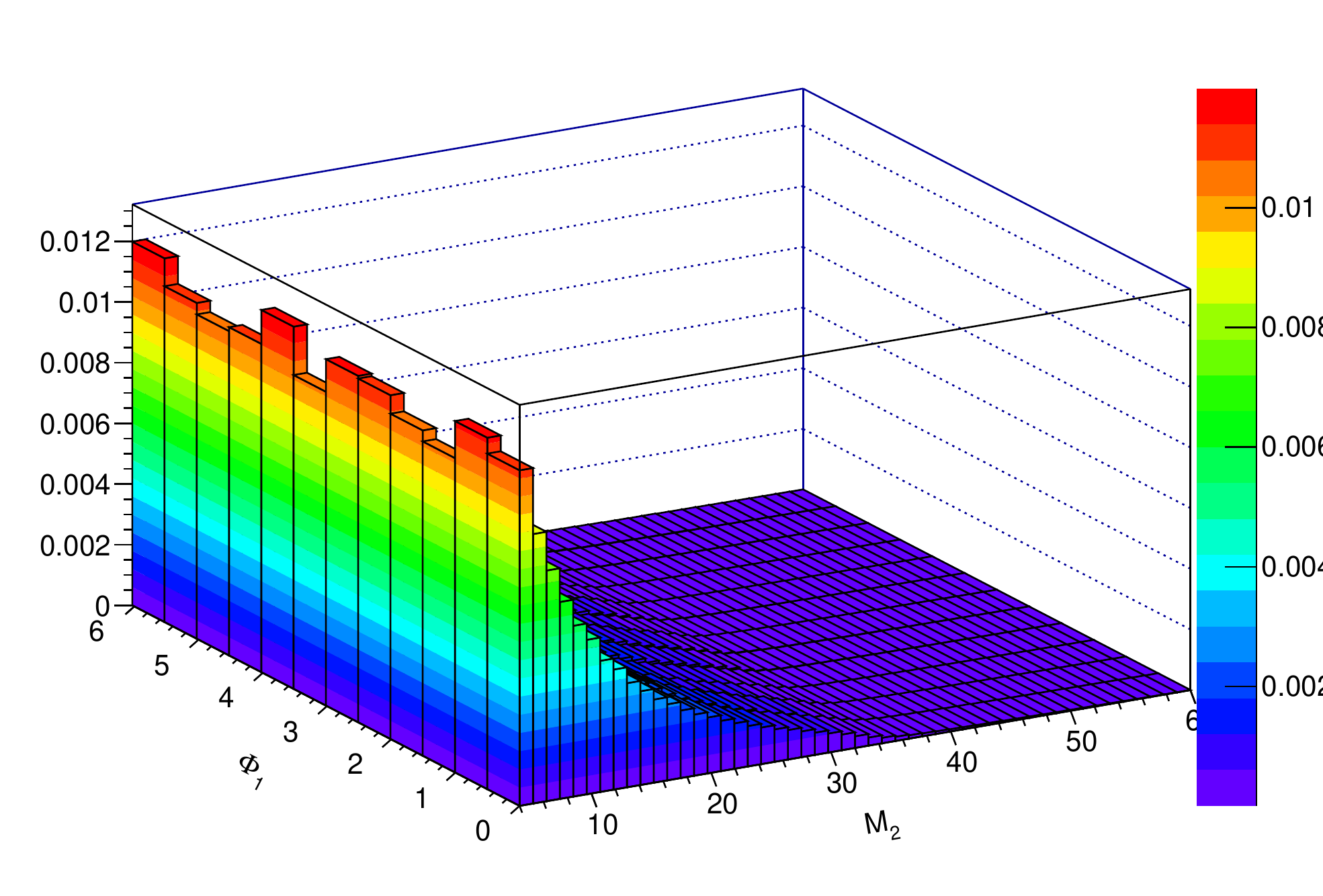}
\includegraphics[width=0.32\textwidth]{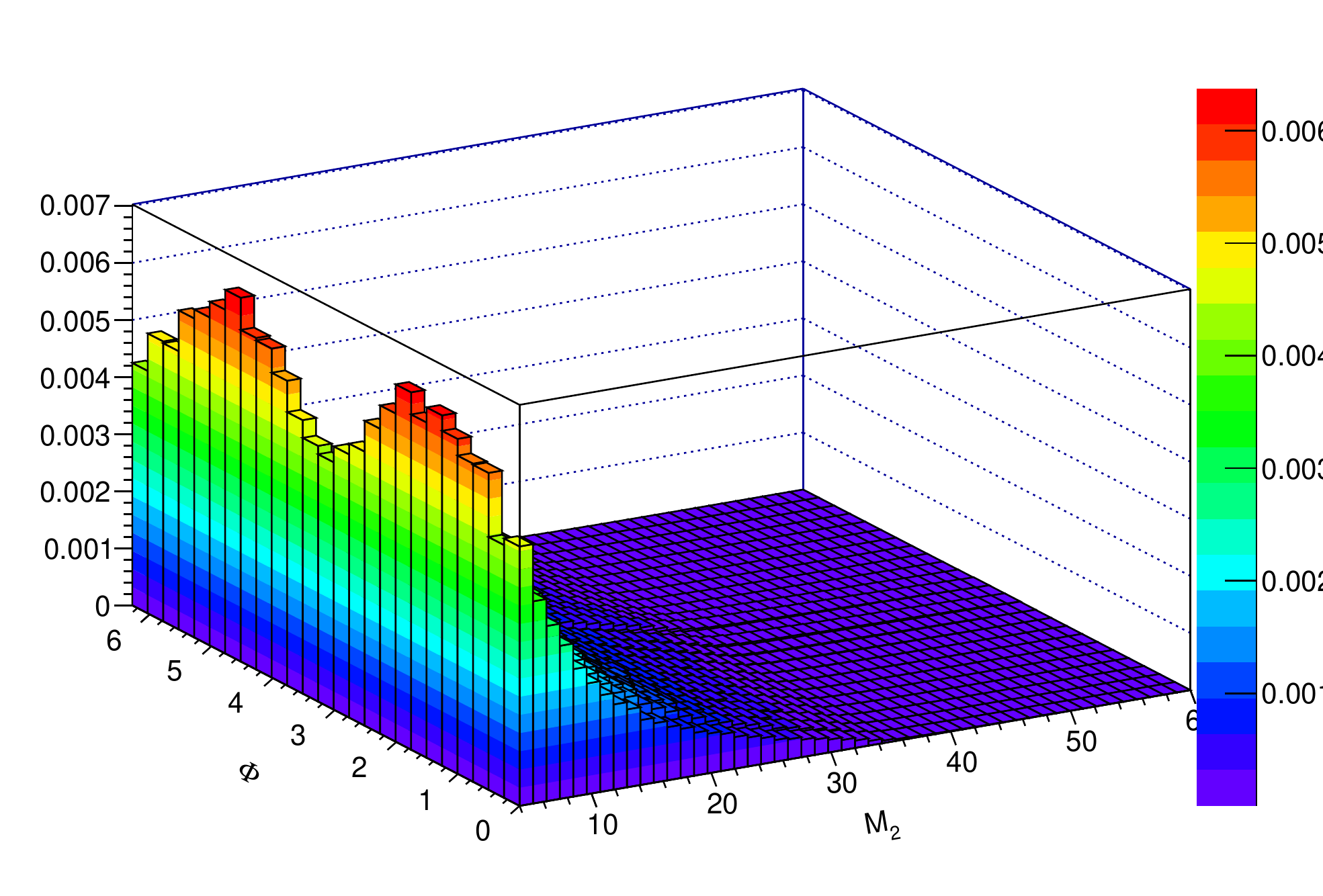}
\includegraphics[width=0.32\textwidth]{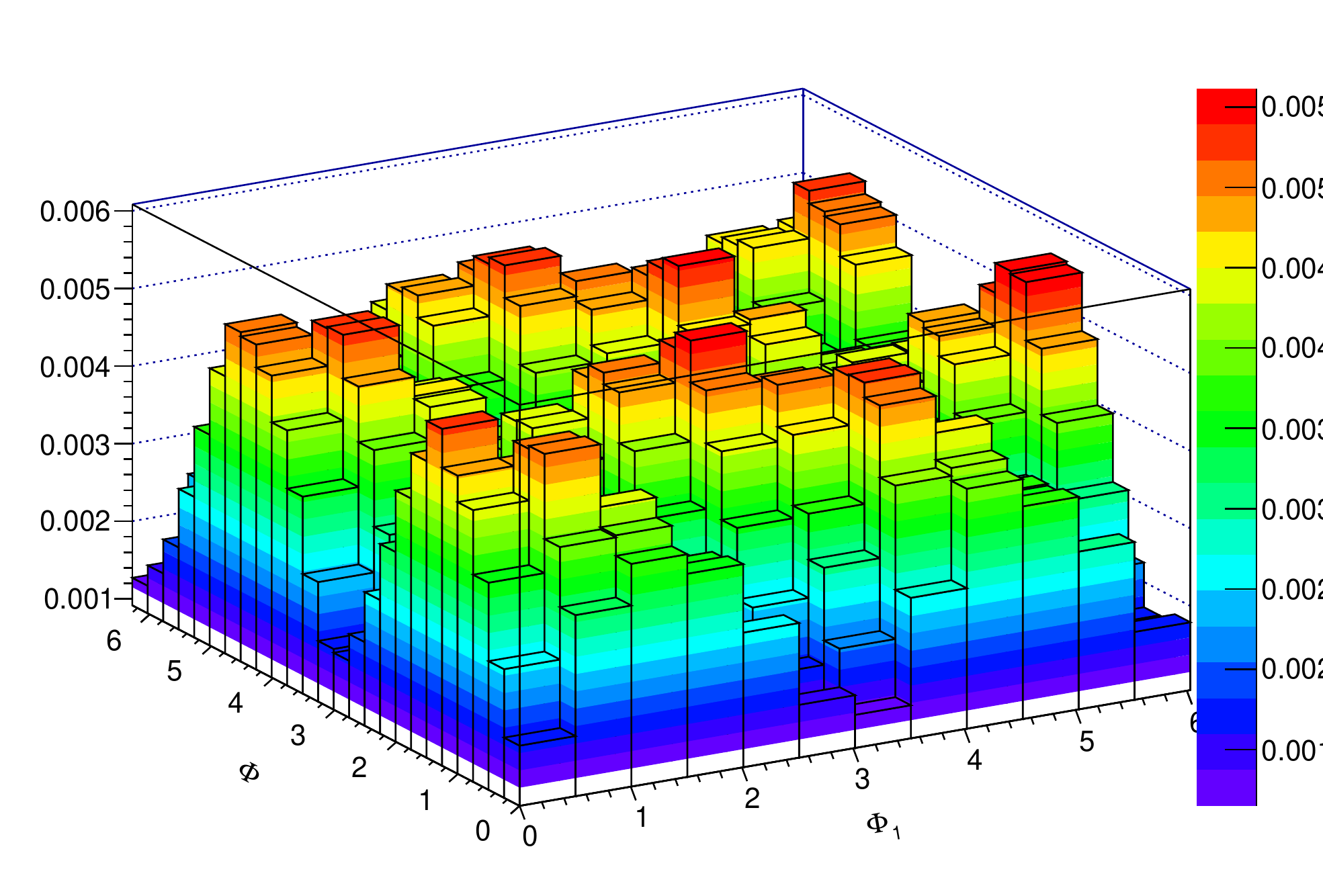}
\caption{The doubly differential spectra for the SM $q\bar{q} \rightarrow 4e$ background.~The phase space is defined as 4~GeV $< M_1 < 120$~GeV and 4~GeV $< M_2 <$ 120~GeV with $\sqrt{s}=125$~GeV and interpreting $M_1 > M_2$ when reconstructing the $4e/4\mu$ final state. We also take $|\eta_\ell | < 2.4$ and $p_{T\ell} > 2~GeV$.~From left to right and top to bottom we show the $(M_1, M_2)$, $(M_1, \Phi_1)$, $(M_1, \Phi)$, $(M_2, \Phi_1)$, $(M_2, \Phi)$, $(\Phi_1, \Phi)$ doubly differential spectra.}
\label{fig:bg_doubdists}
\end{figure*}

\subsection{Validation of Calculations}
\label{sec:validation}
In this section we show a validation of the analytic calculations for the golden channel signal and background.~Both the signal and background are validated against the Madgraph result for the leading order matrix element squared for a large number of random phase space points.~For these comparisons we have generated 100k random phase space points in the range $5~GeV \leq \sqrt{s} \leq 1000~GeV$ so these expressions are valid for essentially any scalar mass and energy range.~We show the validation for the $4e/4\mu$ final state, but as discussed above this is also validates the $2e2\mu$ final state (though it was also explicitly validated in~\cite{Chen:2012jy}) which is simply one term in the $4e/4\mu$ matrix element squared.

We first show in Fig.~\ref{fig:DiffPlotSIG} the validation for the $\varphi \rightarrow 4\ell$ calculation of the matrix element squared obtain in Eq.(\ref{eqn:amp4e_squared}).~We show the validation in two ways.~In the top plot we show the Log($|M|^2$) for a large number of random phase space points and plot the two results on top of one another.~The Madgraph result is shown in red while the analytic result is shown in yellow.~The two results are indistinguishable from one another and thus the two distributions sit on top of each other leading to the orange color seen.

In the bottom plot we show the fractional difference in their matrix elements squared for the same set of phase space points.~The agreement is perfect up to very tiny differences due to numerical precision when evaluating the matrix elements squared for specific phase space points.~To obtain the matrix element squared from Madgraph we have implemented the Lagrangian in Eq.(\ref{eqn:siglag}) (or equivalently vertex in Eq.(\ref{eqn:vertex})) into the FeynRules/Madgraph~\cite{Christensen:2008py,Alwall:2007st} framework.~We have chosen all of the $A_{nij}$ couplings to have random non-zero values for both their real and imaginary parts.~Thus the complete expression including all tensor structures in Eq.(\ref{eqn:vertex}) and their interference has been validated.~One can now easily obtain any expression which includes a subset of the possible tensor structures by simply setting the unwanted $A_{nij}$ to zero.

In Fig.~\ref{fig:DiffPlotBG} we show the same validations for the leading order $q\bar{q} \rightarrow 4\ell$ background again validated against the Madgraph result.~Again we see essentially perfect agreement.~See also~\cite{Chen:2012jy} for how the different components of the $q\bar{q} \rightarrow 4\ell$ depend on $\sqrt{s}$.~We also provide there an analytic expression for the dominant component which has been integrated over all angles and can be used for performing simplified studies.

For further valuations of the signal and background expressions including the phase space, see~\cite{Chen:2012jy},\cite{Chen:2014pia}, and~\cite{TechNote} where the various 1D projections are shown.~In~\cite{WEBSITE} we also provide the various expression available for download.
\begin{figure}
\includegraphics[width=0.48\textwidth]{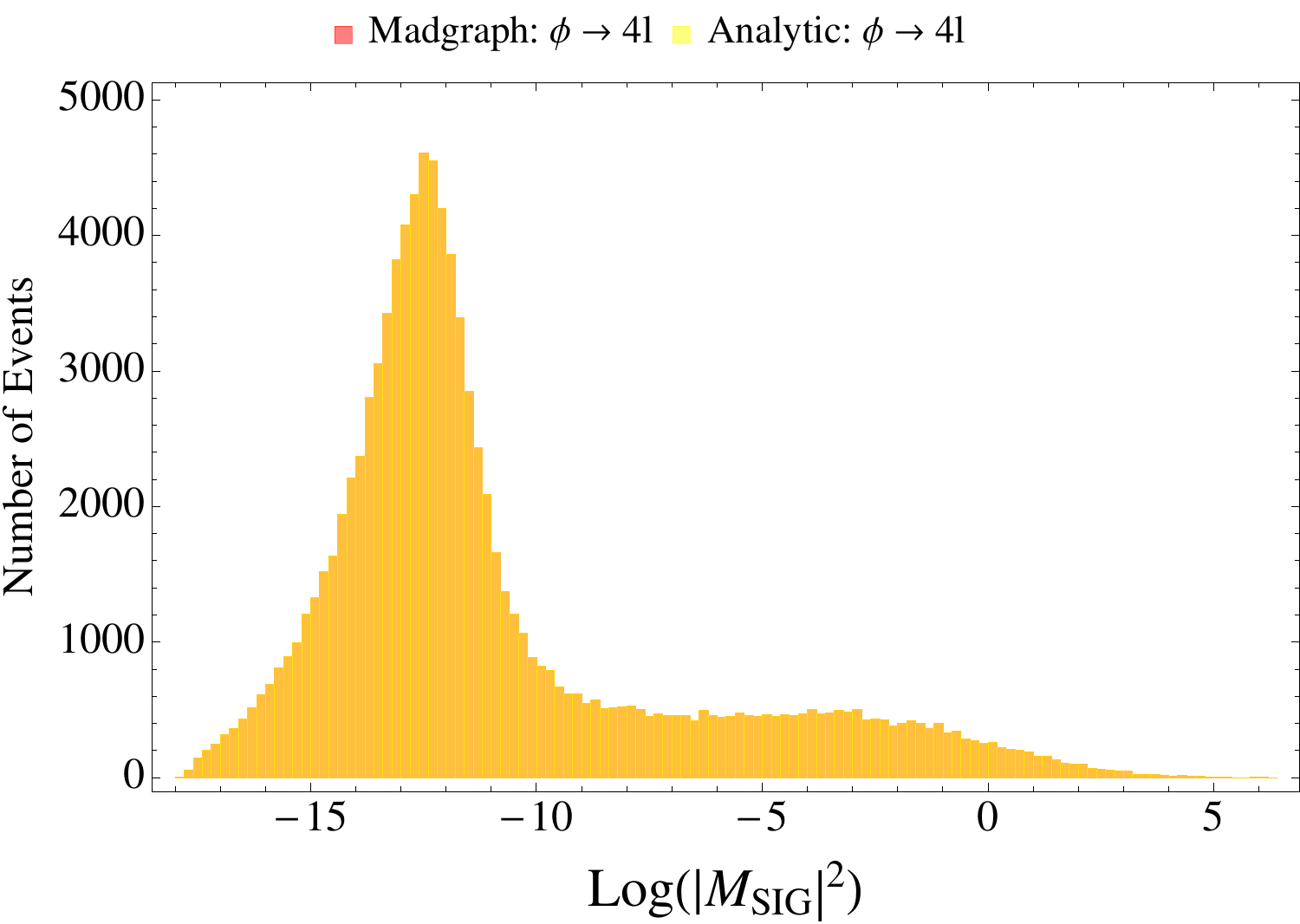}\\
~\\
\includegraphics[width=0.48\textwidth]{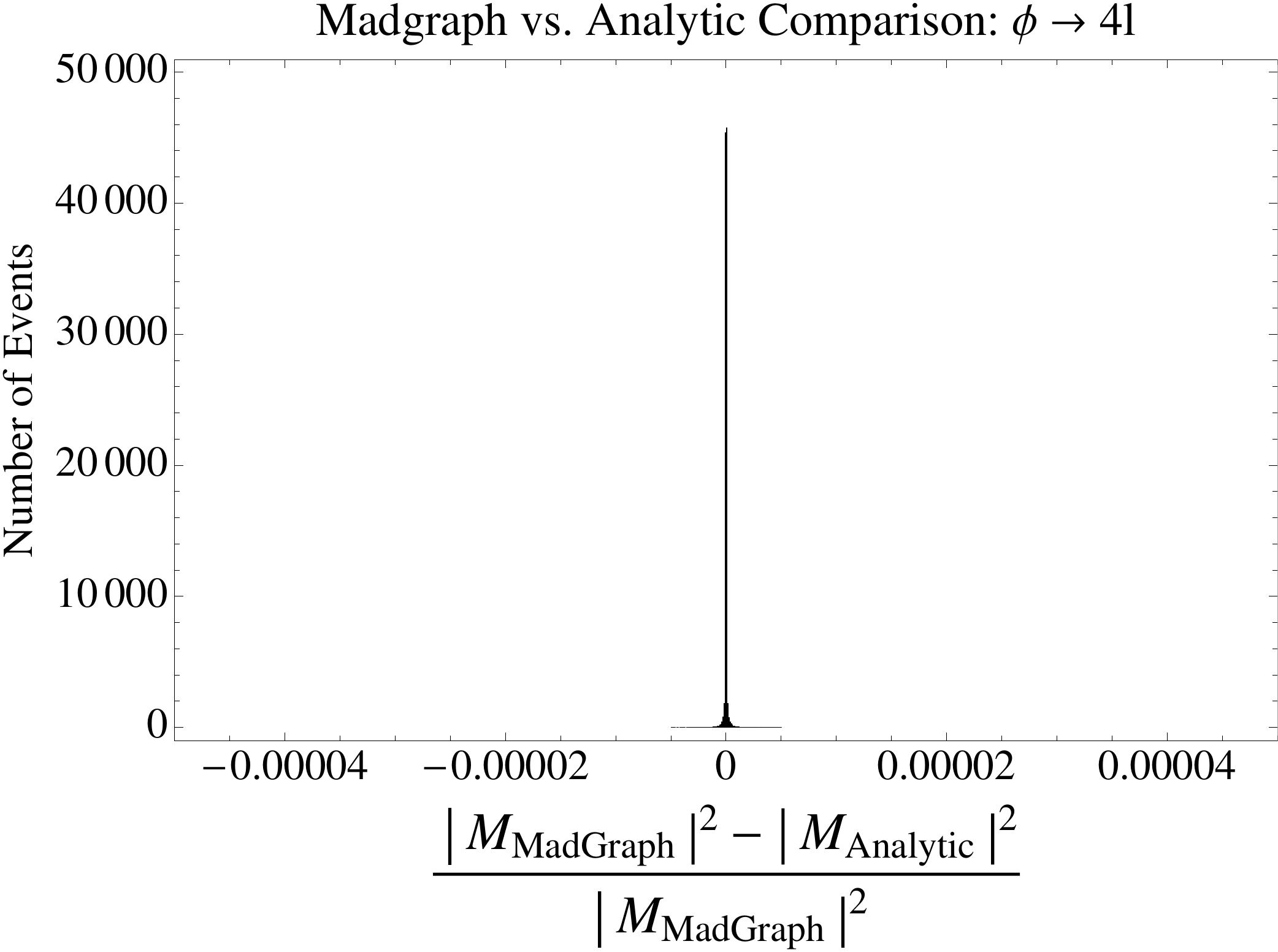}
\caption{Here we compare the Madgraph matrix element squared and the analytic result obtained in Eq.(\ref{eqn:amp4e_squared}) for the $\varphi \rightarrow 4\ell$ signal.~In the top plot shot we show the Log($|M|^2$) for a large number of random phase space points and plot the two results on top of one another.~The Madgraph result is shown in red while the analytic result is shown in yellow.~The two results are indistinguishable from one another and thus the two distributions sit on top of each other leading to the orange color seen.~In the bottom plot we show the fractional difference in their matrix elements squared for the same set of phase space points.~The tiny differences seen are due to numerical precision when evaluating the matrix elements squared for specific phase space points.}
\label{fig:DiffPlotSIG}
\end{figure}

\begin{figure}
\includegraphics[width=0.48\textwidth]{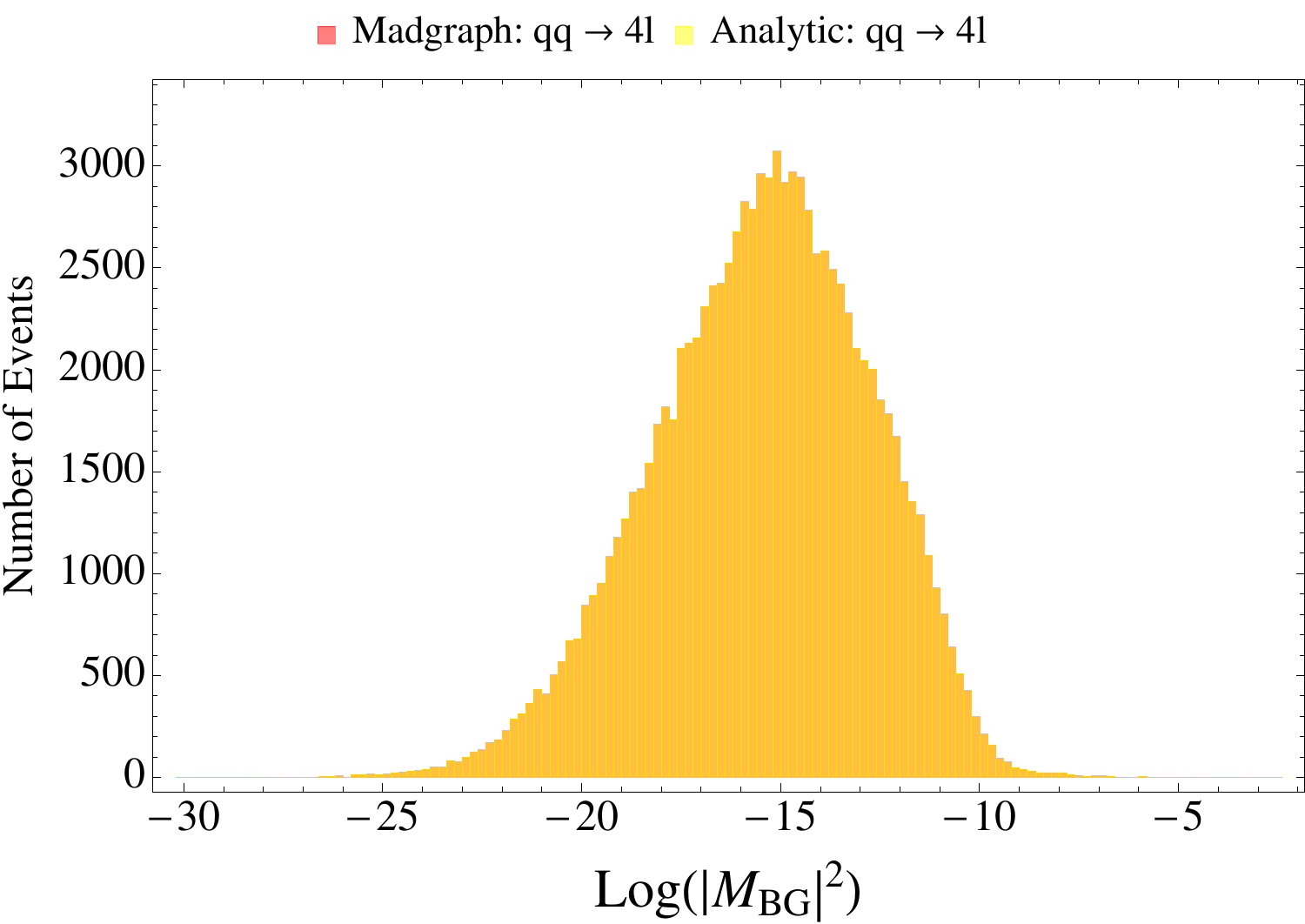}\\
~\\
\includegraphics[width=0.48\textwidth]{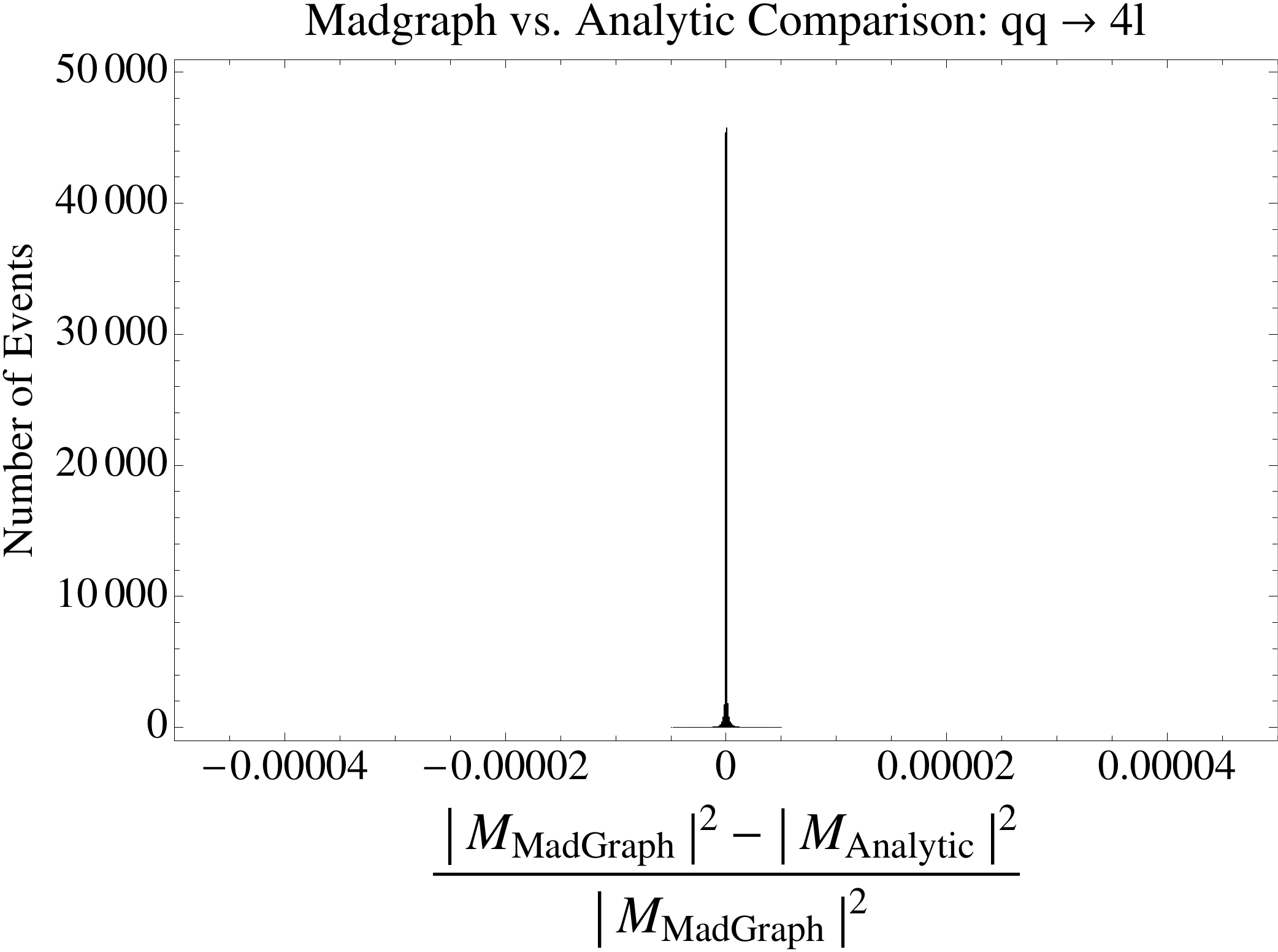}
\caption{Here we compare the Madgraph matrix element squared and the analytic result obtained in Eq.(\ref{eqn:finalbg4eamp2}) for the $q\bar{q} \rightarrow 4\ell$ background.~In the top plot shot we show the Log($|M|^2$) for a large number of random phase space points and plot the two results on top of one another.~The Madgraph result is shown in red while the analytic result is shown in yellow.~The two results are indistinguishable from one another and thus the two distributions sit on top of each other leading to the orange color seen.~In the bottom plot we show the fractional difference in their matrix elements squared for the same set of phase space points.~The tiny differences seen are due to numerical precision when evaluating the matrix elements squared for specific phase space points.}
\label{fig:DiffPlotBG}
\end{figure}

\clearpage

\bibliographystyle{apsrev}
\bibliography{GoldenChannelBib}

\end{document}